\documentclass[twocolumn]{aastex63}

\usepackage{graphicx}
\usepackage{amsmath}	
\usepackage{amssymb}
\usepackage{booktabs}
\usepackage{enumerate}
\usepackage{dirtytalk}
\usepackage{float}
\usepackage{enumitem}
\usepackage{booktabs}
\usepackage{lipsum} 
\usepackage{float}
\usepackage{xcolor}
\usepackage{multirow}
\usepackage{makecell}
\usepackage{blindtext}

\graphicspath{{./}{figures/}}

\accepted{March 11, 2021}

\submitjournal{ApJ}

\shorttitle{Aurora: A Generalised Retrieval Framework}
\shortauthors{Welbanks \& Madhusudhan}

\begin{document}

\title{Aurora: A Generalised Retrieval Framework for Exoplanetary Transmission Spectra}

\author[0000-0003-0156-4564]{Luis Welbanks}
\email{luis.welbanks@ast.cam.ac.uk}
\affil{Institute of Astronomy, University of Cambridge, Madingley Road Cambridge CB3 0HA, UK}

\author[0000-0002-4869-000X]{Nikku Madhusudhan}
\email{nmadhu@ast.cam.ac.uk}
\affil{Institute of Astronomy, University of Cambridge, Madingley Road Cambridge CB3 0HA, UK}

\begin{abstract}
Atmospheric retrievals of exoplanetary transmission spectra provide important constraints on various properties such as chemical abundances, cloud/haze properties, and characteristic temperatures, at the day-night atmospheric terminator. To date, most spectra have been observed for giant exoplanets due to which retrievals typically assume Hydrogen-rich atmospheres. However, recent observations of mini-Neptunes/super-Earths, and the promise of upcoming facilities including JWST, call for a new generation of retrievals that can address a wide range of atmospheric compositions and related complexities. Here we report \textsl{Aurora}, a next-generation atmospheric retrieval framework that builds upon state-of-the-art architectures and incorporates the following key advancements: (a) a generalised compositional retrieval allowing for H-rich and H-poor atmospheres, (b) a generalised prescription for inhomogeneous clouds/hazes, (c) multiple Bayesian inference algorithms for high-dimensional retrievals, (d) modular considerations for refraction, forward scattering, and Mie-scattering, and (e) noise modelling functionalities. We demonstrate \textsl{Aurora} on current and/or synthetic observations of hot Jupiter HD~209458~b, mini-Neptune K2-18b, and rocky exoplanet TRAPPIST-1~d. Using current HD~209458~b spectra, we demonstrate the robustness of our framework and cloud/haze prescription against assumptions of H-rich/H-poor atmospheres, improving on previous treatments. Using real and synthetic spectra of K2-18b, we demonstrate the agnostic approach to confidently constrain its bulk atmospheric composition and obtain precise abundance estimates. For TRAPPIST-1~d, 10 JWST NIRSpec transits can enable identification of the main atmospheric component for cloud-free CO$_2$-rich and N$_2$-rich atmospheres, and abundance constraints on trace gases including initial indications of O$_3$ if present at enhanced levels ($\sim10-100\times$ Earth levels).

\end{abstract}


\section{Introduction}
\label{sec:intro}

The last decade witnessed a revolution in our understanding of exoplanets and the nature of their atmospheres. Since the detection of the first atmosphere of a transiting exoplanet \citep{Charbonneau2002}, spectroscopic observations of exoplanets have provided wide ranging insights into the compositions, temperature structures, and physical processes in their atmospheres \citep[see e.g.,][for a review]{Seager2010, Madhusudhan2019, ZhangXi2020}. Most of the atmospheric observations have been made using transmission spectroscopy which is conducted when an exoplanet transits in front of its host star and light from the star passes through the planet's atmosphere before reaching the observer \citep{Seager2000}. A transmission spectrum, through a wavelength-dependent change in the apparent size of the planet, encodes information about the atmosphere at the day-night terminator of the planet. Particularly, transmission spectroscopy has been key in detecting and quantifying the abundances of multiple molecules and atoms \citep[e.g.,][]{Charbonneau2002, Deming2013, Madhusudhan2014a, Kreidberg2014b, Wyttenbach2015, Sedaghati2017, Nikolov2018, Chen2018, Wakeford2018}, as well as providing important insight into clouds/hazes in exoplanetary atmospheres \citep[e.g.,][]{Pont2008, Sing2016, Nikolov2018, Benneke2019a}. 

Atmospheric spectra of exoplanets are routinely interpreted using retrieval methods. Introduced in \citet{Madhusudhan2009}, atmospheric retrievals of exoplanets aim to solve the inverse problem - to obtain statistical constraints on the atmospheric properties of an exoplanet from an observed spectrum. A retrieval code is composed of a parametric atmospheric model that computes a synthetic spectrum, coupled with an optimisation algorithm that derives the model parameters given the observed spectrum. Although here we focus on retrieval codes for transmission spectroscopy, as discussed below, a plethora of retrieval codes have been developed for other applications \citep[see e.g.,][ for a recent review]{Madhusudhan2018}. Retrieval codes have been developed for the analysis of thermal emission spectra of exoplanets \citep[e.g.,][]{Madhusudhan2009,Madhusudhan2011,Lee2012,Line2013a,Line2014a, Waldmann2015b,Gandhi2018,Brogi2019,Gandhi2019}, phase curves \citep[e.g.,][]{Irwin2020,Changeat2020b,Feng2020}, spectra of directly imaged exoplanets \citep[e.g.,][]{Lee2013, Barstow2014, Lupu2016, Nayak2017, Lavie2017,Damiano2020}, as well as spectra of brown dwarfs \citep[e.g.,][]{Line2014b,Line2015,Burningham2017,Zalesky2019,Kitzmann2020,Piette2020} and solar system planets \citep[e.g.,][]{Rodgers2000, Irwin2001, Irwin2002, Irwin2008, Irwin2014}. 

Retrievals of transmission spectra have become ubiquitous in atmospheric
characterisation studies \citep[see][for a review]{Madhusudhan2018}. The first retrieval code for exoplanetary atmospheres \citep{Madhusudhan2009} performed a grid-based parameter exploration using a large model grid ($\sim$10$^7$ models of 10 parameters each). Subsequent studies adopted more robust statistical optimisation algorithms. The next iteration of retrieval codes used Markov Chain Monte Carlo (MCMC) methods \citep[e.g.,][]{Tegmark2004, Foremanmackey2013}, providing a better parameter exploration of the parameter space but with limitations in calculating the model evidence for model comparison. Retrieval codes utilising MCMC methods include \citet{Madhusudhan2011, Benneke2012, Madhusudhan2014a}, CHIMERA \citep[e.g.,][]{Line2013a, Kreidberg2014b}, MassSpec \citep{DeWit2013}, ATMO \citep[e.g.,][]{Wakeford2017, Evans2017}, BART \citep[e.g.,][]{Cubillos2016, Blecic2016}, PLATON \citep{Zhang2019}, and METIS \citep{Lacy2020}. Concurrently, the retrieval code NEMESIS \citep{Irwin2008} developed for solar system studies using gradient-descent optimisation methods, such as Optimal Estimation (OE), has also seen applications to exoplanetary transmission spectra \citep[e.g.,][]{Barstow2017}.

The next generation of retrieval codes came to light with the implementation of the nested sampling algorithm \citep[e.g.,][]{Skilling2006}, facilitating more efficient parameter space exploration and calculation of model evidence. Transmission retrieval codes like SCARLET \citep[e.g.,][]{Benneke2013,Benneke2019a, Benneke2019b}, $\mathcal{T}$-REx \citep[e.g.,][]{Waldmann2015a}, POSEIDON \citep[e.g.,][]{MacDonald2017}, AURA \citep[e.g.,][]{Pinhas2018}, petitRADTRANS \citep{Molliere2019} amongst others \citep[e.g.,][]{Fisher2018, Fisher2019, Brogi2019, Seidel2020, Min2020} adopted the MultiNest nested sampling algorithm \citep{Feroz2009}. Although MultiNest has been extensively used, other nested sampling algorithms have been implemented like Nestle \citep{nestle} in PLATON, Dynesty \citep{Speagle2020} in PLATON II \citep{Zhang2020}, and PolyChord \citep{Handley2015a} in $\mathcal{T}$-REx III \citep{Al-Refaie2019}. 

The extensive availability of computational methods and packages for statistical inference has made it possible for retrieval codes to update their capabilities and include multiple optimisation algorithms. For instance CHIMERA has used OE, Bootstrap Monte Carlo (BMC), MCMC, as well as MultiNest nested sampling \citep[e.g.,][]{Line2013a, Colon2020}. NEMESIS has been adapted, beyond OE, to use MultiNest nested sampling \citep[e.g.,][]{Krissansen-Totton2018}. Similarly, $\mathcal{T}$-REx through different updates has used MCMC and diverse nested sampling algorithms \citep[e.g.,][]{Waldmann2015b, Al-Refaie2019}.

Parallel efforts are being made towards exploring the viability of machine learning algorithms as a replacement or aid to traditional Bayesian optimisation algorithms. Some studies have used the Random Forest algorithm \citep[e.g.,][]{Breiman1984} to train estimators and predict the parameters that better explain an observed spectrum \citep[e.g.,][]{Marquez-Neila2018,Fisher2020,Guzman-Mesa2020, Nixon2020}. Other studies have used Deep Belief Neural Networks \citep[albeit in studies of emission spectroscopy, e.g.,][]{Waldmann2016}, Generative Adversarial Networks \citep{Zingales2018}, Deep Neural Networks \citep{Soboczenski2018}, and Bayesian Neural Networks \citep{Cobb2019} in efforts to predict atmospheric properties of exoplanets. A complementary approach has been to use machine learning to help inform the priors in a traditional retrieval \citep[e.g.,][]{Hayes2020}. These advancements in retrievals are an active area of research and future work may elucidate on the synergies between traditional retrievals and these novel machine learning techniques.

There have also been developments in model considerations for atmospheric retrievals of transmission spectra. Recent works have investigated the impact of cloud and hazes in atmospheric retrievals \citep[e.g.,][]{Line2016,MacDonald2017,Pinhas2018,Mai2019, Barstow2020a}. Similarly, studies have investigated the relative importance of various model and data considerations, including temperature structures, clouds, and optical data \citep[e.g.,][]{Welbanks2019a} over simpler isobaric, isothermal, semi-analytic model assumptions. Other works have looked into the effect of uncertainties in the system parameters \citep[e.g.,][]{DeWit2013, Fisher2018, Batalha2019,Changeat2020a} or temperature and abundance inhomogeneities \citep{Caldas2019,Changeat2019, MacDonald2020} in the retrieved properties of the atmosphere. Further efforts have investigated the impact of stellar contamination in transmission spectra \citep[e.g.,][]{Pinhas2018,Iyer2020,Bruno2020}. 

While the studies above have focused primarily on retrievals for giant planets with H-rich atmospheres, some studies have also developed retrieval frameworks for smaller planets where the atmosphere may not be assumed to be H-rich a priori. \citet{Benneke2012} investigated an agnostic retrieval framework for super-Earths, which could have a wide range of atmospheric compositions. They highlight that assuming log-uniform priors for the mixing ratios of the chemical species sampled in a retrieval can lead to a highly asymmetric prior for the last species derived using the unit sum constraint, which is unfavourable in the absence of a priori knowledge of the dominant species in the atmosphere, e.g., for super-Earths. To overcome this limitation, \citet{Benneke2012} suggest a reparameterization of the chemical compositions that is applicable to both H-rich and non H-rich atmospheres. The parameterization, based on centered-log-ratio transformations \citep[e.g.,][]{Aitchison1986}, allows for equal prior probability distributions for all the chemical species considered; we discuss this in depth in section~\ref{subsec:reparameterization}. In subsequent work, \citet{Benneke2013} demonstrate the potential of using Bayesian model comparisons along with high-precision transmission spectra of super-Earths/mini-Neptunes to differentiate between cloudy H-rich atmospheres and those of high mean-molecular weight, e.g., H$_2$O-rich. 

After this decade of revolutionary work on retrievals, the next generation of retrieval codes is upon us. Such retrievals must incorporate the lessons learned from atmospheric studies of giant exoplanets and also be adaptable to low-mass planets. In preparation for upcoming observations of temperate mini-Neptunes and super-Earths, the methods for non H-dominated atmospheres must be implemented and updated to be compatible with the latest optimisation algorithms. Upcoming codes should be able to expand their modelling functionalities motivated by data requirements. Lastly, with the increasing model complexity and data quality, new retrieval codes must be prepared for assessing multidimensional, highly degenerate problems. 

We introduce \textsl{Aurora}, a next-generation retrieval framework for the atmospheric characterisation of H-rich and non H-rich planets. Our code incorporates new key features on the previous retrieval code AURA \citep[][]{Pinhas2018}. First, we reparameterise the volume mixing ratios in the atmosphere expanding the previous scope beyond H-rich atmospheres, adapting methods previously used for super-Earths \citep[e.g.,][]{Benneke2012} and other areas of compositional data analysis \citep[e.g.,][]{Aitchison1986, Aitchison2005, Pawlowsky-Glahn2011}. Second, \textsl{Aurora} incorporates the next-generation nested sampling algorithms PolyChord and Dynesty, as well as maintaining compatibility with MultiNest. Third, \textsl{Aurora} includes a new generalised parametric treatment for inhomogeneous clouds and hazes. Compared to previous prescriptions, our new treatment of clouds/hazes is robust against assumptions of whether the atmosphere is H-rich or not. 

Lastly, \textsl{Aurora} incorporates different modular capabilities that enhance the study of transmission spectra using retrievals and forward models. These include assessing stellar heterogeneity \citep[e.g.,][]{Rackham2017,Pinhas2018}, allowing for underestimated variances in the data \citep[e.g.,][]{Foremanmackey2013, Colon2020}, and considering correlated noise using Gaussian processes \citep[e.g.,][]{Rasmussen2006}. Additionally, our forward modelling capabilities can account for light refraction and forward scattering \citep{Robinson2017}, as well as Mie-scattering due to a variety of condensate species \citep{Pinhas2017}. \textsl{Aurora}'s modular capabilities can be incorporated in retrievals should observations require it. 

In what follows, we present our retrieval framework in section \ref{sec:methods}. We benchmark the results of \textsl{Aurora} on current and synthetic observations in section \ref{sec:results}, and present case studies for characterising atmospheres of hot Jupiters, mini-Neptunes, and rocky exoplanets with JWST. We summarise our conclusions in section \ref{sec:conclusions} and discuss the implications of our findings and possible avenues for future developments of \textsl{Aurora}. 

\section{Aurora Retrieval Framework}
\label{sec:methods}

\begin{figure*}
\includegraphics[width=\textwidth]{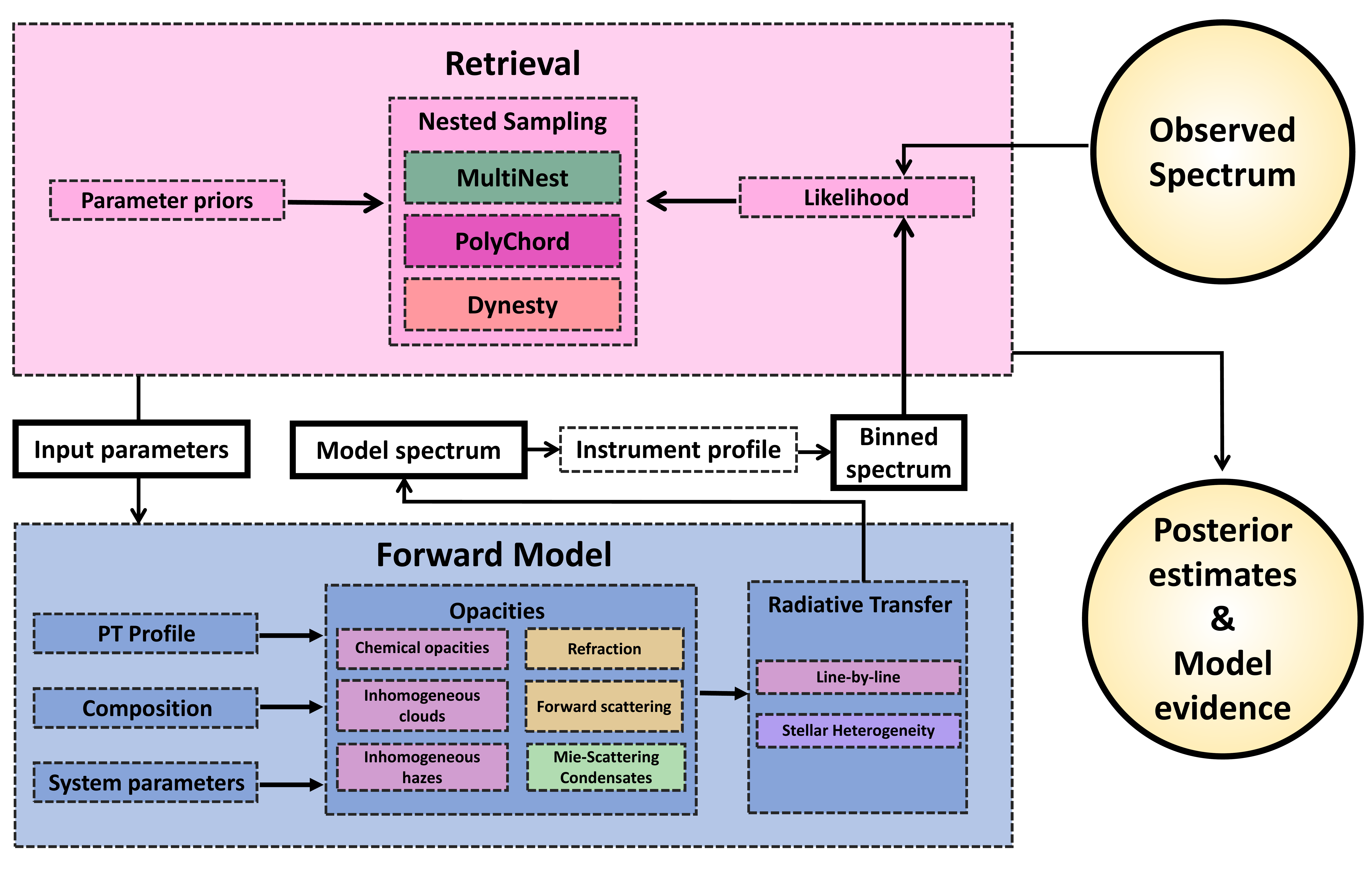}
\centering
\caption[Flow chart]{Schematic of \textsl{Aurora}'s retrieval framework. The retrieval framework combines an atmospheric forward model with a Bayesian inference and parameter estimation algorithm to derive the model evidence and posterior probability distributions of the model parameters given an observed spectrum. \textsl{Aurora} also has modular capabilities for including additional effects in the atmospheric model if required, e.g., light refraction, forward scattering and Mie-scattering.} 
\label{fig:flow}
\end{figure*}

\textsl{Aurora} builds upon the AURA retrieval framework \citep{Pinhas2018} developed in our group and, among other features, expands the retrieval capabilities to rocky exoplanets without the assumption of H-rich atmospheres. The core retrieval methodology for H-rich atmospheres is explained in \citet{Pinhas2018}, with its implementation previously explained in \citet{Welbanks2019a} and employed in different retrieval studies \citep[e.g.,][]{Chen2018, vonEssen2019, Welbanks2019b, Colon2020}. Here, we reintroduce the basic retrieval methodology of the AURA code and discuss the new enhancements made in \textsl{Aurora}. Figure~\ref{fig:flow} shows the schematic diagram of the \textsl{Aurora} framework. 

Like any contemporary retrieval framework \citep{Madhusudhan2018}, AURA and \textsl{Aurora} comprise of a forward model that is interfaced with a Bayesian inference and parameter estimation scheme. The forward model computes a transmission spectrum given a set of parameters for the temperature structure, chemical composition, and presence of clouds/hazes on the planet's atmosphere. The parameter estimation scheme explores the model's parameter space in search of regions of high-likelihood that can explain a set of observations. The Bayesian inference scheme estimates the model evidence and posterior probability distributions of the model parameters, and is performed using the nested sampling algorithm. In what follows, we describe each of these components for our retrieval framework. We highlight the following key advancements introduced in \textsl{Aurora}. 

\begin{itemize}
\item{Generalised inhomogeneous cloud/haze parameterisation}
\item{Generalised considerations for H-poor/H-rich compositions} 
\item{Adaptable Bayesian inference algorithms}
\item{Modular functionalities for considering:}
\begin{itemize}
  \item Refraction and forward scattering
  \item Mie-scattering with a library of condensates
  \item Error inflation and Gaussian processes to treat correlated noise
\end{itemize}
\end{itemize}

We first discuss the standard features that we retain from the retrieval framework of \citet{Pinhas2018}, followed by description of the new features in the \textsl{Aurora} framework built in this work. 

\subsection{Forward Model}
\label{subsec:fwdmodel}

\textsl{Aurora} computes the transmission spectrum of an exoplanet in transit assuming plane-parallel geometry. Our forward model is comprised of a parametric pressure-temperature (P-T) profile; parametric chemical abundances and consideration for multiple sources of opacity including atomic and molecular line opacity, Rayleigh scattering and collision-induced absorption; a treatment for inhomogeneous clouds and hazes; and a line-by-line radiative transfer solver under hydrostatic equilibrium. The forward model can consider light refraction, forward scattering, Mie-scattering and stellar heterogeneity (see section \ref{subsec:modules}).

\subsubsection{Pressure-Temperature Profile}
\label{subsubsec:ptprof}

The temperature in the atmosphere of an exoplanet as a function of pressure is determined by the pressure-temperature (P-T) profile. We follow the parametric prescription of \citet{Madhusudhan2009}. We choose this profile as it is motivated by the profiles observed in the solar system and has been successfully applied to exoplanet studies \citep[e.g.,][]{Madhusudhan2009, Madhusudhan2011, Madhusudhan2014a,Blecic2017}. The equations for temperature in this parameterization divide the model atmosphere into three distinct regions:

\begin{align}
T &= T_0 + \left(\frac{\log(P/P_0)}{\alpha_1}\right)^2, \hspace{0.1cm} P_0 < P < P_1 \label{eq:PT_1}\\
T &= T_2 + \left(\frac{\log(P/P_2)}{\alpha_2}\right)^2, \hspace{0.1cm} P_1 < P < P_3 \label{eq:PT_2}\\
T &= T_2 + \left(\frac{\log(P_3/P_2)}{\alpha_2}\right)^2, \hspace{0.1cm} P > P_3  \label{eq:PT_3}
\end{align}

\noindent where we maintain the empirical choice of \citet{Madhusudhan2009} to set their parameters $\beta_1=\beta_2=0.5$. Here, $T_0$ is the temperature at the top of the model atmosphere $P_0$ (e.g., 10$^{-6}$ bar in this work), $P_1$ is the boundary between the first and second regions, $P_3$ is the boundary between the second and third regions, $P_2$ is the pressure in the parameterization which can capture possible thermal inversions if $P_2> P_1$, and $\alpha_1$ and $\alpha_2$ are the values that determine the curvature of the profile in the different layers. We restrict our temperature profiles to those with $P_2 \leq P_1$, for observations of the day-night terminator where thermal inversions are not expected. \textsl{Aurora} has the option of considering an isothermal profile in which case the free parameter is $T_0$. Then, the temperature at all points in the model atmosphere is assumed to be $T_0$. 

\subsubsection{Sources of Opacity}
\label{subsubsec:opacity}

The presence of different chemical species in the atmosphere of an exoplanet is retrieved by considering their contribution to the star light's extinction. The extinction coefficient $\kappa(\lambda,P,T)$ of the atmosphere is a pressure, temperature, and wavelength dependent quantity that contributes to the differential optical depth 

\begin{align}
d\tau(\lambda,P,T) = \kappa(\lambda,P,T) ds
\label{eq:dtau}
\end{align}

\noindent along the line of sight $s$. The extinction coefficient for each species is given by $\kappa_i(\lambda,P,T)=n_i\,\sigma_i(\lambda,P,T)$, where $n_i$ is the number density and $\sigma_i$ the absorption cross-section of the species. The number density, $n_i$, is parameterised through the volume mixing ratio, $X_{i} = n_i / n_{\text{tot}}$, where $ n_{\text{tot}}$ is the total number density. The volume mixing ratio of each species is a free parameter and assumed to be uniform in the atmosphere. For H-rich atmospheres, \textsl{Aurora} calculates the volume mixing ratio of H$_2$ and He by assuming a particular He/H$_2$ ratio ($X_\mathrm{He}/X_\mathrm{H_2}$) and the following relations

 \begin{align}
 X_\mathrm{H_2} = \frac{ 1 - \displaystyle \sum_{i,i \neq \mathrm{He},\mathrm{H_2}}^{n} X_i} {1+\frac{X_\mathrm{He}}{X_{\mathrm{H_2}}}} , &&
 X_\mathrm{He} = X_\mathrm{H_2} \,\frac{X_\mathrm{He}}{X_\mathrm{H_2}}
 \end{align}
 
\noindent where we adopt a solar value of $X_\mathrm{He}/X_\mathrm{H_2}=0.17$ \citep{Asplund2009} and consider a total of $n$ species in the model atmosphere. The treatment of the volume mixing ratios when a H-rich atmosphere is not assumed a priori is described in section \ref{subsec:reparameterization}.
 
\begin{figure}
\includegraphics[width=0.5\textwidth]{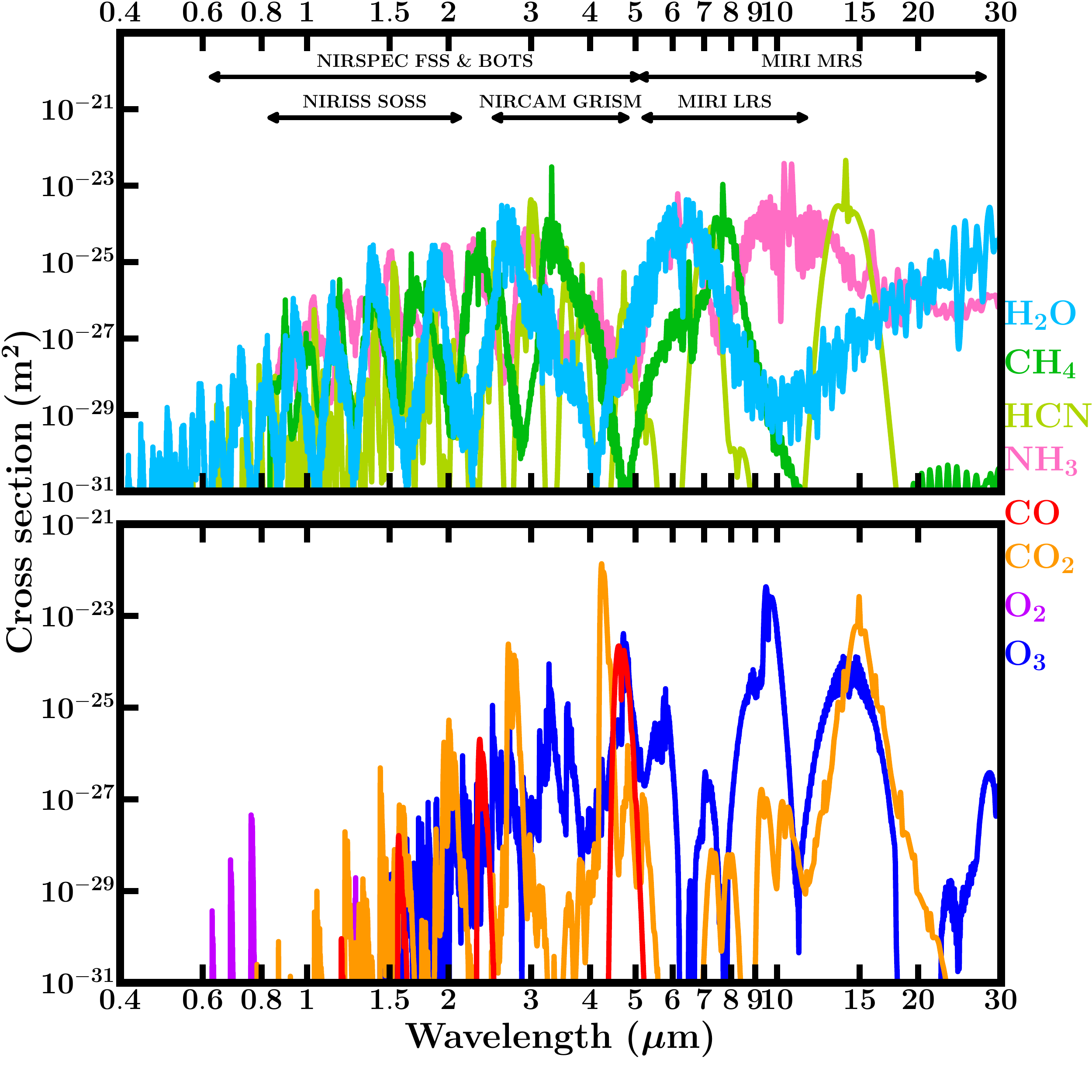}
\centering
\caption[Opacity Sources]{ Top and bottom panels show the absorption cross sections for some of the molecular opacity sources included in this work at a pressure of 1 bar and a temperature of 300~K. The cross sections are shown in the wavelength region where JWST is expected to perform observations. JWST instrument ranges are indicated on the top panel using black lines and arrows.} 
\label{fig:cross_sections}
\end{figure} 

\textsl{Aurora} in general considers the opacity sources expected in the atmospheres of hot Jupiters, mini-Neptunes and temperate rocky planets \citep[e.g.,][]{Madhusudhan2012, Moses2013, Madhusudhan2019}. The opacity sources considered in this work are H$_2$–-H$_2$ and H$_2$–-He collision induced absorption \citep[CIA;][]{Richard2012}, and line opacity due to CH$_4$ \citep{Yurchenko2014}, CO \citep{Rothman2010}, CO$_2$ \citep{Rothman2010}, H$_2$O \citep{Rothman2010}, HCN \citep{Barber2014}, K \citep{Allard2016}, Na \citep{Allard2019}, N$_2$ \citep{Rothman2010}, NH$_3$ \citep{Yurchenko2011}, O$_2$ \citep{Rothman2010}, and O$_3$ \citep{Rothman2010}. The opacities for the chemical species are computed following the methods of \citet{Gandhi2017}, with the updated values of \citet{Gandhi2018}, and with H$_2$-broadened Na and K cross sections as explained in \citet{Welbanks2019b}.

\textsl{Aurora} also incorporates a continually updated library of cross-sections of various other atomic and molecular species \citep{Gandhi2020}. Figure \ref{fig:cross_sections} shows the cross section for most of the molecular opacity sources considered in this work for a pressure of 1 bar and a temperature of 300~K, from 0.4-30~$\mu$m covering the wavelength range expected to be observable by JWST. The Na and K profiles can be seen in Figure 1 of \citet{Welbanks2019b}.

The resulting extinction coefficient is
\begin{align}
&\kappa(\lambda,P,T) = \sum_i X_i n_{\mathrm{tot}}(P,T) \, {\sigma_{i}(\lambda,P,T)} +  X_{\mathrm{H_2}}n_{\mathrm{tot}}^2(P,T) \nonumber  \\
 & \left[ X_{\mathrm{H_2}} \sigma_{\mathrm{H_2}\text{–-}\mathrm{H_2}}(\lambda, T) + X_{\mathrm{He}} \sigma_{\mathrm{H_2}\text{–-}\mathrm{He}}(\lambda, T) \right]
\label{eq:kappa}
\end{align}

\noindent where $\sigma_{\mathrm{H_2}\text{--}\mathrm{H_2}}$ and $\sigma_{\mathrm{H_2}\text{--}\mathrm{He}}$ are the H$_2$–-H$_2$ and H$_2$–-He CIA cross-sections. The extinction coefficient can be amended to remove H$_2$–-He and H$_2$–-H$_2$ CIA and/or include CIA due to other species. Furthermore, the total extinction coefficient can include H$_2$-Rayleigh scattering 

\begin{align}
\kappa_{\mathrm{H_2}\text{-}\mathrm{Rayleigh}}(\lambda,P,T) &= X_{\mathrm{H_2}} n_{\mathrm{tot}}(P,T)  \sigma_{\mathrm{H_2 \, scat}}(\lambda)
\label{eq:kappa_rayleigh}
\end{align}

\noindent where the wavelength dependent cross-section in cgs is given analytically by \citet{Dalgarno1962} as

\begin{align}
    \sigma_{\mathrm{H_2 \, scat}}(\lambda) &=\frac{8.14 \times 10^{-45}}{\lambda^4}+\frac{1.28 \times 10^{-54}}{\lambda^6} \nonumber\\
    &+ \frac{1.61 \times 10^{-64}}{\lambda^8}  + \mathcal{O}(\lambda^{-10})
\end{align}

\noindent and is incorporated up to its third term in \textsl{Aurora}. 

\textsl{Aurora} also includes opacity sources relevant for modelling H-poor atmospheres of rocky planets. This library contains collision induced absorption (CIA) cross-sections of CO$_2$–-CO$_2$, N$_2$–-N$_2$, O$_2$–-O$_2$, O$_2$–-CO$_2$, O$_2$–-N$_2$, amongst others obtained from HITRAN \citep[][]{Karman2019}. These additional CIA cross-sections are generated within the temperature and wavelength limits available in the HITRAN data. The cross-sections for temperatures beyond those limits are set to values at the boundaries. We assume no opacity for wavelengths beyond the database range, as these values are not known. Future efforts, both experimental and theoretical, on extending and revising opacity databases would help obtain cross-sections over the full range of wavelengths and temperatures applicable for such planets. \textsl{Aurora} can also include Rayleigh scattering due to a variety of species including O$_2$, N$_2$, Ar, Ne, CO$_2$, CH$_4$, H$_2$O, CO, and N$_2$O \citep{Rao1977, Sneep2005, Thalman2014}. Rayleigh scattering due to species $i$ is $\kappa_{i\text{-}\mathrm{Rayleigh}}(\lambda,P,T) = X_{i} n_{\mathrm{tot}}(P,T) \sigma_{i\,\mathrm{scat}}(\lambda)$. We include collision induced absorption due to CO$_2$–-CO$_2$, N$_2$–-N$_2$, as well as Rayleigh scattering due to N$_2$, H$_2$O, and CO$_2$ in the H-poor models presented in section \ref{subsubsec:trappist1d}. 

\subsubsection{A New Cloud and Haze Parameterization}
\label{subsubsec:new_cloudsandhazes} 

We introduce a new cloud and haze treatment for inhomogeneous cover that considers a total of four distinct spatial areas (sectors) covering the planet. These four areas are (1) a clear, cloud-free and haze-free, area affected only by Rayleigh scattering, (2) an area covered by hazes only, (3) an area covered by a gray cloud deck with Rayleigh scattering above the cloud deck, and (4) an area covered by a gray cloud deck and hazes above it. The total transit depth is a linear superposition of the transit depths of each sector

\begin{align}
 &\Delta_{\mathrm{planet}}(\lambda)=\phi_{\mathrm{hazes}}\Delta_{\mathrm{hazes}}(\lambda)+\phi_{\mathrm{clouds}}\Delta_{\mathrm{clouds}}(\lambda)\nonumber\\
 &+\phi_{\mathrm{clouds+hazes}}\Delta_{\mathrm{clouds+hazes}}(\lambda)+\phi_{\mathrm{clear}}\Delta_{\mathrm{clear}}(\lambda)
\end{align}

\noindent where the cover fractions are free parameters in the model and $\phi_{\mathrm{clear}}$ is determined by a unit-sum constraint, i.e., $\phi_{\mathrm{clear}}=1-\phi_{\mathrm{hazes}}- \phi_{\mathrm{clouds}}-\phi_{\mathrm{clouds+hazes}}$. 

Hazes, e.g., small size particles resulting from photo-chemical processes, are implemented into our model atmosphere by parameterising their effect on the spectrum as deviations from H$_2$-Rayleigh scattering \citep{Lecavelier2008a}. The parameterization provides a cross-section $\sigma_{\mathrm{hazes}}=a\sigma_0(\lambda/\lambda_0)^\gamma$, where $\gamma$ is the scattering slope, $a$ is the Rayleigh-enhancement factor, and $\sigma_0$ is the H$_2$-Rayleigh scattering cross-section at the reference wavelength $\lambda_0$. We adopt values of $\sigma_0=5.31\times10^{-31}$~m$^2$ and $\lambda_0=350$~nm for consistency with previous works \citep[e.g.,][]{MacDonald2017, Welbanks2019b}. Future observations of non H-rich planets could motivate the use of scattering cross-sections for different species. The extinction due to hazes is $\kappa_{\mathrm{haze}}(\lambda,P,T)=X_{\mathrm{H_2}} n_{\mathrm{tot}}(P,T) \sigma_{\mathrm{hazes}}(\lambda)$. 

\begin{figure}
\includegraphics[width=0.5\textwidth]{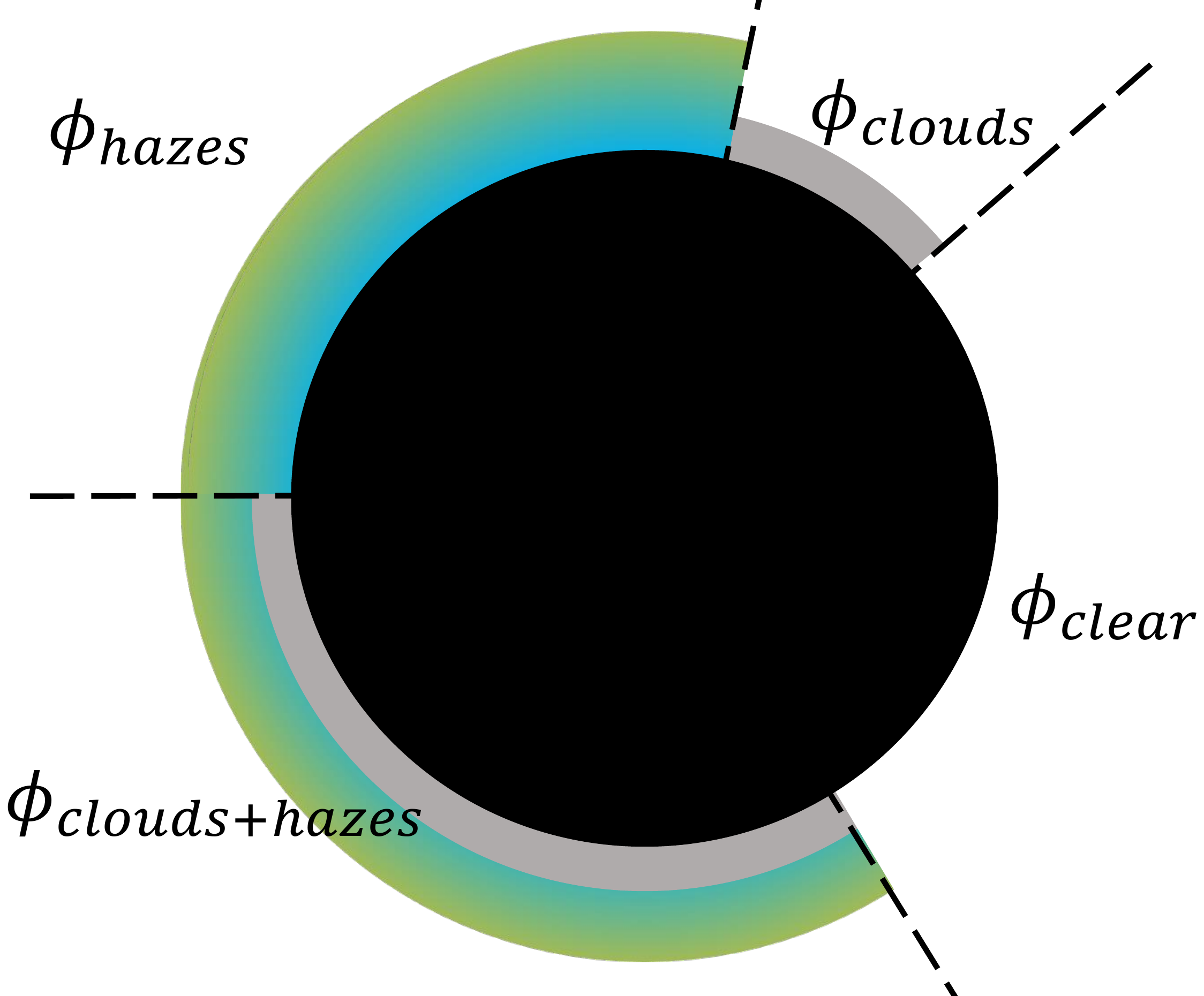}
\centering
\caption[Schematic of clouds]{Schematic of the four-sector generalised parameterization of inhomogeneous clouds and hazes introduced in this work. The planet is enveloped by its atmosphere which is divided into four regions. These are (1) a clear, cloud-free and haze-free, sector, (2) a sector with hazes only, (3) a sector with clouds only, and (4) a sector with clouds and hazes.}
\label{fig:cloud_n_haze_schematic}
\end{figure} 

The regions of the atmosphere covered by a gray cloud deck are included by adopting a parameter for the cloud top pressure $P_{\text{cloud}}$. The optical depth for all pressures higher than $P_{\text{cloud}}$ is considered to be infinite. The extinction coefficient due to the cloud deck $\kappa_{\mathrm{clouds}}(P)$ is infinite for $P>P_{\mathrm{cloud}}$ or zero for $P<P_{\mathrm{cloud}}$.

Previous studies have considered the effects of patchy clouds in transmission spectra \citep[e.g.,][]{Line2016, MacDonald2017, Barstow2020a}. Our model here generalises the approach of previous studies while being able to reduce to previous treatments under specific conditions. If the model prefers to consider the presence of clouds and hazes together, the fractions $\phi_{\mathrm{hazes}}$ and $\phi_{\mathrm{clouds}}$ approach zero and we obtain previous treatments for inhomogeneous cover \citep[e.g.,][]{MacDonald2017,Welbanks2019a}. On the other hand, if the combined fraction is zero (e.g., $\phi_{\mathrm{clouds+hazes}}=0$), our approach allows us to consider the effect of clouds and hazes separately and distinguish whether the contribution to the spectrum is mostly due to deviations from H$_2$-Rayleigh scattering produced by the hazes, or muted features due to a gray cloud deck. Lastly, if the combined fraction is zero and so is the haze only fraction (e.g., $\phi_{\mathrm{hazes}}=\phi_{\mathrm{clouds+hazes}}=0$) we recover the expression for patchy clouds of \citet{Line2016}. By following this approach, we obtain a more robust and flexible treatment compared to our previous prescription that combines the effects of clouds and hazes into one sector \citep[e.g.,][]{MacDonald2017, Welbanks2019a}. We present a schematic of our cloud and haze treatment in Figure \ref{fig:cloud_n_haze_schematic}. 

We find the generalised treatment of clouds and hazes introduced in this work leads to consistent abundance estimates regardless of whether a H-rich atmosphere is assumed or not. In other words, the existing degeneracies between clouds/hazes and composition are treated equally irrespective of the assumption of the bulk atmospheric composition of the planet. On the other hand, combining clouds and hazes into one individual sector as previously performed \citep[e.g.,][]{MacDonald2017, Pinhas2018, Welbanks2019a} can lead to biases and an incomplete exploration of the parameter space that results in distinct solutions when assuming a H-rich atmosphere or not on the same data set. This is mitigated by our new cloud and haze prescription. We discuss these aspects further on the case study of the hot Jupiter HD~209458~b in section \ref{subsec:cloud_validation}.

\subsubsection{Radiative Transfer}
\label{subsubsec:rt}

Our model solves line-by-line radiative transfer in transmission geometry in a plane parallel atmosphere. The model atmosphere is divided into a predetermined number of pressure layers equally spaced logarithmically. The number of layers and their span in pressure space can be arbitrarily established by the user depending on the application. For this work, and based on the empirical results of \citet{Welbanks2019a}, we use 100 pressure layers uniformly spaced in $\log_{10}$(P) from $10^{-6}$ to $10^2$ bar under hydrostatic equilibrium. Our calculation of hydrostatic equilibrium is performed considering the retrieved composition through the atmospheric mean molecular weight (e.g., $\mu=\sum X_{i}m_i$, where $X_{i}$ and $m_i$ are the volume mixing ratio and the atomic/molecular mass of species $i$, respectively), the retrieved pressure-temperature profile, and altitude-dependent gravity.

We solve numerically for the transit depth of the planet

\begin{align}
\Delta_{\lambda} = \frac{R_{p}^2 + 2 \displaystyle\int\limits_{R_p}^{R_p+ H_A} b \left(1 - e^{-\tau_{\lambda}(b)} \right) db - 2 \displaystyle\int\limits_{0}^{R_p} b e^{-\tau_{\lambda}(b)} db}{R_{\text{star}}^2},
\label{eq:td}
\end{align}

\noindent where $R_{\text{star}}$ is the stellar radius, $H_A$ is the maximum height of the planetary atmosphere, $\tau_{\lambda}$ is the total slant optical depth and integral of equation \ref{eq:dtau}, $b$ is the impact parameter, and $R_{p}$ is the radius of the planet. We present equation \ref{eq:td} as a three part expression to highlight the fact that the chosen $R_{p}$ may not correspond to an optically thick surface. If $R_{p}$ corresponds to an optically thick surface, the last integral in equation \ref{eq:td} evaluates to zero. Otherwise, the integral considers the contribution of the non-optically thick parts of the atmosphere, below the arbitrarily chosen position in the planet, to the transit depth. 

The selected value of $R_{p}$, at a given reference pressure $P_{\rm{ref}}$, is used to construct a radial distance grid. The distance and pressure grids follow a one-to-one correspondence determined by hydrostatic equilibrium. It is possible in a retrieval to choose a value of $R_{p}$ for which the $P_{\rm{ref}}$ parameter will be retrieved, to choose a value of $P_{\rm{ref}}$ for which the associated radius $R_{p}$ will be retrieved, or leave both $R_{p}$ and $P_{\rm{ref}}$ as free parameters. \citet{Welbanks2019a} showed that the retrieval results remain mostly unchanged regardless of the choice of free parameter ($R_{p}$ and/or $P_{\rm{ref}}$). In this work we choose to keep $R_{p}$ as our independent variable for which we retrieve $P_{\rm{ref}}$.

\subsection{Considerations for Non H-rich Atmospheres}
\label{subsec:reparameterization}

A core assumption present in most atmospheric retrieval codes for hot Jupiters is that the atmosphere is H-rich. Such assumption can be appropriate for massive planets which, from a formation perspective, captured a gas mixture of predominantly H and He in cosmic proportions from their protoplanetary nebula \citep{Seager2010}. However, when characterising the atmospheres of less massive planets or when pursuing an agnostic approach applicable to atmospheres of general composition, such assumption may need to be relaxed. Instead of assuming a H-rich atmosphere, studies could attempt to retrieve the main gas component of the atmosphere. Such approach would aim to explore a wider range of atmospheric compositions like N$_2$-rich or CO$_2$-rich atmospheres, and not be constrained to H-rich atmospheres only.

However, when pursuing this agnostic approach, the unit-sum constraint, i.e., the requirement that all the volume mixing rations in the atmosphere must add up to one, must be incorporated into the statistical modelling appropriately. Incorporating such constraint is non-trivial and has been the subject of study in a sub-field of statistical analysis called \textit{compositional data analysis} \citep[e.g.,][]{Pearson1897, Tanner1949, Chayes1960, Aitchison1986}. The tools developed by this sub-field of statistics have been implemented in a number of different disciplines like medicine, chemistry, economy, geophysics, amongst many others \citep[see][for a review of the history of compositional data]{Aitchison2005}. The concepts of compositional data analysis were introduced to the exoplanet retrieval literature through the work of \citet{Benneke2012}. 

Implementing the same methods used for the retrieval of H-rich atmospheres to retrievals in which the main atmospheric constituent is not known can result in biased results that do not explore all compositions equally. The traditional method would sample the volume mixing ratios of $n$-1 species (i.e., minor species) and assign the volume mixing ratio of the $n$th species (i.e., H$_2$ in the case of a H-rich atmosphere) following the unit-sum constraint. \citet{Benneke2012} highlight that following this approach will result in a highly asymmetric prior \citep[see Fig. 1 of][]{Benneke2012} for the $n$th species. Under these circumstances, the retrieval is not truly agnostic and the resulting atmospheric composition will be dependent on which molecule was chosen to be the $n$th species. 

To circumvent this problem one must allow for all species to have the same prior probability density in a permutation-invariant prescription. If the prior probability for all species is identical, it is safe for the retrieval to sample over the parameter space of all $n$ species. The solution is the centered-log-ratio transformation, defined as

\begin{equation}
   z_i=log\left( \frac{X_i}{g(X)}\right)
\end{equation}

\noindent where g(X) is the geometric mean $g(X)=(X_1...X_N)^{1/N}$ \citep{Aitchison1986}. The transformed $z_i$ values, also called compositional parameters, treat all part of the gas symmetrically.

In \textsl{Aurora}, when not assuming a H-rich atmosphere we reparameterise the volume mixing ratios ($X_i$) using the centered-log-ratio transformation and obtain the compositional parameters ($z_i$). We assume that the combination of H$_2$ and He is one single part ($z_\mathrm{(H_2+He)}$) which we then use to determine the separate H$_2$ and He volume mixing ratios using a He/H$_2$ ratio. Then, we sample over the entire transformed space for all $n$ gas components with the assumption that one of those is a mixture of H$_2$ and He in solar proportion.

Once sampling is performed in the space of the centered-log-ratio transformation, and to maintain the descriptions above about the treatment of different opacity sources, the inverse transformation \citep[][]{Pawlowsky-Glahn2011} 

\begin{equation}
X_j= \frac{\exp(z_j)}{\sum\limits_{i=1}^N \exp(z_i)}
\end{equation}

\noindent is calculated and the volume mixing ratios $X_i$'s are used in our calculations.

It is important to highlight that the compositional parameters ($z_i$) have slightly different properties than their counterparts, the volume mixing ratios ($X_i$). While the typical prior range for $X_i$ is 10$^{-12}<$ $X_i<1$, the limits for $z_i$ is $-\infty<z_i<\infty$ where $-\infty$ is the limit of a species not being present and $\infty$ means the species is the only one in the atmosphere. While a straightforward expression for the scenario in which all volume mixing ratios are equal is not available, the compositional parameters are present in equal parts when all $z_i=0$. Lastly, the unit-sum constraint for the volume mixing ratios is $\sum X_i=1$, and transforms to $\sum z_i=0$ for the compositional parameters. 

\subsection{Multialgorithmic Statistical Inferences} 
\label{subsec:Bayesian}

The strength in the retrieval approach when assessing the properties of an exoplanet's atmosphere resides in its ability to provide robust statistical estimates of the parameters and models used to explain the observations. As explained in the section \ref{sec:intro}, many statistical approaches exist in exoplanetary atmospheric retrievals: grid-based searches \citep[e.g.,][]{Madhusudhan2009}, MCMC \citep[e.g.,][]{Madhusudhan2011,Benneke2012,Line2013a,DeWit2013,Madhusudhan2014a, Cubillos2016, Wakeford2017, Zhang2019}, non-linear optimal estimators \citep[e.g.,][]{Lee2013, Barstow2017}, amongst others \citep[see][for a review]{Madhusudhan2018}. Of the different approaches available, Bayesian inference tools ease the comparison of models while providing estimates of the posterior distributions of the model parameters. One of these methods, nested sampling \citep{Skilling2006} has been successfully incorporated into exoplanetary retrieval literature \citep[e.g.,][]{Benneke2013, Line2013a, Waldmann2015a, MacDonald2017, Gandhi2018, Pinhas2018, Krissansen-Totton2018,Molliere2019,Zhang2020} due to its ability to handle high dimensionality problems, sample the complete parameter space of the model, and use prior information on the model parameters. An overview of the Bayesian approach to inference problems is available in \citet{Sivia2006, Trotta2008,Trotta2017}. 

The likelihood of observing the data ($\mathcal{D}$) given a specific set of model parameters ($\theta_\mathcal{M}$) for a model ($\mathcal{M}$) is

\begin{equation}
\label{eq:likelihood}
   \mathcal{L}=P(\mathcal{D}|\theta_\mathcal{M}, \mathcal{M}).
\end{equation}

\noindent Considering the Bayesian approach, where the degree of belief on the model assumptions must be accounted for, one must incorporate the prior distribution ($\pi$) on the model parameters $\pi=P(\theta_\mathcal{M}|\mathcal{M})$. The marginalised likelihood, also known as evidence, is obtained by integrating the likelihood over the full parameter space 

\begin{equation}
\label{eq:evidence}
\mathcal{Z}=\int P(\mathcal{D}|\theta_\mathcal{M}, \mathcal{M}) P(\theta_\mathcal{M}|\mathcal{M}) d\theta_\mathcal{M}= P(\mathcal{D}, \mathcal{M}).
\end{equation}

The model evidence is the quantity we are interested in evaluating when comparing different models. This is also the quantity different nested sampling algorithms aim to provide. Furthermore, using Bayes theorem it is possible to obtain the posterior probability distribution for each parameter given the data 

\begin{equation}
\label{eq:posterior_probability}
   P(\theta_\mathcal{M}|\mathcal{D}, \mathcal{M})=\frac{P(\mathcal{D}|\theta_\mathcal{M}, \mathcal{M}) P(\theta_\mathcal{M}|\mathcal{M})}{P(\mathcal{D}, \mathcal{M})}.
\end{equation}

\textsl{Aurora} uses a likelihood function for data with independently distributed Gaussian errors

\begin{equation}
\mathcal{L}\left(\mathcal{D}| \theta_\mathcal{M} \, , \mathcal{M}_i \right) = \prod_{k}^{N} \frac{1}{\sqrt{2\pi}\sigma_{k}} \mathrm{\exp}\left(-\frac{[\mathcal{D}_{k} - \mathcal{M}_{i,k}]^2}{2\sigma_{k}^2}\right)
	\label{eq:likelihood_function}
\end{equation}

\noindent for a data set of length $N$ and computed for each model realisation $\mathcal{M}_i$. \textsl{Aurora} follows the same binning strategy as AURA \citep[see section 2.1.6 in][]{Pinhas2018} where a model spectrum at a much higher resolution than the data is convolved with the point spread function (PSF) of the instrument with which the observations were obtained and then binned down to the spectral resolution of the data. 

The prior distributions employed in this study are shown in Table \ref{table:priors} in the appendix. The priors for the parameters are mostly standard prescriptions adopted from previous studies \citep[e.g.,][]{Pinhas2019,Welbanks2019b}. The priors for the molecular abundances generally span the complete detectable range unless stated otherwise, with the prior distribution either log-uniform for the volume mixing ratios for H-rich retrievals or uniform in the corresponding compositional parameters ($z_i$), discussed in section \ref{subsec:reparameterization}, for non H-rich retrievals. The priors for the parameters associated with other physical properties, e.g., pressure-temperature profile and cloud/haze parameters, are also uniform or log-uniform and span the corresponding physically plausible ranges. 

\subsubsection{Next-generation Bayesian Inference Algorithms}
\label{subsubsec:samplers}

The main functionality of a nested sampling algorithm is to obtain the model evidence ($\mathcal{Z}$) while also deriving the posterior probability distributions of the model parameters as a by-product. A full description of the nested sampling algorithm is available in \citet{Skilling2004,Skilling2006,Feroz2009}. In \textsl{Aurora} we implement three different algorithms, MultiNest \citep{Feroz2009, Feroz2013} through its implementation PyMultiNest \citep{Buchner2014}, PolyChord \citep{Handley2015a, Handley2015b} through its implementation PyPolyChord, and Dynesty \citep{Speagle2020}. Each nested sampling algorithm is different and the in-depth description for each implementation is available in their release papers listed above. 

Generally, a nested sampling algorithm generates a number of live points drawn from the prior distribution, which sample the parameter space \citep{Feroz2009}. In each iteration, the point with lowest likelihood is replaced by a new one which ought to have a larger likelihood. This means that the live points sample the prior volume using continuously shrinking iso-likelihood contours, which with every iteration converge to the highest likelihood regions of the parameter space. At each step, every sampled value creates a model realisation that results in an evaluation of the likelihood function. The process finishes when a termination condition, like a pre-set fractional change in the the model likelihood, is met. Upon completion, the combination of all sampled points can be used to estimate the model evidence. The procedure to generate new live points can vary between different implementations of the nested sampling algorithm which are briefly discussed below. 

MultiNest has been previously implemented in exoplanet retrievals \citep[e.g.,][]{Benneke2013, Line2013a, Waldmann2015a, MacDonald2017, Gandhi2018, Pinhas2018, Krissansen-Totton2018, Molliere2019, Zhang2020}. To draw unbiased samples from the likelihood-constrained prior, MultiNest uses what is called an ellipsoidal rejection sampling scheme. The basis for this scheme is that the replacement point is sought from within the set of ellipsoids described by the full set of live points at any iteration \citep{Feroz2019}. With each iteration the ellipsoids described by the iso-likelihood contours shrink. This procedure is optimal for a small number of parameters but has an exponential scaling with dimensionality. 

PolyChord, on the other hand, uses what is called slice-based sampling. In this procedure, the algorithm samples uniformly within the parameter space for which the posterior probability is higher than a given probability level or `slice'. Unlike the exponential scaling problem with MultiNest at higher dimensions, PolyChord's scaling is $\sim \mathcal{O}(D^3)$ \citep{Handley2015b}. This makes MultiNest preferred for low dimensionality problems, while PolyChord is preferred at higher dimensionalities \citep[see Figure 7 in][]{Handley2015b}. 

Lastly, Dynesty \citep{Speagle2020} uses a generalisation of nested sampling, in which the number of live points is variable, called dynamic nested sampling \citep{Higson2019}. In dynamic nested sampling, an initial run with a constant number of live points is used by the algorithm to approximate areas in prior space of the highest likelihood. Then, the algorithm proceeds to iteratively calculate the range of likelihoods where a larger number of live points will have the greatest result in accuracy. In dynamic nested sampling the number of live points is \textit{dynamically} allocated to control the resolution at which the prior space is sampled. This would allow for runs that focus on sampling the posterior distribution or better estimate the model evidence. Dynesty allows for both dynamic and static nested sampling. Furthermore, Dynesty has four main approaches to generating samples: uniform sampling (including from ellipsoids like MultiNest), random walks, multivariate slice sampling (similar to PolyChord), and Hamiltonian slice sampling. Each approach has its benefits and impediments, and can be better suited for different problem dimensionalities. \cite{Speagle2020} offers an extensive overview of each feature available in Dynesty. 

Every algorithm for nested sampling offers different capabilities. While Dynesty is able to handle both static and dynamic sampling, it comes at the cost of multiple tuning parameters that can affect the behaviour of a given run. PolyChord is able to handle problems of higher dimensionality more efficiently than MultiNest, but MultiNest still outperforms PolyChord in the number of likelihood evaluations required for problems in low dimensions \citep[$\lesssim$80 dimensions,][]{Handley2015b}. \textsl{Aurora} offers the tools to perform retrievals optimising for evaluation of the model evidence, parameter posterior distributions, or both. The user has the freedom to choose the correct sampling algorithm for their needs depending on the complexity of the problem and its dimensionality.

\subsubsection{Model Comparison and Detection Significance}

The difference in evidence ($\mathcal{Z}$) between models can be used to derive an equivalent detection significance (DS), a figure of merit traditionally used to compare different models. The detection significance is traditionally expressed in units of `sigma' ($\sigma$) and corresponds to the number of standard deviations away from the mean of a normal distribution \citep{Trotta2008}. Expressing a result in `sigmas' does not necessarily mean the detection of new physics or a species in the spectrum of a planet. Instead, it is a useful way to translate the odds in favour of a more complex model into a frequentist metric. The relevance of a model preference can be somewhat arbitrary and different authors suggest different categories for expressing them. For instance, \citet{Trotta2008} suggests that a difference of 2.0 to 2.6$\sigma$ is weak at best, while \citet{Kass1995} suggests that the equivalent of $\sim 2.1\sigma$ is positive evidence. A way to transform the difference in model evidences to a detection significance was proposed by \citet{Benneke2013}. We perform the comparison of our models by solving equation 11 in \citet{Benneke2013}, and obtaining a detection significance as 

\begin{equation}
    DS=\sqrt{2}\,\operatorname{erfc}^{-1}\left[\exp \left(\mathcal{W}_{-1} \left(-\frac{1}{B \, e} \right)\right)\right]
\end{equation}

\noindent where $\operatorname{erfc}$ is the complementary error function, $\mathcal{W}_{-1}$ is the Lambert $\mathcal{W}$ function in its lower branch (i.e., $k=-1$ branch), $e$ is Euler's number, and B is the Bayes factor defined as $B=\mathcal{Z}_1/\mathcal{Z}_2$, with the set requirement of $\mathcal{Z}_1\geq\mathcal{Z}_2$ so the Bayes factor is greater than or equal to unity. 

\subsection{Modular Capabilities}
\label{subsec:modules}

\textsl{Aurora}'s design is modular ensuring that future capabilities can be easily incorporated into the existing retrieval framework. As part of this modular structure, we include in \textsl{Aurora} preexisting features in AURA \citep{Pinhas2018} such as the functionality to retrieve stellar properties from a transmission spectrum. Furthermore, we introduce new modular capabilities that aid in the analysis of transmission spectra in the context of retrievals and forward models. These key additions include new considerations for noise modelling and forward models considering light refraction, forward scattering, and Mie-scattering.

\subsubsection{Stellar Heterogeneity}
\label{subsubsec:stellar_het}
One of the main features of AURA \citep{Pinhas2018} was to retrieve stellar properties embedded in the transmission spectrum as well as the planetary properties. Inhomogeneities in the stellar photosphere were modelled by retrieving the areal fraction of the projected stellar disc covered by heterogeneities ($\delta$), hot faculae or cool spots, the heterogeneity temperature ($T_{\mathrm{het}}$), and the photospheric temperature ($T_{\mathrm{phot}}$). \textsl{Aurora} inherits this capability but we do not include it in the present study.

\subsubsection{New Noise Modelling Modules}
\label{subsubsec:noise}
\textsl{Aurora} has the capability to treat noise models different from the traditionally assumed white noise. \textsl{Aurora} can consider the possibility of underestimated variance in the data by retrieving an error-bar inflation free parameter \citep{Foremanmackey2013}. This approach assumes that the variance is underestimated by a fractional amount $f$. The variance of the data is 

\begin{equation}
    \mathrm{S}^2=\sigma_{\mathrm{obs}}^2+f^2 \, \Delta_{\mathrm{mod}}^2
\end{equation}

\noindent where $\sigma_{\mathrm{obs}}$ is the error in the observations and $\Delta_{\mathrm{mod}}$ is the model's transit depth. This term replaces the variance term in equation \ref{eq:likelihood_function}. This functionality has been tested on the recent spectroscopic observations of KELT-11b \citep{Colon2020}.

\textsl{Aurora} also has the capability to consider correlated noise in the data being analysed. To do so, we have incorporated \texttt{Celerite} \citep{celerite} and \texttt{George} \citep{george} to model the covariance function and compute the likelihood of a Gaussian Process (GP) model \citep{Rasmussen2006}. The effects of a GP in transmission spectra fall beyond the scope of this paper and we reserve it to a future study.

\subsubsection{Refraction and Forward Scattering}
\label{subsubsec:refraction_scattering}

In \textsl{Aurora} we have incorporated the analytic descriptions for forward scattering and refraction of transit spectra proposed by \citet{Robinson2017}. These prescriptions have been incorporated in the context of producing forward models and synthetic observations.

\begin{figure}
\includegraphics[width=0.45\textwidth]{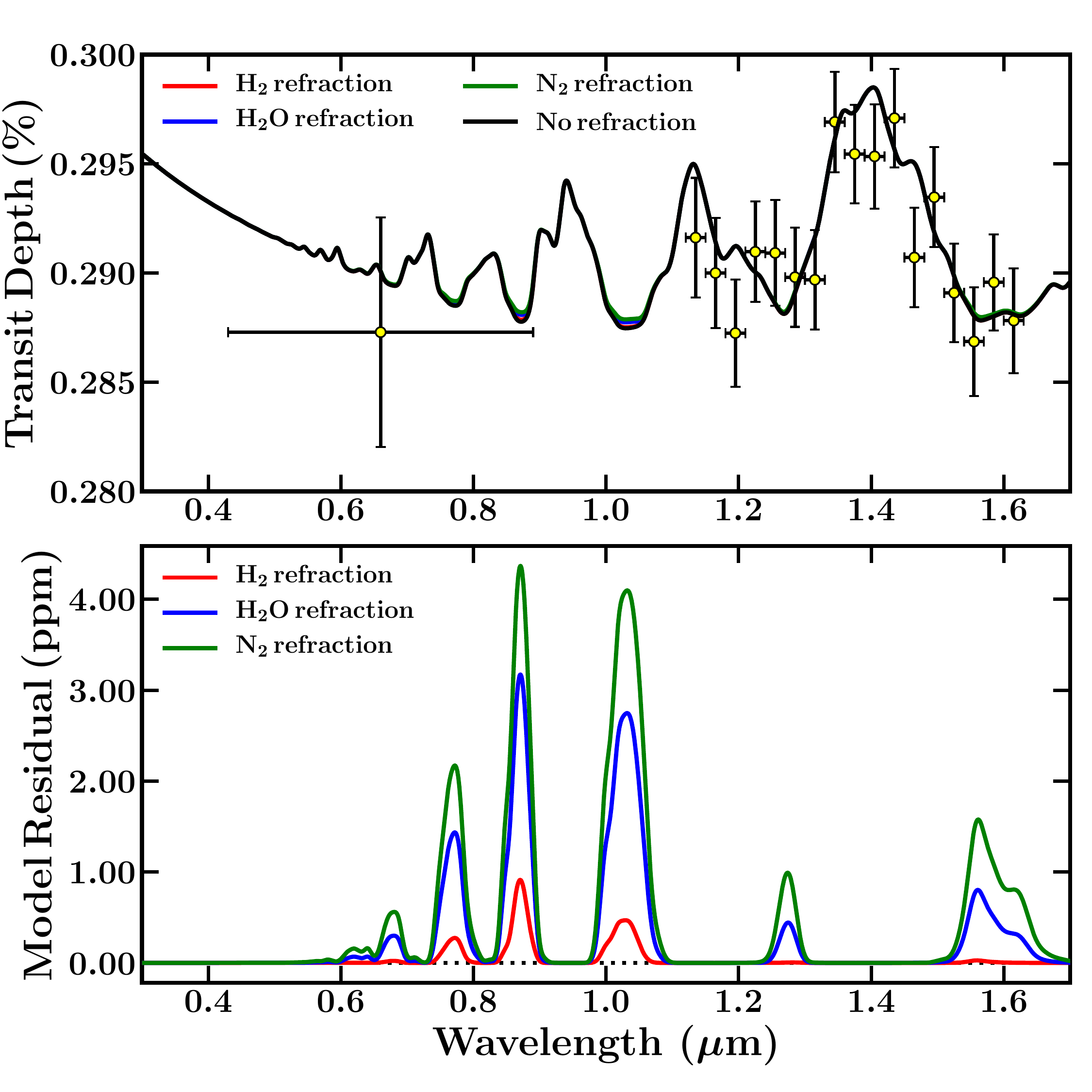}
\centering
\caption[refraction]{Forward models including the effects of wavelength-dependent refraction for the mini-Neptune K2-18b. Top: No refraction model shown in black while the red, blue, and green lines show the effects of H$_2$, H$_2$O, and N$_2$ refraction respectively. Black error bars with yellow markers show current K2 and HST-WFC3 observations for reference. Bottom: Residual of the refraction models relative to the non-refraction model.} 
\label{fig:refraction}
\end{figure}  

Refraction effects are calculated using the prescription for the maximum pressure at which the effect of refraction is large enough to cause a light ray from one side of the planet to come from the far limb (i.e., opposite side) of the host star \citep{Robinson2017}. We incorporate the wavelength-dependent refractivity \citep{Robinson2017}, and use them to calculate the maximum pressure probed ($P_\mathrm{max}$) at each wavelength following equation 15 of \citet{Robinson2017}. The optical depth for pressures higher than $P_\mathrm{max}$ is set to infinity. Figure \ref{fig:refraction} shows the effect of considering refraction in forwards models of K2-18b. For these forward models we consider refraction due to H$_2$, H$_2$O, or N$_2$. The forward models are determined by the median retrieved parameters in section \ref{subsec:samplers_results}. Figure \ref{fig:refraction} shows that the effects of refraction are almost negligible, $\sim 4$~ppm. Additional models considering the effect of refraction for a rocky exoplanet are shown in appendix \ref{app:trappist_fwd_models}.

\begin{figure}
\includegraphics[width=0.45\textwidth]{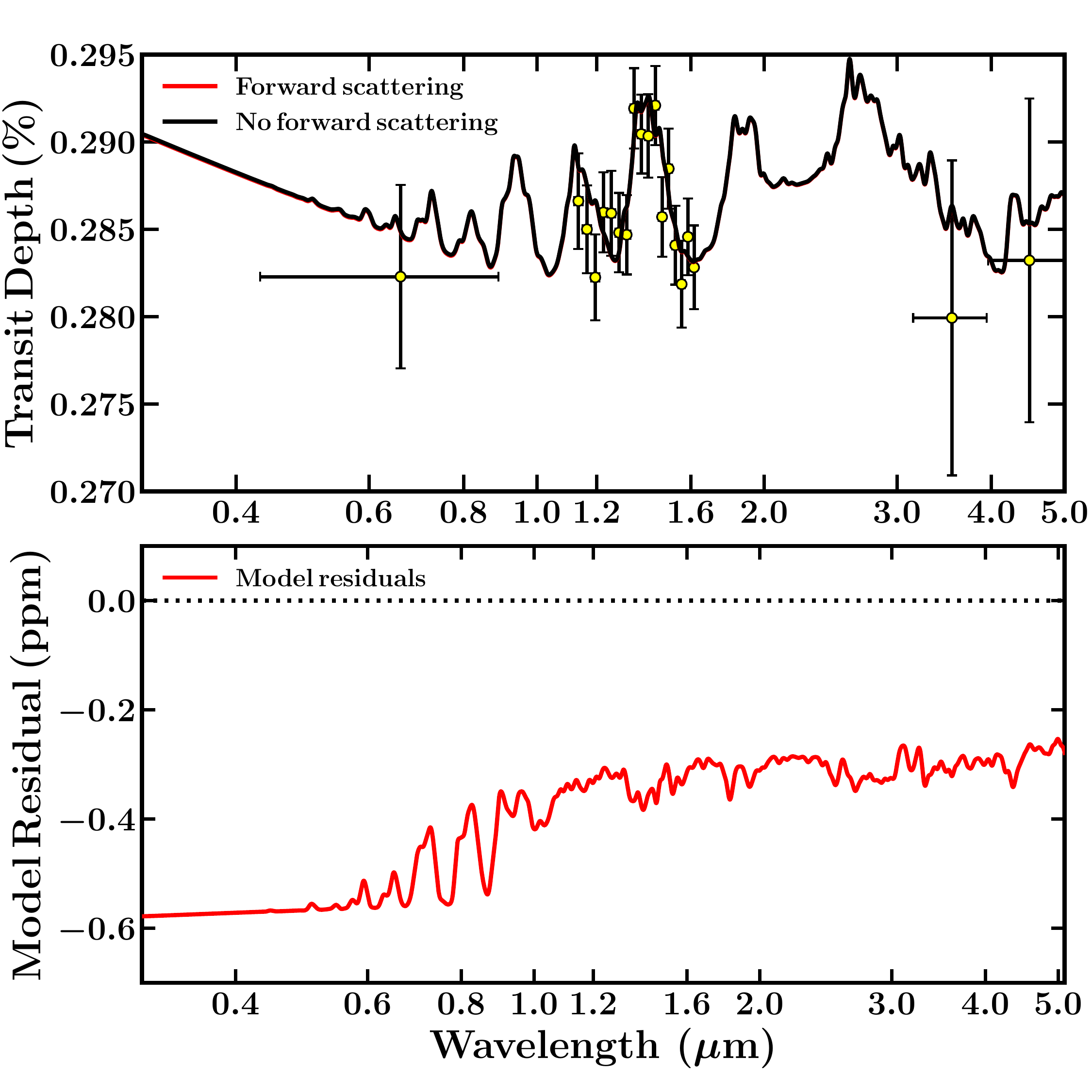}
\centering
\caption[Scattering]{Forward scattering models for the atmosphere of the mini-Neptune K2-18b. Top: Forward scattering model shown in red, and in black the model without forward scattering. Black error bars with yellow markers show current K2, HST-WFC3, and \textit{Spitzer} observations for reference. Bottom: Residual of the forward scattering mode relative to the non-forward scattering model. } \label{fig:scattering}
\end{figure}

The standard forward model in \textsl{Aurora} combines the absorption and scattering optical depths into the total optical depth as seen in equation \ref{eq:dtau}. However, it is possible that a portion of the scattered light in the planet's atmosphere will be directed towards the observer. This portion of light is said to be forward scattered. The additional fraction of light reaching the observer results in an attenuation to the transit depth. In \textsl{Aurora}, we can model this by correcting the effective optical depth for the effects of forward scattering. The modified optical depth is $d\tau_{\text{eff}}=d\tau_\lambda (1-f\tilde{\omega}_o)$, where $f$ is the forward scattering correction factor and $\tilde{\omega}_o$ is the single scattering albedo \citep{Robinson2017}. We calculate the correction factor $f$ using the analytic correction expressed in equation 6 of \cite{Robinson2017} for the Henyey-Greenstein phase function \citep{Henyey1941}. The correction proposed by \cite{Robinson2017} is a function of the stellar radius, the planet-star physical separation, and the asymmetry parameter $g$. Figure \ref{fig:scattering} shows the decrease in transit depth due to considering forward scattering, in the same H$_2$-rich forward model for K2-18b described above, assuming $g=0.95$ and $\tilde{\omega}_o=1$. The effect of incorporating forward scattering in the model of K2-18b results in a difference of less than 1~ppm. Models considering forward scattering for a rocky exoplanet are shown in appendix \ref{app:trappist_fwd_models}.

The detectability of these secondary effects remains to be confirmed. Current observations using HST, and ground based observatories do not posses the precision necessary to identify them. In the meantime \textsl{Aurora} possesses the capabilities to model these effects in transmission spectra of exoplanets in the context of forward models. The implementation of these models in the context of retrievals remains a possibility for future studies should the data require so. 

\subsubsection{Mie-Scattering Forward Models}
\label{subsubsec:miescattering}
\textsl{Aurora} includes Mie-scattering in the forward models due to condensates with different particle sizes and compositions adopted from \citet{Pinhas2017}. The effective cross-section for these species is calculated using their extinction and scattering coefficients, along with the corresponding asymmetry parameter $g$ following equation 11 of \citet{Pinhas2017}, obtained using Mie theory.

\begin{figure}
\includegraphics[width=0.45\textwidth]{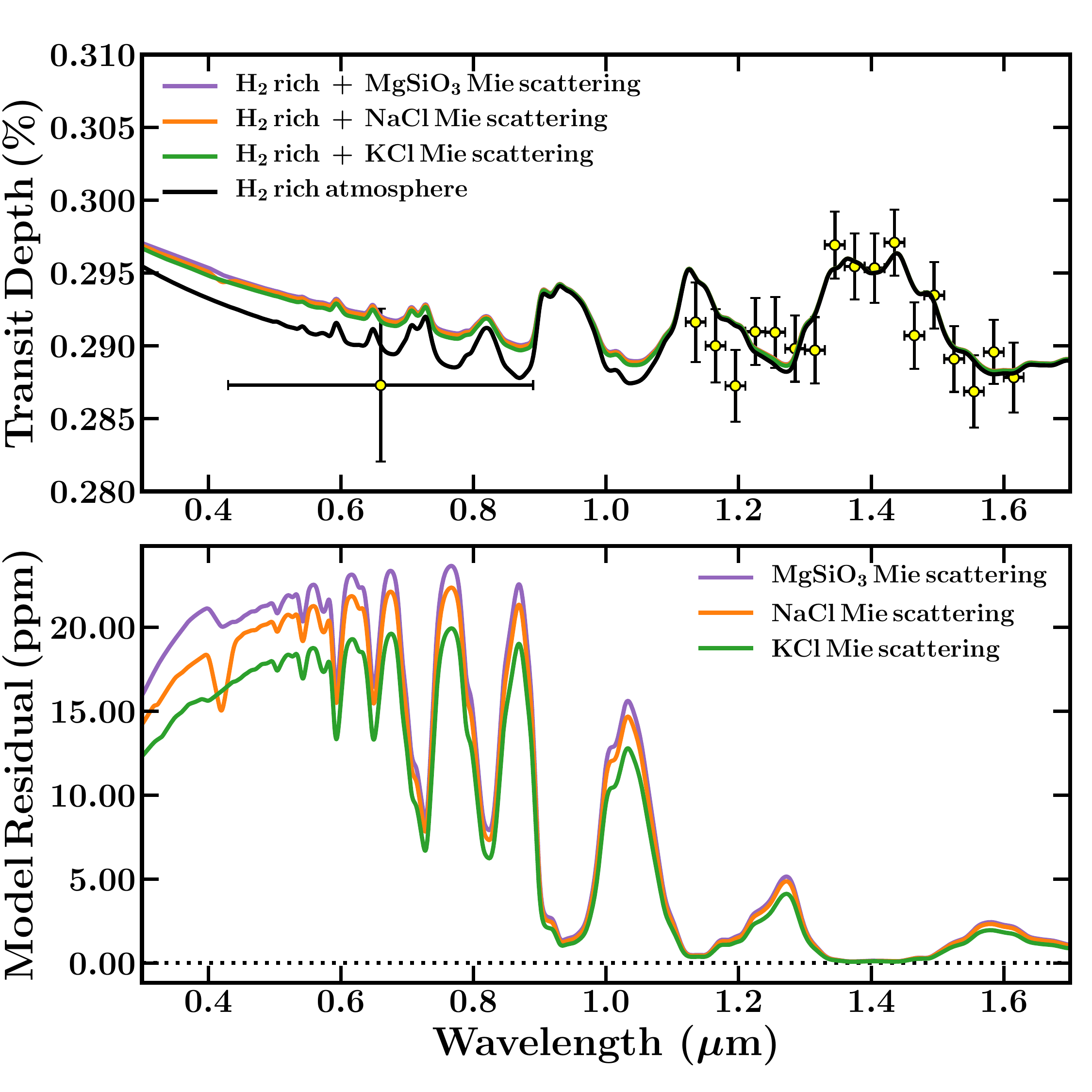}
\centering
\caption[Mie-scattering]{Top: Forward models including the effects of Mie-scattering for the mini-Neptune K2-18b. The condensates shown are MgSiO$_3$ (purple), NaCl (orange), and KCl (green). In black the H$_2$-rich only model is shown. Black error bars with yellow markers show current K2 and HST-WFC3 observations for reference. Bottom: Model residuals. } \label{fig:mie_scattering}
\end{figure} 

Figure \ref{fig:mie_scattering} shows the spectroscopic features of Mie-scattering for different compositions in the H$_2$-rich atmosphere of K2-18b. The models assume the retrieved chemical abundances from the results in section \ref{subsec:samplers_results}. The models shown include H$_2$-Rayleigh scattering and H$_2$–-H$_2$ and H$_2$–-He CIA. In black we show the H$_2$-rich atmosphere only, while in purple, orange, and green the effects of Mie-scattering due to MgSiO$_3$, NaCl, and KCl are shown respectively. The assumed abundances for the condensate species is 10$^{-16}$, similar to expectations for NaCl and KCl from equilibrium chemistry calculations \citep[e.g.,][]{Woitke2018} for the equilibrium temperature of the planet of T$_\mathrm{eq.}\sim290$~K \citep[e.g.,][]{Welbanks2019b}, with a particle size of 4.89$\times10^{-2}$ $\mu$m \citep[$<0.1\mu$m, e.g.,][]{Adams2019,Lavvas2019}. As shown in the bottom panel of Figure \ref{fig:mie_scattering}, the maximum difference between the clear H$_2$-rich model and the models considering Mie-scattering is $\sim25$~ppm, within the precision limits of current observations. Future observatories with high-precision measurements in the optical wavelengths may be able to distinguish the effects of these condensate species in the atmospheres of exoplanets.

\section{Results}
\label{sec:results}

We validate \textsl{Aurora}'s new retrieval features on real and synthetic spectro-photometric observations. First we validate our H-rich and non H-rich approaches as well as the new prescription for inhomogeneous cloud and haze cover on the prototypical hot Jupiter HD~209458~b \citep{Henry2000, Charbonneau2000} using observations from \citet{Sing2016}. Next we test the different nested sampling algorithms included in \textsl{Aurora} using the most recent observations of K2-18b \citep{Foremanmackey2015} from \citet{Benneke2019b}, and investigate the robustness of the retrieved abundance estimates comparing them to previous works \citep[e.g.,][]{Benneke2019b, Welbanks2019b, Madhusudhan2020}. Lastly, we investigate future atmospheric constraints of mini-Neptunes and rocky exoplanets using synthetic observations.

\begin{figure*}
\includegraphics[width=\textwidth]{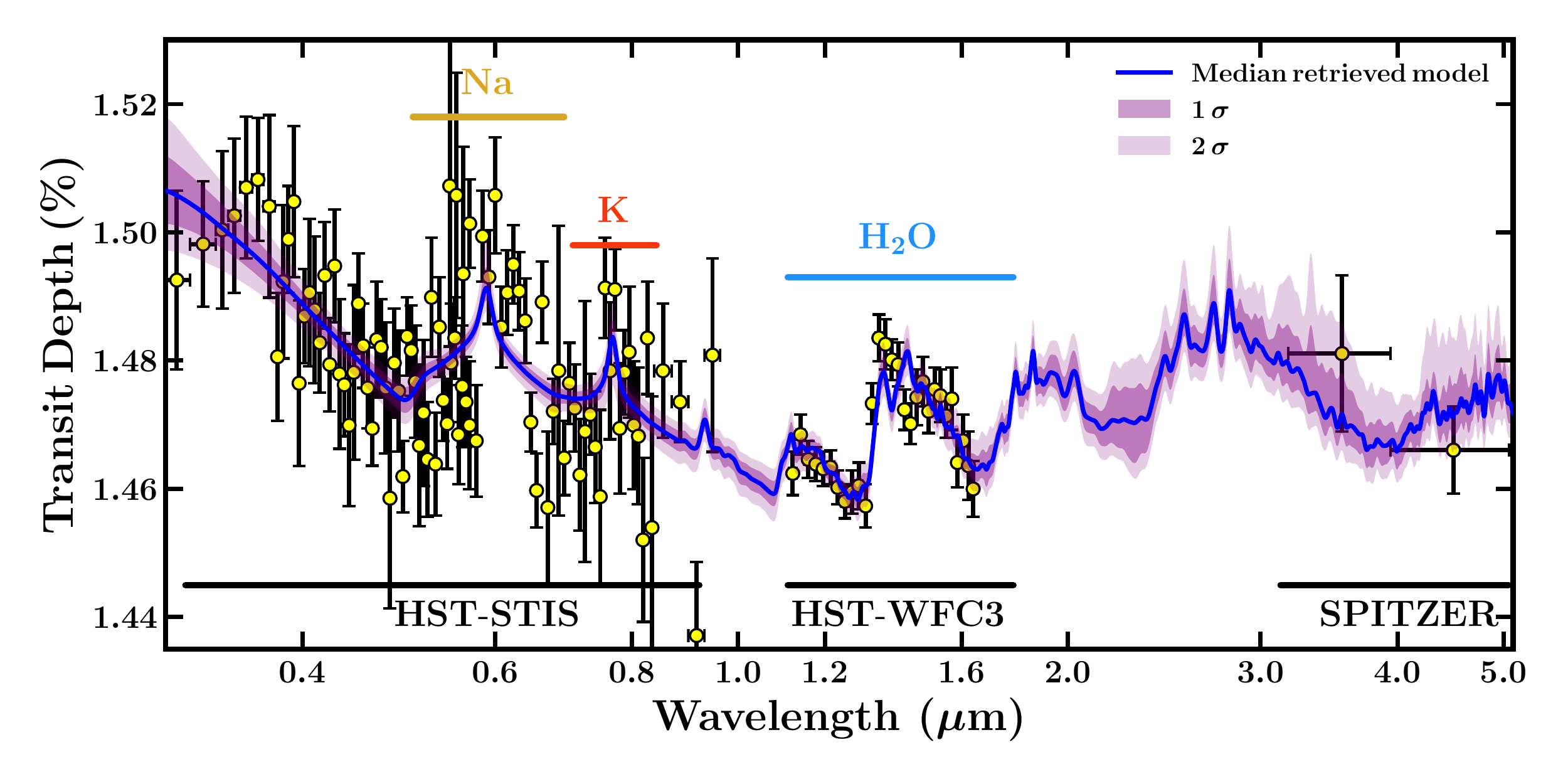}
\centering
\caption[Cloud validation]{Validation of \textsl{Aurora}'s retrieval framework on the spectrum of HD~209458~b. Retrieved median model (blue line), 1 and 2$\sigma$ confidence intervals (purple shaded regions), and spectroscopic observations (black error bars with yellow markers) for the highest evidence model (model 3, see Table \ref{table:HD209485b_validation}). Horizontal lines at the bottom of the figure show the wavelength coverage of HST-STIS, HST-WFC3, and \textit{Spitzer}. Yellow, orange, and blue horizontal lines show the approximate wavelength regions where Na, K, and H$_2$O spectral features are expected, respectively.}\label{fig:hd209_spectrum}
\end{figure*}  

\subsection{Validation of Aurora on hot Jupiter HD~209458~b.}
\label{subsec:cloud_validation}

We perform a series of retrievals on the transmission spectrum of HD~209458~b from \citet{Sing2016}, composed of spectro-photometric observations with HST-STIS, HST-WFC3, and \textit{Spitzer} . We use the standard model set-up described in \citep{Welbanks2019a, Pinhas2019, Welbanks2019b}. Our sources of opacity include H$_2$–-H$_2$ and H$_2$–-He CIA, H$_2$-Rayleigh scattering, and line opacity due to H$_2$O, Na, K, CH$_4$, NH$_3$, HCN, CO, and CO$_2$. We conduct a range of retrievals with different cloud and haze prescriptions, and assumptions of whether the atmosphere is H-rich or not. 

We perform retrievals using four models with different considerations for clouds and hazes allowed by our generalised prescription explained in section \ref{subsubsec:new_cloudsandhazes}. Model 0 considers a clear atmosphere (i.e., $\phi_{\mathrm{clouds}}=\phi_{\mathrm{hazes}}=\phi_{\mathrm{clouds+hazes}}=0$). Model 1 considers one sector for a clear atmosphere and one sector for the combined effects of clouds and hazes (i.e., $\phi_{\mathrm{clouds}}=\phi_{\mathrm{hazes}}=0$). Model 2 considers one sector for a clear atmosphere, one sector for the presence of clouds only, and one sector for the presence of hazes only (i.e., $\phi_{\mathrm{clouds+hazes}}=0$). Model 3 considers one sector for a clear atmosphere, one sector for clouds only, one sector for hazes only, and one sector for the combined presence of clouds and hazes (i.e., the full inhomogeneous cloud and haze prescription introduced in this work). For each cloud and haze model above, we perform a retrieval assuming a H-rich atmosphere and a retrieval relaxing such assumption. In summary we perform eight retrievals in this section with the models above, four assuming a H-rich atmosphere and four not assuming a H-rich atmosphere.

\subsubsection{A Generalised Cloud and Haze Prescription}

We consider a generalised cloud and haze prescription in order to explore a larger parameter space than available when restricting the presence of clouds and hazes to one sector only (i.e., model 1). We find that assuming a H-rich atmosphere or not can result in different solutions when restricting the clouds and hazes to the same region, as in model 1 \citep[e.g.,][]{MacDonald2017}. This is not the case for any of the other models in this section (i.e., model 0, model 2, or model 3). When assuming a H-rich atmosphere we find that, using any of the models for inhomogeneous cloud/haze cover, the spectrum of HD~209458~b can be explained by two possible scenarios. The first is the known solution with median values of sub-solar\footnote{We clarify that in this context we refer to abundances of H$_2$O as sub-solar by assessing them relative to expectations from thermochemical equilibrium for solar elemental abundances \citep{Asplund2009}. For a solar composition, the expectation is a H$_2$O abundance of $\log_{10}(X_{\text{H}_2\text{O}})\sim-3.3$ for a planet with the equilibrium temperature of HD~209458~b \citep{Madhusudhan2012}.} H$_2$O of $\log_{10}(X_{\text{H}_2\text{O}})\sim-4.5$, $\log_{10}(X_{\text{Na}})\sim-5.2$, $\log_{10}(X_{\text{K}})\sim-7.0$ and a cloud and haze cover of roughly 50\% \citep[e.g.,][]{MacDonald2017,Pinhas2019, Welbanks2019a, Welbanks2019b}; the second is a physically implausible solution with high Na abundances that make up for $\sim$20\% of the atmosphere's composition and an atmosphere fully covered by clouds and hazes (e.g., $\log_{10}(X_{\text{H}_2\text{O}})\sim-2.5$, $\log_{10}(X_{\text{Na}})\sim-0.7$, $\log_{10}(X_{\text{K}})\sim-2.4$). Both modes are simultaneously retrieved by model 2 and model 3, regardless of whether or not a H-rich composition is assumed. However, when treating clouds and hazes together (i.e., model 1), while assuming a H-rich atmosphere results in the two modes as discussed above, relaxing the H-rich assumption results in only the high Na abundance solution. Therefore, model 1 may be susceptible to potential biases in retrieved solutions when the dominant atmospheric composition may not be assumed to be H-rich a priori. On the other hand, models 2 and 3 provide more generalised parameterisations that do not depend strongly on the H-rich assumption. In what follows we restrict the prior space of the $\log_{10}$ abundances of Na, K and CO to an upper limit of -1.5, consistent with assumptions in previous studies \citep[e.g.,][]{MacDonald2017,Pinhas2019, Welbanks2019a, Welbanks2019b}. We implement this upper limit by rejecting the unphysical solutions. 

\begin{figure}
\includegraphics[width=0.5\textwidth]{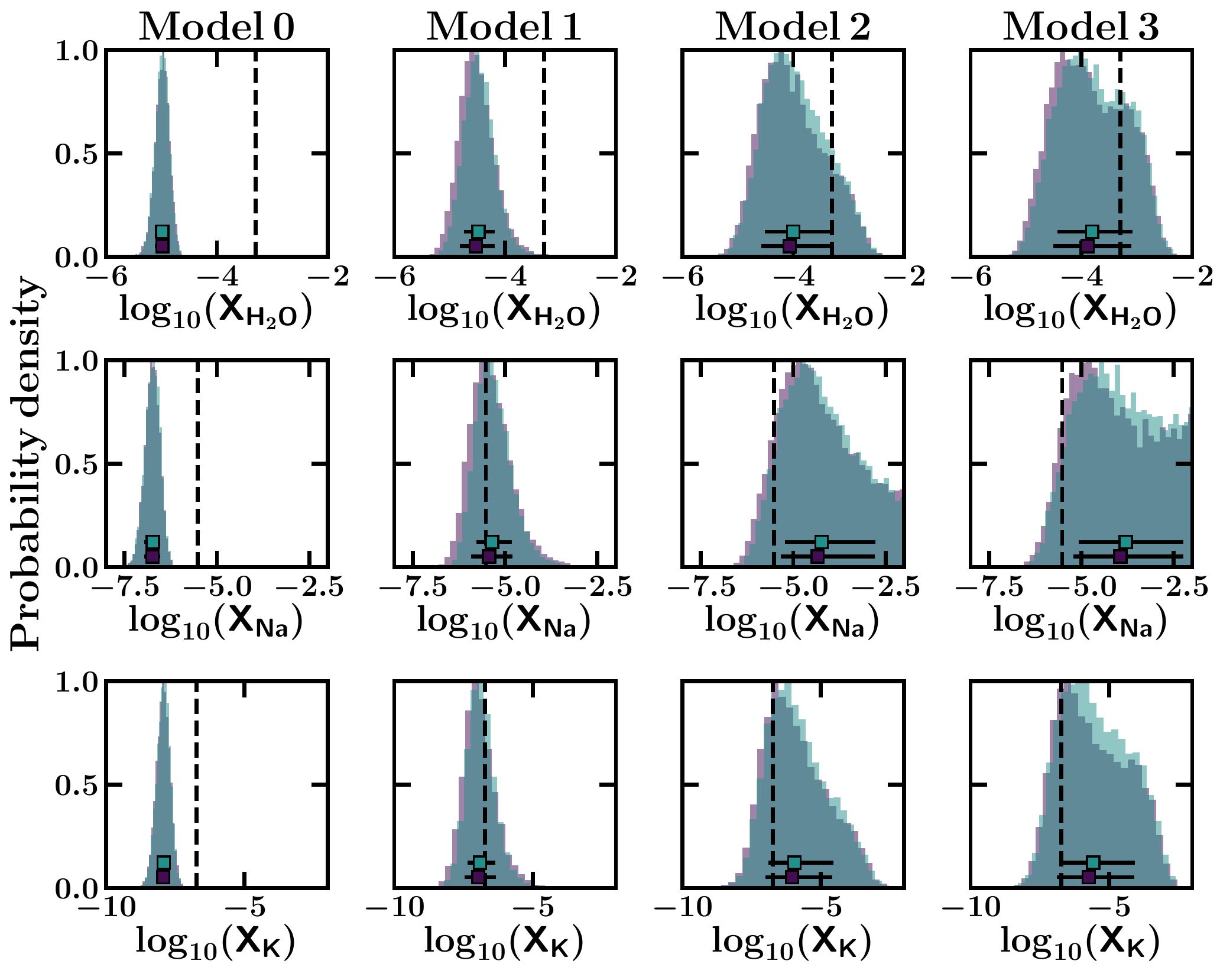}
\centering
\caption[Posteriors of clouds]{Posterior distributions for H$_2$O, Na and K abundances under the four different cloud and haze models. The purple (blue) distributions correspond to retrievals (not) assuming a H-rich atmosphere. Solar abundance expectations are shown using vertical black dashed lines. The markers in the posterior distributions show the median value (square marker) and the 1$\sigma$ (error bars) range covering $\sim$68.27\% of the samples around the median value.}
\label{fig:cloud_n_haze_posteriors}
\end{figure}

\subsubsection{Effect of Cloud and Haze Prescriptions}

We re-run all eight cases with the new constraints on the abundances of Na, K and CO. We present the complete set of retrieved parameters for the 4 cloud and haze models assuming a H-rich atmosphere and not assuming a H-rich atmosphere in Table \ref{table:HD209485b_validation} included in the appendix. Figure \ref{fig:hd209_spectrum} shows the median retrieved spectrum for model 3 which results in the highest model evidence, while figure \ref{fig:cloud_n_haze_posteriors} shows the H$_2$O, Na, and K posterior distributions for all 4 cloud and haze models.

Considering a cloud-free atmosphere (model 0) results in tight H$_2$O abundance constraints with precisions smaller than 0.5~dex. Regardless of the treatment for the main gas constituent in the atmosphere, both cloud free retrievals result in sub-solar H$_2$O abundances with abundance estimates smaller than the models considering clouds and hazes. These low abundances are the consequence of having a larger observable atmosphere (i.e., larger atmospheric column), unocculted by clouds, in which a small abundance of H$_2$O can contribute enough to explain the observations \citep[see e.g.,][]{Welbanks2019a}. 

Contrary to the cloud-free solutions, the cloudy and hazy scenarios (i.e., models 1, 2 and 3) result in higher H$_2$O abundances although with still generally sub-solar values. Models 1, 2 and 3 are consistent in their retrieved parameters when assuming a H-rich atmosphere and when relaxing this assumption. The retrieved H$_2$O abundances are consistent with each other and within 1$\sigma$ between all three cloud and haze models. The same is true for the Na and K abundances. Model 1, consisting of one fraction combining clouds and hazes as in \citet{MacDonald2017}, results in tighter constraints relative to models 2 and 3. These tighter constraints indicate that part of the parameter space explored by the other two prescriptions was not considered in model 1. The increase in model evidence for model 2 and model 3 relative to model 1 indicate that the increased parameter space contains previously unsampled regions of high likelihood. The retrieved P-T profiles are consistent between models 1, 2, and 3, with a retrieved temperature near the photosphere for model 3 assuming a H-rich atmosphere of T$_{100\text{mbar}}=1308^{+345}_{-278}$~K, consistent with previous studies \citep[e.g.,][]{Welbanks2019a,Welbanks2019b}. On the other hand, the retrieved P-T profile for model 0 is tightly constrained at colder temperatures (e.g., T$_{100 \text{mbar}}=852^{+20}_{-12}$~K for the retrieval assuming a H-rich atmosphere) and inconsistent with the planet's equilibrium temperature \citep[T$_\mathrm{eq.}\sim$1450~K, e.g.,][]{Welbanks2019b}. 

When comparing the retrievals assuming H-rich atmospheres, model 3 has the highest model evidence with a value of $\log(\mathcal{Z})=958.40$. Using model 3 as our reference, model 0 is strongly disfavoured at 4.6$\sigma$; model 1 is disfavoured at 1.8$\sigma$; and model 2 is weakly disfavoured at 1.4$\sigma$. A similar interpretation is available when comparing the non H-rich retrievals amongst themselves. 

\subsubsection{H-rich vs. Non H-rich Assumptions}

We also compare retrievals assuming a H-rich atmosphere against retrievals not assuming a H-rich atmosphere. Relaxing the assumption of a H-rich atmosphere requires an additional parameter to retrieve the volume mixing ratio of a mixture of H$_2$ and He in solar proportion. This additional parameter results in a decrease in model evidence relative to retrievals assuming a H-rich atmosphere. Retrievals using model 3 favour assuming a H-rich atmosphere at 2.82$\sigma$ over not assuming a H-rich atmosphere. Despite this decrease in evidence, retrievals not assuming a H-rich atmosphere find that 99.9\% of the atmosphere is made up of H$_2$ and He. By not assuming a H-rich atmosphere a priori, our models are able to robustly confirm that the data corresponds to the atmosphere of a H-rich planet. Our results indicate that assuming a H-rich atmosphere is appropriate for the spectrum of HD~209485b as expected. These results demonstrate for gas giants that both approaches, assuming a H-rich atmosphere or not, are consistent and that the retrieved parameter estimates are robust against either methodology.

\begin{deluxetable*}{cl|llll}
\tablecaption{Retrieved parameters for the spectrum of K2-18b using different nested sampling algorithms as explained in section \ref{subsec:samplers_results}. 
\label{table:samplers}
}
\tablecolumns{6}
\tabletypesize{\small}
\tablehead{
 \colhead{}& \colhead{Parameter}  & \colhead{MultiNest} & \colhead{PolyChord} & \colhead{Dynesty static} & \colhead{Dynesty dynamic}
 }
\startdata
  \multirow{5}{*}{\parbox[t]{6mm}{{\rotatebox[origin=c]{90}{\makecell{Chemical\\ Species} }}}}& $\log_{10} \left(X_{\textnormal{H}_2\textnormal{O}}\right)$& $-2.28 ^{+ 1.16 }_{- 1.15 }$  & $-2.21 ^{+ 1.24 }_{- 1.20 }$ & $ -2.29 ^{+ 1.20 }_{- 1.15 }$ & $-2.32 ^{+ 1.20 }_{- 1.12 }$ \\
&$\log_{10} \left(X_{\textnormal{CH}_4}\right)$ &  $-8.33 ^{+ 2.63 }_{- 2.35 }$  & $-8.11 ^{+ 2.64 }_{- 2.65 }$ & $-8.09 ^{+ 2.63 }_{- 2.51 }$  & $-8.18 ^{+ 2.65 }_{- 2.49 }$  \\
 &$\log_{10} \left(X_{\textnormal{NH}_3}\right)$ &  $ -8.82 ^{+ 2.27 }_{- 2.09 }$  & $-8.73 ^{+ 2.37 }_{- 2.22 }$ & $-8.67 ^{+ 2.26 }_{- 2.15 }$  & $-8.73 ^{+ 2.27 }_{- 2.14 }$   \\
 &$\log_{10} \left(X_{\textnormal{CO}}\right)$&  $-6.89 ^{+ 3.53 }_{- 3.33 }$  & $-6.63 ^{+ 3.58 }_{- 3.71 }$ & $-6.70 ^{+ 3.54 }_{- 3.42 }$  & $-6.71 ^{+ 3.42 }_{- 3.36 }$ \\
  &$\log_{10} \left(X_{\textnormal{CO}_2}\right)$&  $-7.49 ^{+ 3.42 }_{- 2.97 }$  & $-7.31 ^{+ 3.28 }_{- 3.15 }$ & $-7.25 ^{+ 3.23 }_{- 3.12 }$  & $-7.21 ^{+ 3.17 }_{- 3.13 }$ \\
 \hline
&$T_0$ (K) &  $179.70 ^{+ 57.86 }_{- 44.47 }$  & $185.27 ^{+ 64.87 }_{- 50.45 }$ & $182.81 ^{+ 59.85 }_{- 49.08 }$  & $181.45 ^{+ 57.28 }_{- 46.58 }$   \\
 \hline
 &$\log_{10}$($P_{\rm{ref}}$) (bar)&  $-0.86 ^{+ 0.37 }_{- 0.44 }$& $-0.89 ^{+ 0.39 }_{- 0.44 }$ & $ -0.86 ^{+ 0.37 }_{- 0.44 }$ & $-0.86 ^{+ 0.37 }_{- 0.43 }$ \\
  \hline
\enddata
\end{deluxetable*}

\subsubsection{Assessing the Highest Evidence Model}

The highest evidence model (i.e., model 3, H-rich assumption) results in retrieved abundance estimates for H$_2$O, Na, and K that are consistent with previous results \citep[e.g.,][]{MacDonald2017,Barstow2017,Pinhas2019,Welbanks2019a,Welbanks2019b}. However, their precisions are wider with abundances of $\log_{10}(X_{\text{H}_2\text{O}})=-3.89 ^{+ 0.78 }_{- 0.62 }$, $\log_{10}(X_{\text{Na}})=-3.95 ^{+ 1.71 }_{- 1.27 }$, $\log_{10}(X_{\text{K}})=-5.73 ^{+ 1.65 }_{- 1.15 }$. The wider estimates result in a median H$_2$O abundance still sub-solar based on expectations of thermochemical equilibrium, but consistent with a solar value to within 1$\sigma$. Importantly, while the H$_2$O abundance is largely subsolar, both Na and K abundances are significantly super-solar, implying a relative depletion in H$_2$O compared to Na and K as found in \citet{Welbanks2019b}. The retrieved cloud and haze parameters indicate a non-clear atmosphere covered by clouds and hazes with a cloud deck located above the expected photosphere. The retrieved fractions are $\phi_{\mathrm{clouds}}=0.34 ^{+ 0.18 }_{- 0.20 }$, $\phi_{\mathrm{hazes}}=0.27 ^{+ 0.09 }_{- 0.10 }$, and $\phi_{\mathrm{clouds+hazes}}=0.24 ^{+ 0.19 }_{- 0.16 }$. 

Noteworthy too are the retrieved values for the Rayleigh-enhancement factor. Model 3 (H-rich) retrieves a Rayleigh-enhancement of $\log_{10}$(a)$=3.28 ^{+ 1.01 }_{- 1.13 }$, while model 2 (H-rich) retrieves $\log_{10}$(a)$=2.88 ^{+ 0.91 }_{- 0.85 }$. Both retrieved Rayleigh-enhancement factors have median values and upper limits smaller than the retrieved median value for model 1 (e.g., $\log_{10}$(a)$=4.35 ^{+ 0.71 }_{- 1.01 }$ for the H-rich case). This may indicate a tendency to over estimate the Rayleigh-enhancement factor in the hazes when using model 1 \citep[e.g.,][]{MacDonald2017}. If true, this possibility must be accounted for when studying the nature of super-Rayleigh slopes as performed in recent studies \citep[e.g.,][]{Ohno2020}. Similarly, although consistent with each other, the retrieved median value for the scattering slope $\gamma$ is higher for model 1 than for models 2 and 3. The constraints from the H-rich retrievals are $\gamma=-14.04 ^{+ 4.53 }_{- 3.94 }$ for model 1, $\gamma=-16.57 ^{+ 3.15 }_{- 2.37 }$ for model 2, and $\gamma=-16.15 ^{+ 3.36 }_{- 2.60 }$ for model 3. We note that the interpretation of the Rayleigh-enhancement factor ($a$) should be done in conjunction with the value for the scattering slope ($\gamma$) as these parameters are correlated. Lastly, the retrieved cloud-top pressure for the gray cloud deck is consistent within 1$\sigma$ between all approaches with a retrieved value of $\log_{10}$($P_{\text{cloud}}$)=$-4.72 ^{+ 0.99 }_{- 0.84 }$ for the model with the highest evidence. 

Finally, we compare model 2 to model 3. The retrieved parameters are consistent between the two approaches and have similar precisions. Due to their similar performance and relatively small difference in model evidence, we consider both approaches interchangeable for the effects of this work. In what follows we consider model 2 (i.e., one sector for a clear atmosphere, one sector for clouds only and one sector for hazes only) as our preferred model due to its smaller number of parameters and similar performance to model 3 (i.e., full inhomogeneous cloud and haze prescription). We utilise model 2 as our approach for inhomogeneous cloud and hazes in the remaining of the results section unless otherwise stated. 
 
\subsection{Testing Multiple Nested Sampling Algorithms}
\label{subsec:samplers_results}

We validate the different nested sampling algorithms in \textsl{Aurora} by performing retrievals using the same model and the same data. We discuss three nested sampling algorithms in section \ref{subsubsec:samplers}. Four retrievals are performed, one using MultiNest, one using PolyChord, one using Dynesty in its static nested sampling mode, and one using Dynesty in its dynamic nested sampling mode. We use the observed transmission spectrum of K2-18b from \citet{Benneke2019b} including K2 band photometry, HST-WFC3 G141 grism spectra, and \textit{Spitzer} IRAC photometric observations. The model considers an isothermal and clear atmosphere. We assume a H-rich atmosphere and consider the absorption due to H$_2$–-H$_2$ and H$_2$–-He CIA, H$_2$O, CH$_4$, NH$_3$, CO and CO$_2$. In total, the model has 7 free parameters: 5 molecular species, 1 parameter for the temperature of the isotherm, and 1 parameter for the reference pressure. The retrieved parameters are used to produce the forward models in sections \ref{subsubsec:refraction_scattering} and \ref{subsubsec:miescattering}.

When initialising the nested sampling algorithms, different parameters responsible for the algorithm's settings can be modified. Examples of such parameters are the maximum number of iterations in the sampling algorithm (PyMultiNest), parameters to increase the number of posterior samples produced (PyPolyChord), or the maximum number of likelihood evaluations before terminating (Dynesty). We keep most settings for the nested sampling algorithms to their default values. We only modify parameters needed for a direct comparison, e.g., the number of live points used to sample the prior distributions.

MultiNest was setup with 2000 live points. PolyChord was also setup with 2000 live points and 7 repeats. The number of repeats is specific to PolyChord's settings and it corresponds to the length of the sampling chain used to generate a new live point. The longer the chain, the less correlated the live points and the more reliable the evidence inference is, however, the run takes longer to be completed. The default value for the number of repeats used by PolyChord is 5$\times$ the number of dimensions in the problem; that is 5$\times$ the number of model parameters. Since we do not need an estimate for the model evidence in this exercise, as we are not comparing the model evidence between samplers, we do not need to choose a significantly larger number of repeats. We find that for our atmospheric model with 7 free parameters (i.e., 7 dimensions), our choice of 7 repeats (i.e., 1$\times$ the number of dimensions) is sufficient. 

For Dynesty, two separate runs were performed: one using static nested sampling, and the other using dynamic nested sampling. For the static Dynesty run we used 2000 live points. Similarly, the dynamic Dynesty run had an initial number of 2000 live points. The Dynesty runs were set up to generate the new live points using multi-ellipsoidal decomposition with uniform sampling, this is so that their sampling methods were similar to MultiNest.

\begin{figure*}
\includegraphics[width=\textwidth]{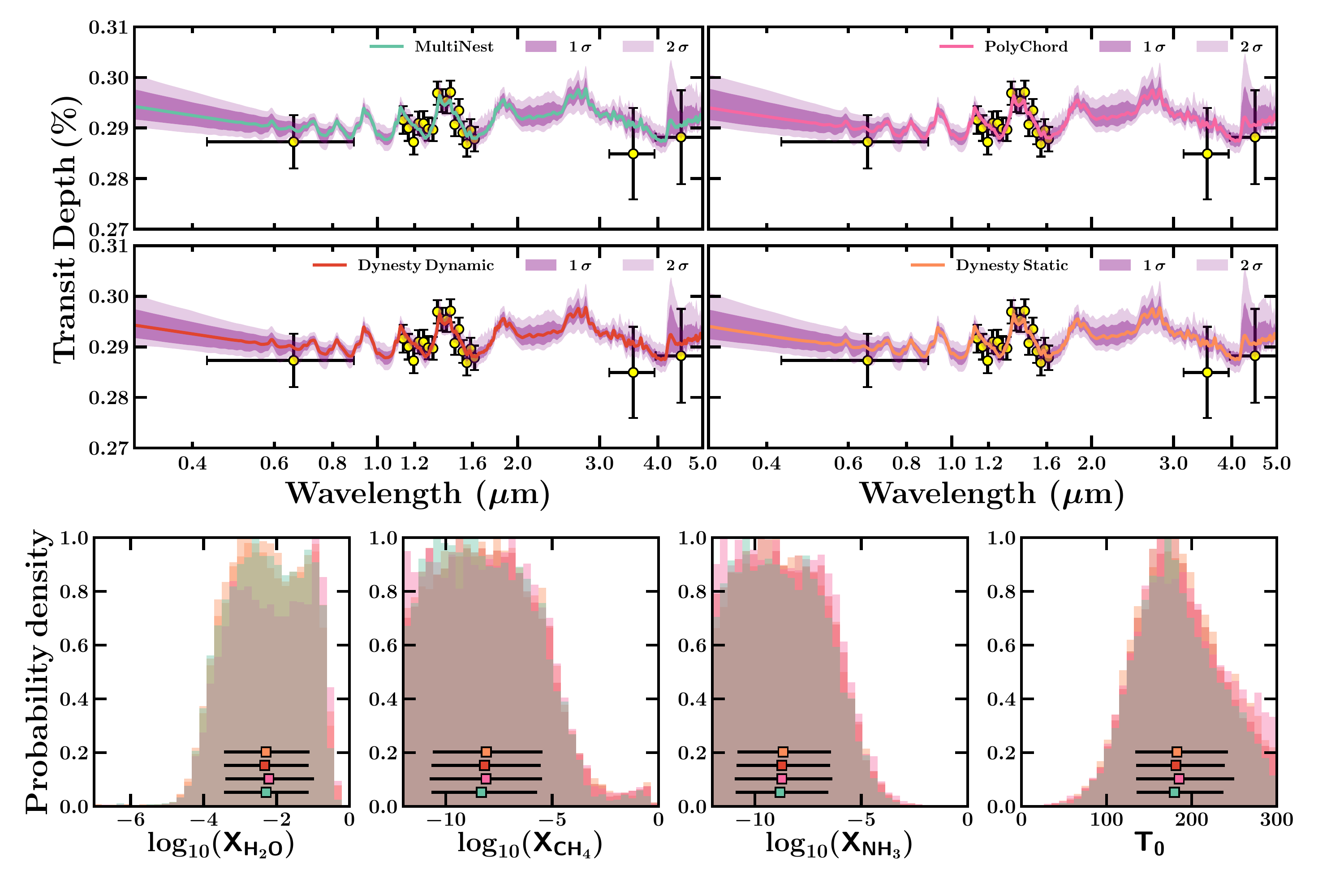}
\centering
\caption[Samplers spectra and posteriors]{Comparison between retrievals with different nested sampling algorithms. Top four panels show the retrieved median model (coloured line), 1 and 2$\sigma$ confidence intervals (shaded purple regions), and K2, HST-WFC3, and \textit{Spitzer} observations (black error bars with yellow markers). Bottom row shows the posterior distributions for H$_2$O, CH$_4$, NH$_3$, and $T_0$ the isothermal temperature. The posterior distributions and coloured lines are shown in green for Multinest, pink for PolyChord, red for Dynesty in its dynamic mode, and coral for Dynesty in its static mode. All the retrievals are consistent with each other.} \label{fig:samplers}
\end{figure*} 

Figure \ref{fig:samplers} presents the retrieved spectra for the data of K2-18b when using each of the different nested sampling algorithms in \textsl{Aurora}, along with the posterior distributions for the parameters of interest when comparing to previous works \citep[e.g.,][]{Benneke2019b, Welbanks2019b, Madhusudhan2020}. The complete list of retrieved parameters is shown in Table \ref{table:samplers}. 

All four nested sampling algorithms included in \textsl{Aurora} retrieve values consistent with each-other and with previous works. The retrieved H$_2$O abundance for MultiNest is $\log_{10} \left(X_{\textnormal{H}_2\textnormal{O}}\right)=-2.28 ^{+ 1.16 }_{- 1.15 }$, for PolyChord it is $\log_{10} \left(X_{\textnormal{H}_2\textnormal{O}}\right)=-2.21 ^{+ 1.24 }_{- 1.20 }$, for Dynesty in its static run is $\log_{10} \left(X_{\textnormal{H}_2\textnormal{O}}\right)=-2.29 ^{+ 1.20 }_{- 1.15 }$, and for Dynesty in its dynamic run is $\log_{10} \left(X_{\textnormal{H}_2\textnormal{O}}\right)=-2.32 ^{+ 1.20 }_{- 1.12 }$. These H$_2$O, abundances are consistent within 1$\sigma$ with the works of \citet{Benneke2019b, Welbanks2019b, Madhusudhan2020}. In agreement with previous works, the retrieved CH$_4$ and NH$_3$ median abundances are lower than expectations from chemical equilibrium. Comparing between the four sampler setups, the posterior distributions obtained for the parameters studied are largely consistent with each other. The retrieved precision on the model parameters is consistent between samplers, with a precision on the H$_2$O abundance of $\sim1.2$~dex. 

Our comparison shows that despite the differences in sampling algorithms, the parameter posterior distributions are mostly independent of the method employed for a problem of this dimensionality (i.e., 7 model parameters) and with current data quality. We assess the different run times for each of the nested sampling algorithms by performing these retrievals on one thread of a 4 core Intel Core i7-4700MQ CPU at 2.40GHz. The fastest Nested algorithm under these conditions was MultiNest, followed by the static run of Dynesty ($\sim3\times$ longer), the dynamic run of Dynesty ($\sim5\times$ longer) and Polychord ($\sim6\times$ longer). These run times are not necessarily representative of the full potential of each sampler. Different parameters in the setup of each algorithm can result in different run times.

In general, we have shown that \textsl{Aurora} includes the capabilities to retrieve the atmospheric properties of an exoplanet spectrum using different nested sampling algorithms. For current data and models, MultiNest is an optimal tool that retrieves the posterior distribution of our parameters efficiently. As data increases and possible degeneracies between the parameters in our models are exacerbated, or as the dimensionality of our problems and models increases, PolyChord and Dynesty are tools that offer alternative ways to perform retrievals. The user in \textsl{Aurora} has the freedom to choose the optimal tool for the problem at hand.

\newpage
\subsection{Application to Mini-Neptune K2-18b}
\label{subsec:retrieval_miniNeptunes}

We validate \textsl{Aurora}'s retrieval framework on current observations of K2-18b \citep{Benneke2019b} including K2, HST-WFC3, and \textit{Spitzer}, spectro-photometric data. Unlike the previous section, we perform retrievals not assuming a H-rich atmosphere. Using a full Bayesian approach, we test the validity of previous assumptions of a H-rich atmosphere when analysing the most recent transmission spectrum of this mini-Neptune. Then, we perform retrievals on HST-STIS and JWST-NIRSpec synthetic observations of the same planet and assess the constraints on the chemical abundances and cloud/haze properties.

\subsubsection{Case Study: Current Observations of K2-18b}
\label{subsubsec:k218b_xhrich}

\begin{deluxetable*}{cl|cccc}
\tablecaption{Retrieved parameters for current observations of K2-18b, assuming and not assuming a H-rich atmosphere. 
\label{table:k218b}
}
\tablecolumns{2}
\tabletypesize{\small}
\tablehead{
 \colhead{}& \colhead{Parameter}  & \colhead{H-rich retrieval} & \colhead{H-rich retrieval} & \colhead{No H-rich retrieval} & \colhead{No H-rich retrieval} \\
 \colhead{}& \colhead{}          & \colhead{ w/o O$_2$, O$_3$}& \colhead{w O$_2$, O$_3$}& \colhead{w/o O$_2$, O$_3$}& \colhead{w O$_2$, O$_3$}
 }
\startdata
&$\log_{10} \left(X_{\textnormal{H}_2+\textnormal{He}}\right)$& N/A & N/A  &$-0.09 ^{+ 0.09 }_{- 5.71 }$ &$-0.14 ^{+ 0.13 }_{- 6.64 }$  \\
&$\log_{10} \left(X_{\textnormal{H}_2\textnormal{O}}\right)$&$-2.10 ^{+ 1.07 }_{- 1.20 }$ &$-2.12 ^{+ 1.09 }_{- 1.24 }$ &$-1.20 ^{+ 1.15 }_{- 1.81 }$ &$-1.22 ^{+ 1.18 }_{- 2.03 }$  \\
 \multirow{5}{*}{\parbox[t]{6mm}{{\rotatebox[origin=c]{90}{\makecell{Chemical\\ Species} }}}}&$\log_{10} \left(X_{\textnormal{N}_2}\right)$&$-6.29 ^{+ 3.38 }_{- 3.45 }$&$-6.59 ^{+ 3.49 }_{- 3.40 }$  &$-5.53 ^{+ 3.24 }_{- 3.23 }$  &$-5.66 ^{+ 3.28 }_{- 3.25 }$  \\
 &$\log_{10} \left(X_{\textnormal{CH}_4}\right)$&$-8.16 ^{+ 2.54 }_{- 2.41 }$ &$-8.20 ^{+ 2.60 }_{- 2.31 }$&$-6.71 ^{+ 3.84 }_{- 2.57 }$&$-6.59 ^{+ 5.10 }_{- 2.79 }$  \\
 &$\log_{10} \left(X_{\textnormal{HCN}}\right)$&$-8.07 ^{+ 2.53 }_{- 2.46 }$&$-8.03 ^{+ 2.51 }_{- 2.45 }$ &$-6.78 ^{+ 2.89 }_{- 2.48 }$ &$-6.81 ^{+ 3.05 }_{- 2.61 }$  \\
 &$\log_{10} \left(X_{\textnormal{NH}_3}\right)$ &$-8.73 ^{+ 2.16 }_{- 2.03 }$&$-8.64 ^{+ 2.11 }_{- 2.08 }$&$-7.58 ^{+ 2.30 }_{- 2.02 }$ &$-7.61 ^{+ 2.39 }_{- 2.10 }$ \\
 &$\log_{10} \left(X_{\textnormal{CO}}\right)$&$-6.64 ^{+ 3.23 }_{- 3.34 }$&$-6.64 ^{+ 3.18 }_{- 3.24 }$&$-5.76 ^{+ 3.11 }_{- 3.10 }$  &$-5.78 ^{+ 3.14 }_{- 3.13 }$ \\
 &$\log_{10} \left(X_{\textnormal{CO}_2}\right)$&$-7.16 ^{+ 3.00 }_{- 2.99 }$&$-7.25 ^{+ 2.97 }_{- 2.99 }$&$-6.16 ^{+ 2.87 }_{- 2.91 }$  &$-6.27 ^{+ 2.95 }_{- 2.88 }$  \\
 &$\log_{10} \left(X_{\textnormal{O}_2}\right)$& N/A  &$-6.33 ^{+ 3.40 }_{- 3.39 }$& N/A&$-5.72 ^{+ 3.28 }_{- 3.21 }$\\
 &$\log_{10} \left(X_{\textnormal{O}_3}\right)$& N/A&$-7.44 ^{+ 2.96 }_{- 2.84 }$& N/A&$-6.26 ^{+ 3.06 }_{- 2.92 }$  \\
 \hline
\multirow{6}{*}{\parbox[t]{6mm}{{\rotatebox[origin=c]{90}{P-T Parameters }}}} &$T_0$ (K) &$162.12 ^{+ 73.73 }_{- 67.26 }$ &$166.50 ^{+ 72.47 }_{- 67.41 }$&$192.49 ^{+ 68.43 }_{- 77.90 }$  &$194.86 ^{+ 67.88 }_{- 79.54 }$ \\
 & $\alpha_1$&$1.32 ^{+ 0.43 }_{- 0.48 }$ &$1.32 ^{+ 0.43 }_{- 0.48 }$  &$1.20 ^{+ 0.49 }_{- 0.52 }$ &$1.17 ^{+ 0.50 }_{- 0.51 }$ \\
 &$\alpha_2$ &$1.20 ^{+ 0.50 }_{- 0.54 }$ &$1.19 ^{+ 0.51 }_{- 0.54 }$ &$1.17 ^{+ 0.51 }_{- 0.55 }$&$1.16 ^{+ 0.51 }_{- 0.55 }$\\
 &$\log_{10}$($P_{1}$) (bar)&$-1.64 ^{+ 1.61 }_{- 1.65 }$&$-1.74 ^{+ 1.62 }_{- 1.63 }$&$-1.77 ^{+ 1.58 }_{- 1.61 }$ &$-1.69 ^{+ 1.56 }_{- 1.64 }$ \\
 &$\log_{10}$($P_{2}$) (bar) &$-4.00 ^{+ 1.74 }_{- 1.32 }$&$-4.03 ^{+ 1.70 }_{- 1.27 }$  &$-4.11 ^{+ 1.65 }_{- 1.22 }$&$-4.09 ^{+ 1.65 }_{- 1.24 }$\\
 &$\log_{10}$($P_{3}$) (bar)  &$0.56 ^{+ 0.95 }_{- 1.33 }$ &$0.52 ^{+ 0.97 }_{- 1.29 }$&$0.50 ^{+ 0.97 }_{- 1.29 }$  &$0.58 ^{+ 0.93 }_{- 1.28 }$ \\
 \hline
 &$\log_{10}$($P_{\rm{ref}}$) (bar)&$-1.27 ^{+ 0.54 }_{- 0.62 }$ &$-1.23 ^{+ 0.54 }_{- 0.61 }$&$-1.27 ^{+ 0.72 }_{- 0.58 }$  &$-1.19 ^{+ 0.76 }_{- 0.62 }$ \\
 \hline
\multirow{5}{*}{\parbox[t]{6mm}{{\rotatebox[origin=c]{90}{\makecell{Cloud-Haze \\ \,Parameters \,}}}}} &$\log_{10}$(a)&$0.76 ^{+ 4.28 }_{- 2.98 }$  &$1.03 ^{+ 4.40 }_{- 3.19 }$&$1.55 ^{+ 4.54 }_{- 3.46 }$ &$1.55 ^{+ 4.39 }_{- 3.46 }$\\
&$\gamma$ &$-10.17 ^{+ 6.91 }_{- 6.39 }$&$-9.97 ^{+ 6.78 }_{- 6.31 }$ &$-10.06 ^{+ 6.98 }_{- 6.16 }$&$-9.88 ^{+ 6.71 }_{- 6.18 }$ \\
 &$\phi_{\mathrm{hazes}}$&$0.27 ^{+ 0.25 }_{- 0.17 }$&$0.27 ^{+ 0.24 }_{- 0.18 }$&$0.28 ^{+ 0.25 }_{- 0.18 }$ &$0.29 ^{+ 0.26 }_{- 0.19 }$\\
 &$\log_{10}$($P_{\text{cloud}}$) (bar)&$-1.57 ^{+ 1.99 }_{- 2.46 }$ &$-1.38 ^{+ 1.89 }_{- 2.21 }$ &$-1.23 ^{+ 1.88 }_{- 2.16 }$&$-1.19 ^{+ 1.88 }_{- 2.19 }$\\
 & $\phi_{\mathrm{clouds}}$ &$0.30 ^{+ 0.24 }_{- 0.19 }$ &$0.29 ^{+ 0.25 }_{- 0.18 }$&$0.30 ^{+ 0.26 }_{- 0.19 }$&$0.29 ^{+ 0.25 }_{- 0.19 }$ \\
  \hline
& $\log$($\mathcal{Z}$)&179.15 & 179.08&176.84&176.49 \\
  \hline
\enddata
 \tablecomments{All retrievals were computed using MultiNest. N/A means that the parameter in question was not considered in the model by construction.}
\end{deluxetable*}

\begin{figure*}
\includegraphics[width=\textwidth]{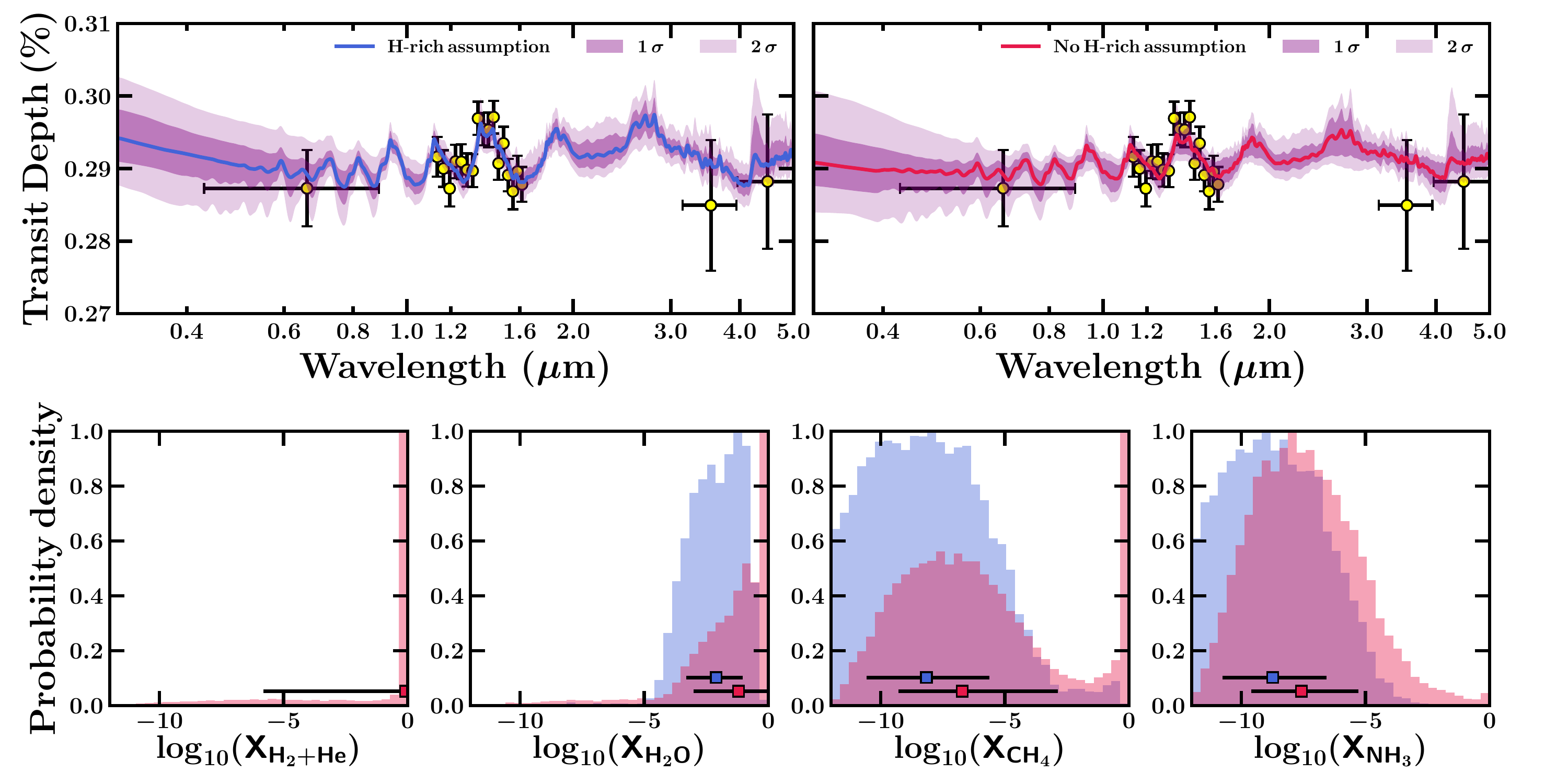}
\centering
\caption[extant estimates]{Retrieval of current K2-18b observations with and without the assumption of a H-rich atmosphere. Top: Retrieved model spectra and observations for the two cases. Each panel shows the retrieved median model (blue for H-rich and red for non H-rich), 1 and 2$\sigma$ confidence intervals (shaded purple regions), and the K2, HST-WFC3, and \textit{Spitzer} observations (black error bars with yellow markers). These cases do not consider O$_2$/O$_3$ as these molecules were not preferred by the data (see Table \ref{table:k218b}). Bottom row shows the posterior distributions for H$_2$+He, H$_2$O, CH$_4$, and NH$_3$ for the two retrievals.} \label{fig:k218b_posteriors}
\end{figure*} 

While previous studies have assumed that K2-18b has a H-rich atmosphere \citep[e.g.,][]{ Tsiaras2019, Benneke2019b, Welbanks2019b, Madhusudhan2020}, the robustness of this assumption has remained untested. Here, we apply the new functionality of \textsl{Aurora}, to retrieve atmospheric properties of an exoplanet without assuming a H-rich atmosphere to the broadband transmission spectrum of K2-18b. As discussed in sections \ref{subsec:samplers_results} and \ref{subsec:retrieval_miniNeptunes}, we include spectroscopic observations from HST-WFC3 G141 grism (1.05-1.7 $\mu$m), as well as photometric data in the \textit{Spitzer} IRAC 3.6 and 4.5$\mu$m bands, and K2 optical bands (0.43-0.89$\mu$m) from \citet{Benneke2019b}. We also redo an analysis assuming the planet has a H-rich atmosphere and with more model considerations than the results in section \ref{subsec:samplers_results}. By performing both sets of retrievals we can compare their model evidences and assess if the assumption of a H-rich atmosphere is preferred by our retrievals. Furthermore, we expand on previous studies and consider the possibility of O$_2$ and O$_3$ absorption for illustration. 

The retrievals on the full broadband spectrum of K2-18b consider absorption due to H$_2$O, N$_2$, CH$_4$, HCN, NH$_3$, CO, CO$_2$ and H$_2$–-H$_2$ and H$_2$–-He CIA. A second set of retrievals expands the number of absorbers included by considering O$_2$ and O$_3$ absorption. Our models employ a full parametric P-T profile, include the presence of H$_2$-Rayleigh scattering, and follow our new inhomogeneous cloud and haze treatment using two distinct cloud/haze sectors (i.e., model 2 in section \ref{subsec:cloud_validation}). We perform retrievals by assuming the atmosphere is H-rich as well as relaxing this assumption. The retrieved parameters are shown in Table \ref{table:k218b}. Figure \ref{fig:k218b_posteriors} shows the retrieved spectra and a subset of the retrieved posterior distributions for the highest evidence models assuming a H-rich atmosphere and not assuming a H-rich atmosphere.

We first assess the retrievals when assuming a H-rich atmosphere. The retrieved H$_2$O abundances are consistent when considering the possibility of O$_2$ and O$_3$ absorption and when not. With retrieved H$_2$O abundances of $\log_{10}(X_{\text{H}_2\text{O}})=-2.10 ^{+ 1.07 }_{- 1.20 }$ (not considering O$_2$ or O$_3$) and $\log_{10}(X_{\text{H}_2\text{O}})=-2.12 ^{+ 1.09 }_{- 1.24 }$ (considering O$_2$ and O$_3$), the results are in agreement with the estimates of \citet{Welbanks2019b, Benneke2019b, Madhusudhan2020}. Likewise, both retrievals find a depletion of CH$_4$ and NH$_3$ despite the strong absorption of these species in the HST WFC3 and \textit{Spitzer} bands, in agreement with retrievals in previous studies. The limited spectral information in the optical wavelengths results in weak constraints on the cloud and haze parameters. The use of a more complex cloud and haze parameterization (i.e., more parameters) relative to previous studies \citep[e.g.,][]{Welbanks2019b,Madhusudhan2020}, does not result in better constraints on the presence of clouds and hazes. The derived parameters are largely consistent with a clear atmosphere, i.e., small retrieved cloud cover fractions and haze cover fractions, and cloud deck top pressures mostly near or below the photosphere. 

The retrieved parameters when not assuming a H-rich atmosphere are consistent, within 1$\sigma$, with the retrieved parameters when assuming a H-rich atmosphere discussed above. Although consistent, this second approach results in wider and higher abundance estimates for all the chemical species considered. The retrieved H$_2$O abundances have median values almost 1~dex higher than those obtained when assuming a H-rich atmosphere. These retrieved abundances are $\log_{10}(X_{\text{H}_2\text{O}})=-1.20 ^{+ 1.15 }_{- 1.81 }$ (not considering O$_2$ or O$_3$) and $\log_{10}(X_{\text{H}_2\text{O}})=-1.22 ^{+ 1.18 }_{- 2.03 }$ (considering O$_2$ and O$_3$). 

Despite the higher H$_2$O abundance estimates, the retrievals indicate that the main component of the atmosphere is H$_2$ and He with retrieved abundances of $\log_{10}$($X_{\text{H}_2+\text{He}}$)=$-0.09 ^{+ 0.09 }_{- 5.71 }$ (not considering O$_2$ or O$_3$) and $\log_{10}$($X_{\text{H}_2+\text{He}}$)=$-0.14 ^{+ 0.13 }_{- 6.64 }$ (considering O$_2$ and O$_3$). The retrieved H$_2$+He abundance estimates correspond to a median of $\sim$72-81\%, and allow for H$_2$+He abundances of less than 1\% within 1$\sigma$ as shown in Figure \ref{fig:k218b_posteriors}. Assuming a solar He/H$_2$ ratio of 0.17 \citep{Asplund2009}, the retrieved median H$_2$+He abundance estimate of $\log_{10}$($X_{\text{H}_2+\text{He}}$)=$-0.09$ (H$_2$ and He volume mixing ratio of $\sim$81\%) indicates a $\log_{10}$($X_{\text{H}_2}$)=$-0.16$ (H$_2$ volume mixing ratio of $\sim$69\%) and $\log_{10}$($X_{\text{He}}$)=$-0.93$ (He volume mixing ratio of $\sim$12\%). 

All other chemical abundances are poorly constrained with most uncertainties greater than 3~dex. Similarly, the cloud and haze parameters remain unconstrained. Overall, retrieving the main gas constituent in the atmosphere of K2-18b using current observations results in a H$_2$ and He-rich atmosphere ($\sim$72-81\% median volume mixing ratio) with strong H$_2$O absorption ($\sim$6\% median volume mixing ratio), consistent with previous retrieval studies \citep[e.g.,][]{Benneke2019b, Welbanks2019b, Madhusudhan2020}.

The highest model evidence corresponds to the retrieval assuming a H-rich atmosphere and not considering absorption due to O$_2$ or O$_3$. Neither approach, assuming a H-rich atmosphere or not, favours the presence of O$_2$ and O$_3$ absorption in the atmosphere of K2-18b. In the H-rich approach, the additional parameter space due to considering the presence of these extra two absorbers dilutes the model evidence to a 1.17$\sigma$ equivalent level. Likewise, the non H-rich approach results in a decrease in model evidence equivalent to a 1.54$\sigma$ level when considering absorption due to O$_2$ or O$_3$.

\begin{figure*}
\includegraphics[width=\textwidth]{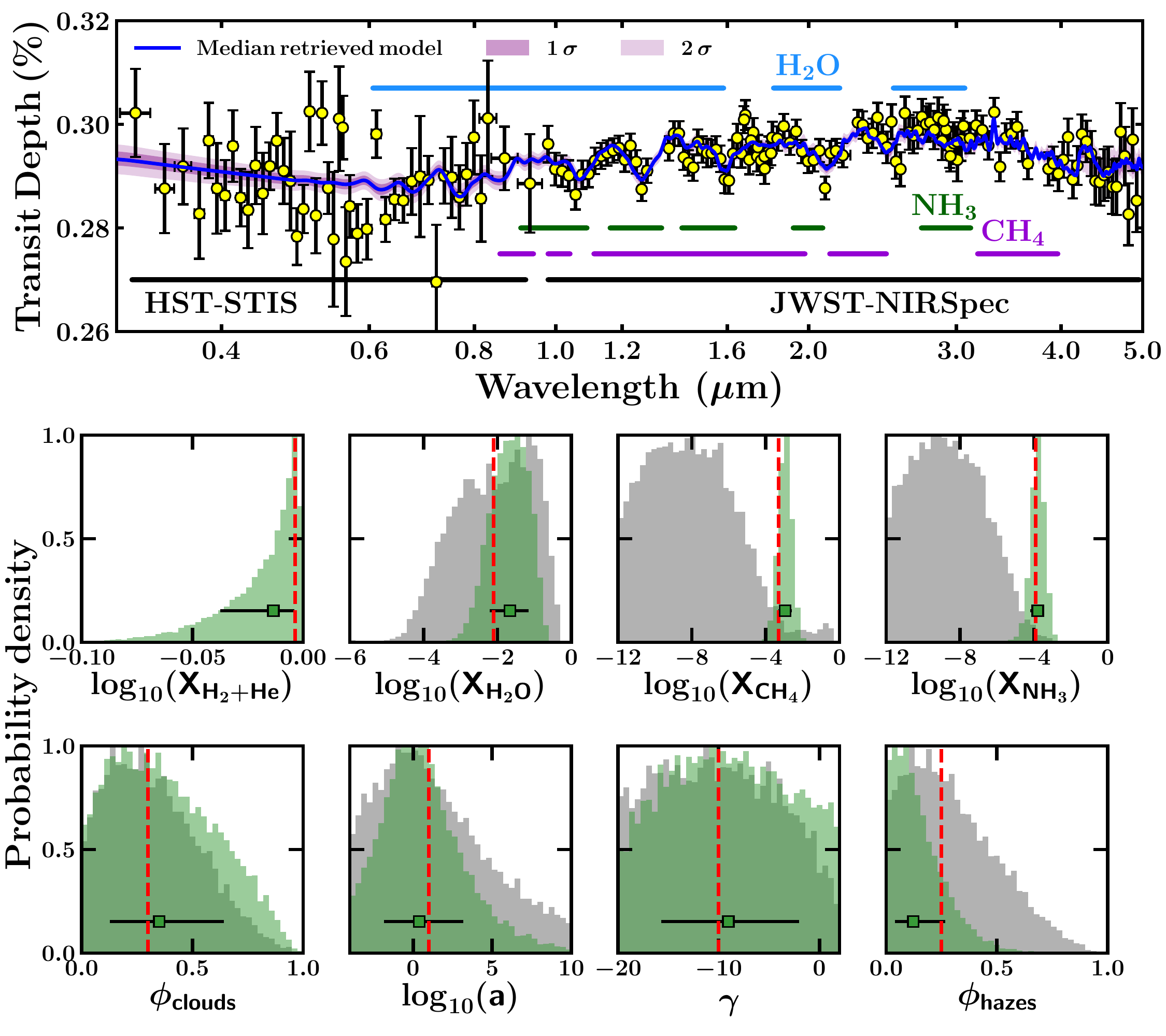}
\centering
\caption[JWST]{Retrieval of synthetic observations of K2-18b. Top: Synthetic observations (black error bars with yellow markers) and median retrieved model (dark blue) for STIS G430L and G750L, and NIRSpec G140H, G235H and G395H. The 1 and 2$\sigma$ confidence intervals (shaded purple areas) are not entirely visible as their span is much smaller than the spread in the data. Horizontal lines at the bottom of the figure show the wavelength coverage of HST-STIS and JWST-NIRSpec. Blue, green, and purple horizontal lines show, respectively, the approximate wavelength regions where H$_2$O, NH$_3$, and CH$_4$ spectral features are expected. Bottom: Retrieved posterior distributions for H$_2$+He, H$_2$O, CH$_4$, and NH$_3$ abundances, and relevant cloud and haze parameters. The green posterior distributions in the foreground correspond to the retrieval using synthetic observations. The vertical red dashed lines indicate the true model input values for this synthetic retrieval. The gray posterior distributions in the background show estimates obtained with existing real observations  of K2-18b as discussed in section \ref{subsubsec:k218b_xhrich}, illustrating constraints possible with current data. We note that the red vertical lines are unrelated to the gray posteriors. 
\label{fig:K218b_stis_jwst}}
\end{figure*} 

Increasing the parameter space to retrieve the abundance of H$_2$ and He results in a decrease in model evidence. When not considering O$_2$ and O$_3$ absorption, the model evidence for the H-rich assumed retrieval is higher at a 2.68$\sigma$ level compared to the retrieval without a priori assumptions on the bulk composition of the atmosphere. Similarly, the H-rich assumption is preferred at a 2.79$\sigma$ level over the non H-rich assumed model when considering absorption due to O$_2$ and O$_3$. This preference of almost 3$\sigma$ for the H-rich model should not be interpreted as a detection of a H-rich atmosphere on K2-18b but instead must be understood as an indication that the additional parameter space explored by the non H-rich approach is of lower likelihood. On the other hand, the fact that both retrieval approaches infer a H-rich atmosphere can be interpreted as a demonstration that current data favours a H-rich atmosphere on K2-18b.

\subsubsection{Future Spectroscopic Observations: K2-18b}
\label{subsubsec:future_k218b}

In order to investigate the abundance constraints that may be possible with future observations, we generate synthetic HST-STIS and JWST-NIRSpec observations of K2-18b based on the retrieved median H$_2$O abundance for our highest evidence model in section \ref{subsubsec:k218b_xhrich}. We choose abundances for CH$_4$ and NH$_3$ that are $\sim1\times$ solar \citep[$\log_{10}(X_{\text{CH}_4})=-3.3$, $\log_{10}(X_{\text{NH}_3})=-3.9$, e.g.,][]{Woitke2018,Madhusudhan2020}, consistent with their apparent depletion relative to the retrieved $\sim10\times$ solar H$_2$O abundance \citep[see section \ref{subsubsec:k218b_xhrich}, e.g.,][]{Madhusudhan2020}. The input model also includes absorption due to HCN, CO, and CO$_2$, with an input nominal abundance of 1~ppm. We generate a model spectrum at a constant spectral resolution of R=5000 between 0.3 and 5.5$\mu$m. Given that current observations of K2-18b do not place strong constraints on the presence of clouds and hazes, we use input values for the cloud and haze prescription that fall within 1$\sigma$ of the retrieved parameters in section \ref{subsubsec:k218b_xhrich}. These input parameters are a Rayleigh enhancement factor $a=10$, a slope $\gamma=-10$, a gray cloud deck with a top pressure in bar of $\log_{10}$($P_{\text{cloud}}$)=-1.6, and a 25\% cover due to the hazes and 30\% cover due to clouds. The input P-T profile is set by the retrieved parameters for the highest evidence model in section \ref{subsubsec:k218b_xhrich}.

The synthetic JWST observations are generated using PANDEXO \citep{Batalha2017b}. We generate observations for a transmission spectrum of K2-18b observed with JWST-NIRSpec using its three high-resolution gratings (G140H/F100LP, G235H/F170LP, and G395H/F290LP) in the subarray SUB2048 mode, i.e., a total of 3 transits. Further details about the model inputs to PANDEXO are described in appendix \ref{app:pandexo_k218b}. We also model synthetic HST-STIS observations covering the optical wavelengths from $\sim0.3-1.0\mu$m. Comparing an observed HST-WFC3 transmission spectrum of K2-18b \citep{Benneke2019b} with that of HD~209458~b \citep{Deming2013} it is seen that 9 transits of K2-18b provide data of comparable quality, in terms of precision per spectral bin, to 1 transit of HD~209458~b. Since there are no HST-STIS observations of K2-18b available, we derive a synthetic HST-STIS spectrum of K2-18b by scaling the uncertainties and resolution from an observed HST-STIS spectrum of HD~209458~b \citep{Sing2016} in the same proportion as that of the HST-WFC3 spectra between the two planets. We note that the resulting synthetic observations of K2-18b would require a significantly larger number of transits with HST-STIS than the 9 observed with HST-WFC3. Nevertheless, we consider this optimistic scenario as a test case to demonstrate our retrievals. Our synthetic observations have Gaussian-distributed uncertainties with a mean precision of $\sim$72~ppm in the STIS G430L band and $\sim$71~ppm in the STIS G750L band.

The resulting synthetic observations are shown in Figure \ref{fig:K218b_stis_jwst}. The synthetic HST-STIS and JWST-NIRSpec observations provide a spectral range of $\sim0.3$-$5.0\mu$m, encoding information about the presence of clouds, hazes and absorption due to different species like H$_2$O, CH$_4$, and NH$_3$. We perform a retrieval on these observations considering absorption due to H$_2$O, CH$_4$, HCN, NH$_3$, CO, CO$_2$, N$_2$, O$_2$, O$_3$, and H$_2$–-H$_2$ and H$_2$–-He CIA. We employ the same cloud and haze prescription employed in section \ref{subsubsec:k218b_xhrich}. We do not assume the bulk composition of the atmosphere and retrieve it instead. 

Figure \ref{fig:K218b_stis_jwst} shows the retrieved posterior probability distributions for H$_2$+He, the detected species H$_2$O, CH$_4$, and NH$_3$, and some relevant cloud/haze parameters. The full posterior distribution is shown in the appendix \ref{app:k218b_posterior}. The bottom half in Figure \ref{fig:K218b_stis_jwst} also shows (in gray) the probability distributions obtained with current K2, HST-WFC3, and \textit{Spitzer} spectro-photometric observations (first column in Table \ref{table:k218b}). Comparing both gray and green probability distributions it is possible to appreciate that abundance estimates will be largely improved using JWST observations. For the assumed synthetic model and data considerations, we retrieve the abundances to be consistent with the input values at: $\log_{10} \left(X_{\textnormal{H}_2\textnormal{O}}\right)=-1.66 ^{+ 0.50 }_{- 0.55 }$, $\log_{10} \left(X_{\textnormal{CH}_4}\right)=-2.94 ^{+ 0.35 }_{- 0.37 }$ and $\log_{10} \left(X_{\textnormal{NH}_3}\right)=-3.79 ^{+ 0.36 }_{- 0.40 }$. The corresponding detection significances of the molecules are $\sim$5$\sigma$, $\sim$7$\sigma$, and $\sim$3$\sigma$ for H$_2$O, CH$_4$ and NH$_3$ respectively. In principle, even better abundance precisions and detection significances may be attained by combining with other observations (e.g., JWST-MIRI, JWST-NIRISS, HST-WFC3), or considering data of higher resolution. We also note that these precisions and detection significances are dependent on the input model assumptions: $\sim$10$\times$solar H$_2$O and $\sim$1$\times$solar CH$_4$ and NH$_3$. Nevertheless, these results demonstrate the capability of \textsl{Aurora} to precisely retrieve the true input values of a mini-Neptune like K2-18b. 

Furthermore, the retrieval on synthetic data demonstrates \textsl{Aurora}'s ability to retrieve the bulk atmospheric composition of a mini-Neptune like K2-18b. With a retrieved abundance of $\log_{10}(X_{\text{H}_2+\text{He}})=-0.013 ^{+ 0.010 }_{- 0.024 }$, \textsl{Aurora} correctly identifies H$_2$ and He as the bulk composition of the atmosphere as shown in Figure \ref{fig:K218b_stis_jwst}. The retrieved median abundance indicates that H$_2$ and He account for more than 97\% of the atmosphere's composition, consistent with the input model. Future observations with JWST will make it possible to unequivocally retrieve the bulk gas composition in the atmosphere of K2-18b improving on present day estimates derived using current K2, HST-WFC3, and \textit{Spitzer} observations. 

The cloud and haze parameters in the input model are motivated by current constraints on K2-18b using existing data, as discussed above, which indicate a relatively cloud/haze free atmosphere. Under these cloud/haze assumptions, the retrieved abundance estimates and their detection significances are not strongly affected when only JWST-NIRSpec observations are considered in our retrievals. The retrieved cloud and haze parameters are mostly unconstrained, consistent with the cloud/haze free input model, and similar to constrains obtained with current data (e.g., gray posterior distributions in Figure \ref{fig:K218b_stis_jwst}). Even when both HST-STIS and JWST-NIRSpec observations are considered, the cloud/haze constraints are only marginally improved, as shown in Figure \ref{fig:K218b_stis_jwst} and appendix \ref{app:k218b_posterior}, as expected considering the low cloud/haze cover in the input model. Regardless of the cloud/haze constraints, however, the chemical abundances are still derived precisely as discussed above.

In principle, further spectroscopic observations, including those with other JWST instruments like NIRISS and MIRI, could help obtain better constraints than those reported here. At the same time, it could also be important to revisit the model assumptions in present retrievals (e.g., by considering higher dimensional models, temporal variability, etc.) when confronted with observations of higher quality (e.g., higher-resolution, better signal-to-noise, broader wavelength coverage) expected in the near future. We discuss these implications and the prospect for future works in section \ref{sec:conclusions}.

\begin{figure}
\includegraphics[width=0.5\textwidth]{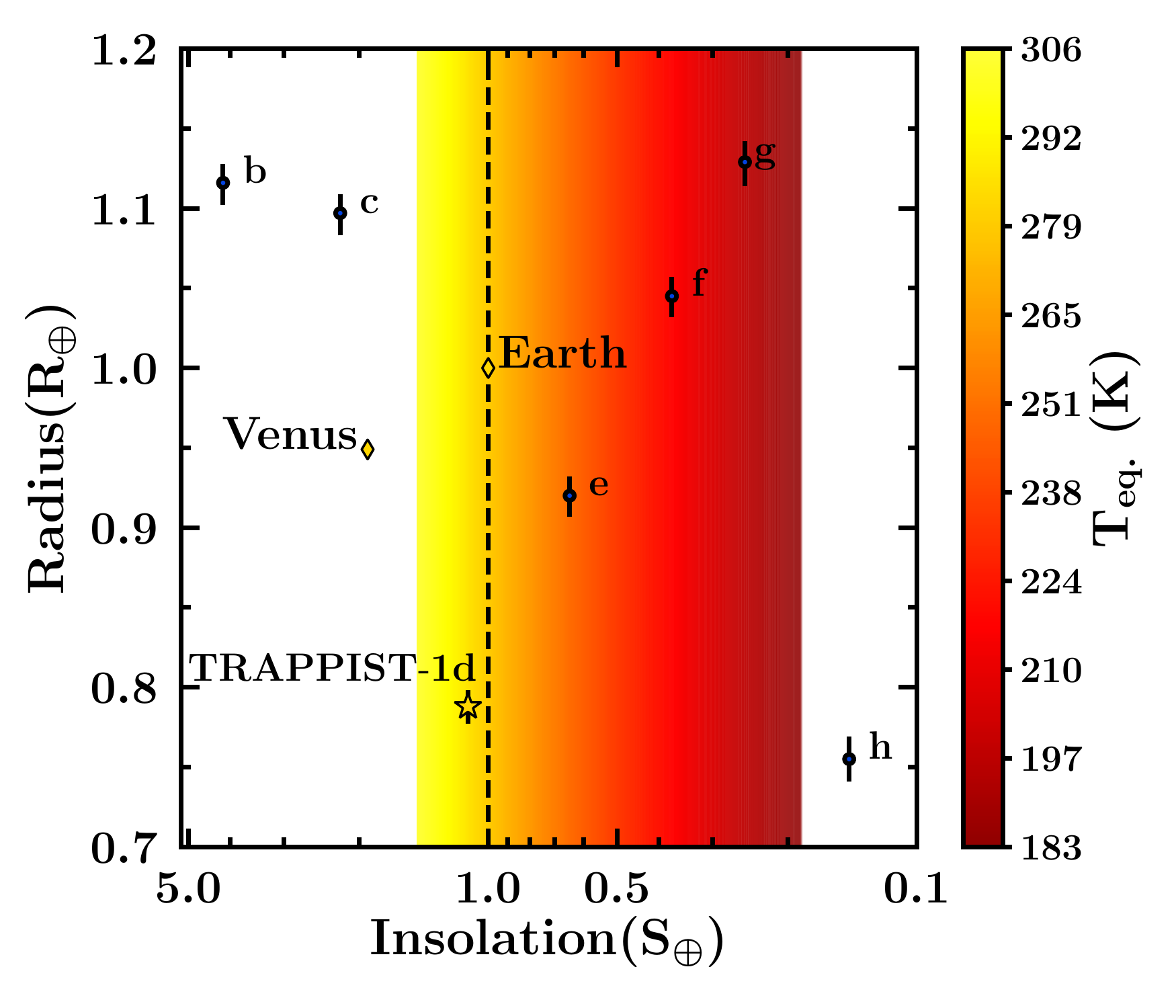}
\centering
\caption[Trappist hab]{Planet radius versus stellar insolation for the TRAPPIST-1 system. The seven TRAPPIST-1 planets \citep[values from ][]{Agol2020} are shown alongside Earth and Venus (yellow diamonds). The shaded region represents the optimistic habitable zone for an M dwarf at the temperature of TRAPPIST-1 \citep{Kopparapu2013}. TRAPPIST-1d (gold star) is at the inner edge of the optimistic habitable zone of TRAPPIST-1, and is the closest to Earth in terms of insolation. Equilibrium temperature calculated assuming full redistribution and zero albedo.} 
\label{fig:trappist_hab}
\end{figure}

\begin{figure*}
\includegraphics[width=\textwidth]{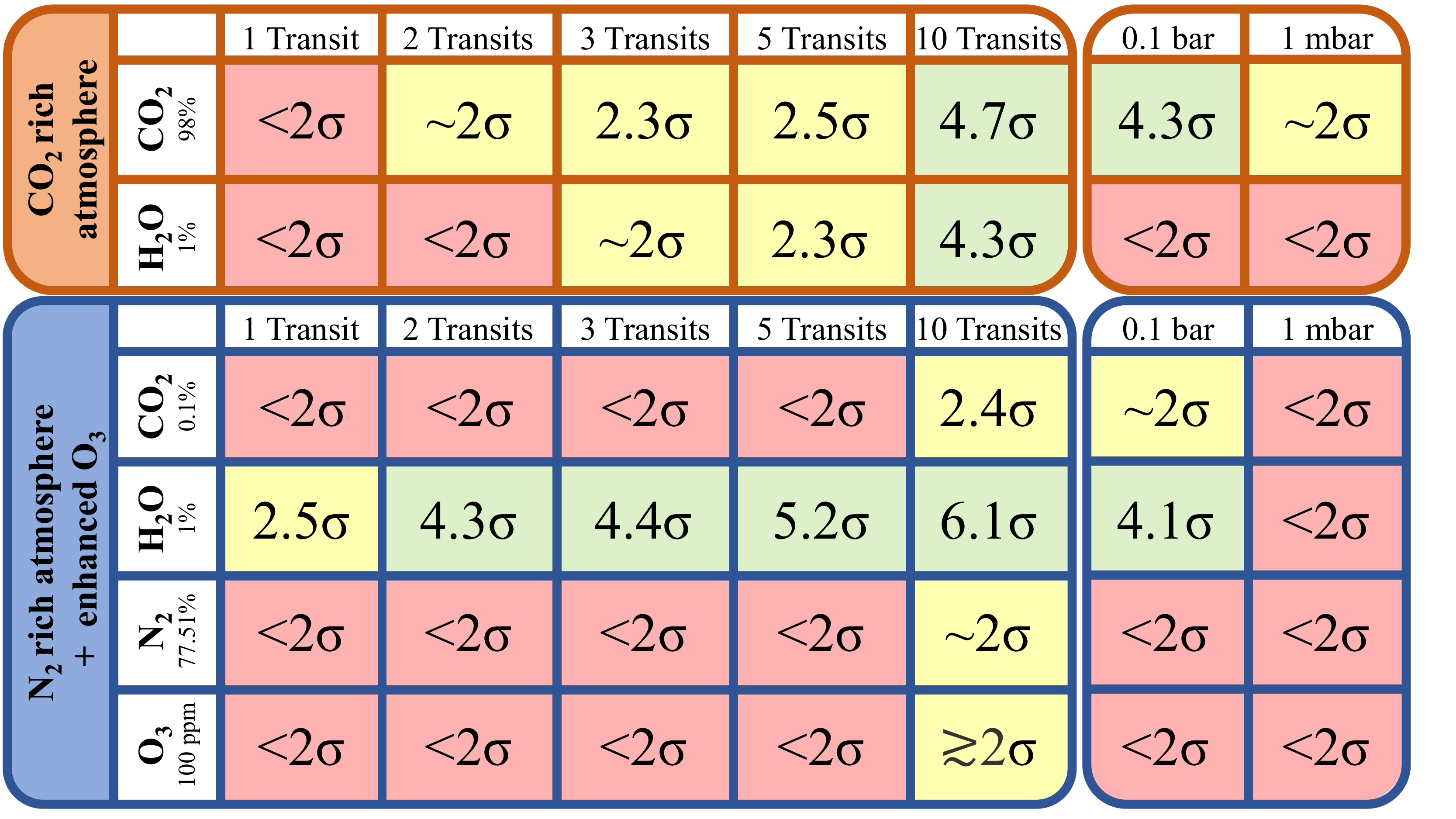}
\centering
\caption[DS_matrix]{Detection significance for different chemical species in the atmosphere of TRAPPIST-1~d as a function of observed number of transits with JWST-NIRSpec. Top: CO$_2$-rich atmosphere. Bottom: N$_2$-rich atmosphere, with enhanced O$_3$. Left boxes show the results for a cloud-free atmosphere observed with 1, 2, 3, 5, and 10 transits. Right boxes show the results for observing 10 transits of a fully cloudy atmosphere with a cloud deck top pressure of 0.1 bar and 1 mbar.} \label{fig:detection_matrix}
\end{figure*}

\subsection{Application to Rocky Exoplanets.}
\label{subsubsec:trappist1d}

We demonstrate \textsl{Aurora}'s capabilities to identify the composition of atmospheres which are not H-rich. We use synthetic JWST-NIRSpec observations of the rocky exoplanet (i.e., terrestrial-size exoplanet) TRAPPIST-1~d \citep[R$_\text{p}=0.788$~R$_\oplus$, M$_\text{p}=0.388$~M$_\oplus$,][]{Gillon2017, Agol2020} to validate \textsl{Aurora}'s retrieval capability for H-poor atmospheres. Of all seven TRAPPIST-1 planets, TRAPPIST-1~d is the closest to Earth in terms of insolation (see Figure \ref{fig:trappist_hab}). This makes TRAPPIST-1~d an appealing candidate for characterisation with JWST, especially in the context of planets residing in their habitable zone. This opportunity has been recognised by the JWST Guaranteed Time Observations (GTO) program by planning to observe 2 transits of the planet using NIRSpec prism (GTO 1201, PI: David Lafreniere).

Figure \ref{fig:trappist_hab} shows the planet radius versus stellar insolation for the planets in the TRAPPIST-1 system. When compared to planets in the solar system, TRAPPIST-1~d falls between Venus and Earth in terms of insolation. As such, we consider two possible model configurations for our application of \textsl{Aurora}: a CO$_2$-rich atmosphere (e.g., loosely similar to Venus' atmosphere), and an N$_2$-rich atmosphere (e.g., loosely similar to Earth's atmosphere with enhanced O$_3$). The CO$_2$-rich atmosphere is composed of 98\% CO$_2$, 1\% H$_2$O, 11.7~ppm H$_2$+He, and N$_2$ in the remaining percentage ($\sim0.99$\%). The N$_2$-rich atmosphere is composed of 77.51\% N$_2$, 21.38\% O$_2$, 1\% H$_2$O, 0.1\% CO$_2$, and 0.01\% O$_3$. We note that this O$_3$ abundance is $\sim10-100\times$ higher than present day Earth's atmospheric abundance in the stratosphere \citep[e.g.,][]{Anderson1987, Barstow2016b}. The atmospheres are modelled to follow an isotherm at 250~K. We considered three cases for each composition: (1) a clear atmosphere, (2) an atmosphere with a gray cloud deck covering the entire planet at a cloud top pressure $P_{\text{cloud}}$=0.1 bar, and (3) a gray cloud deck covering the entire planet at a cloud top pressure $P_{\text{cloud}}$=1 mbar. The model spectra are generated at a constant resolution of R=5000 between $0.3$-$5.5\mu$m.

\begin{figure*}
\includegraphics[width=\textwidth]{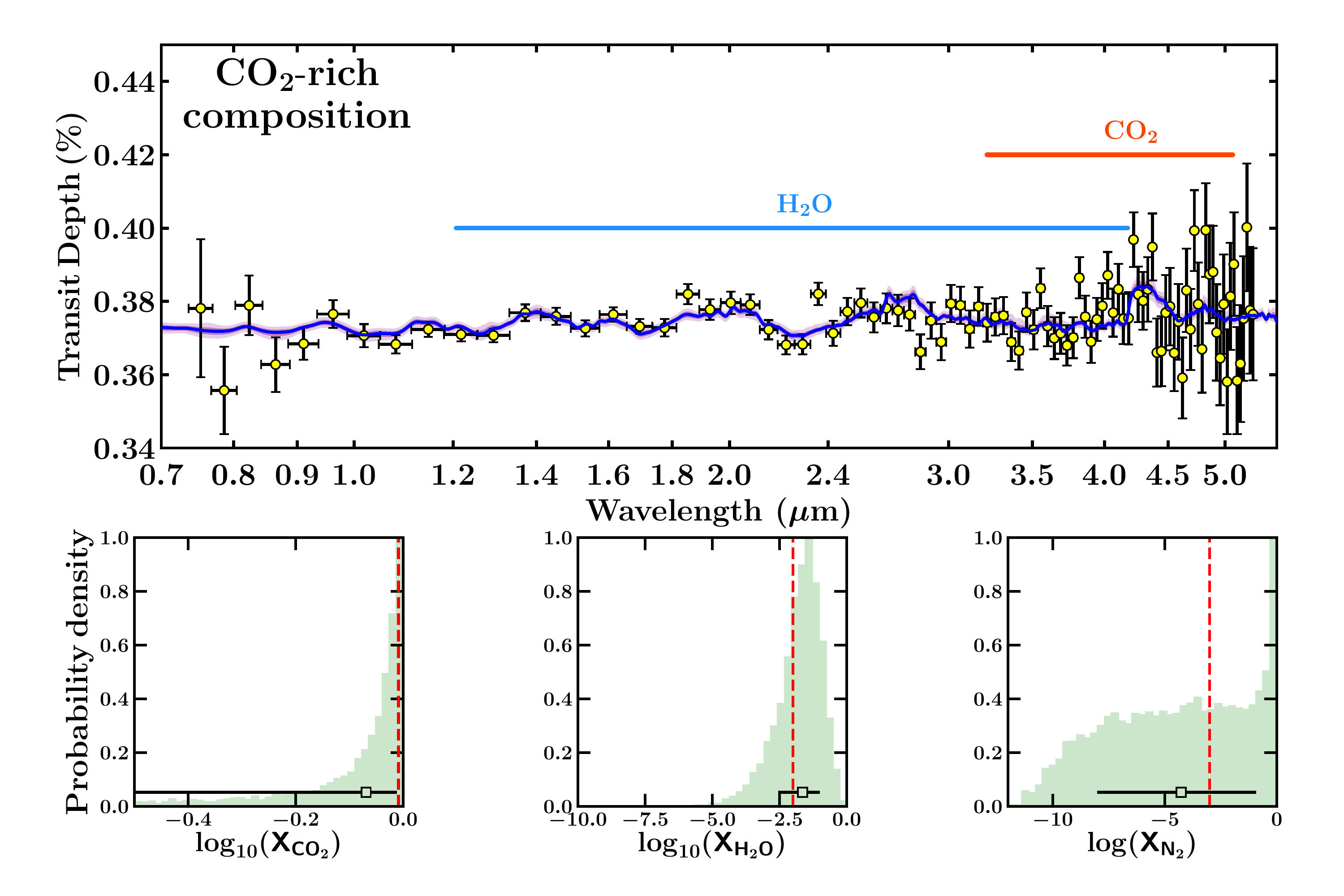}
\centering
\caption[TRAPPIST_CO2_rich]{Retrieval of synthetic TRAPPIST-1~d observations for a CO$_2$-rich atmosphere. Top: Synthetic observations (gold markers, black error bars) and median retrieved model (blue line) for NIRSpec prism. Shaded purple areas indicate 1 and 2$\sigma$ confidence intervals. Blue and red horizontal lines show, respectively, the approximate wavelength regions where H$_2$O and CO$_2$ spectral features are expected. Bottom: Posterior distributions for CO$_2$, H$_2$O, and N$_2$. Vertical red dashed lines show the true input values.} \label{fig:trapco2}
\end{figure*} 

\begin{figure*}
\includegraphics[width=\textwidth]{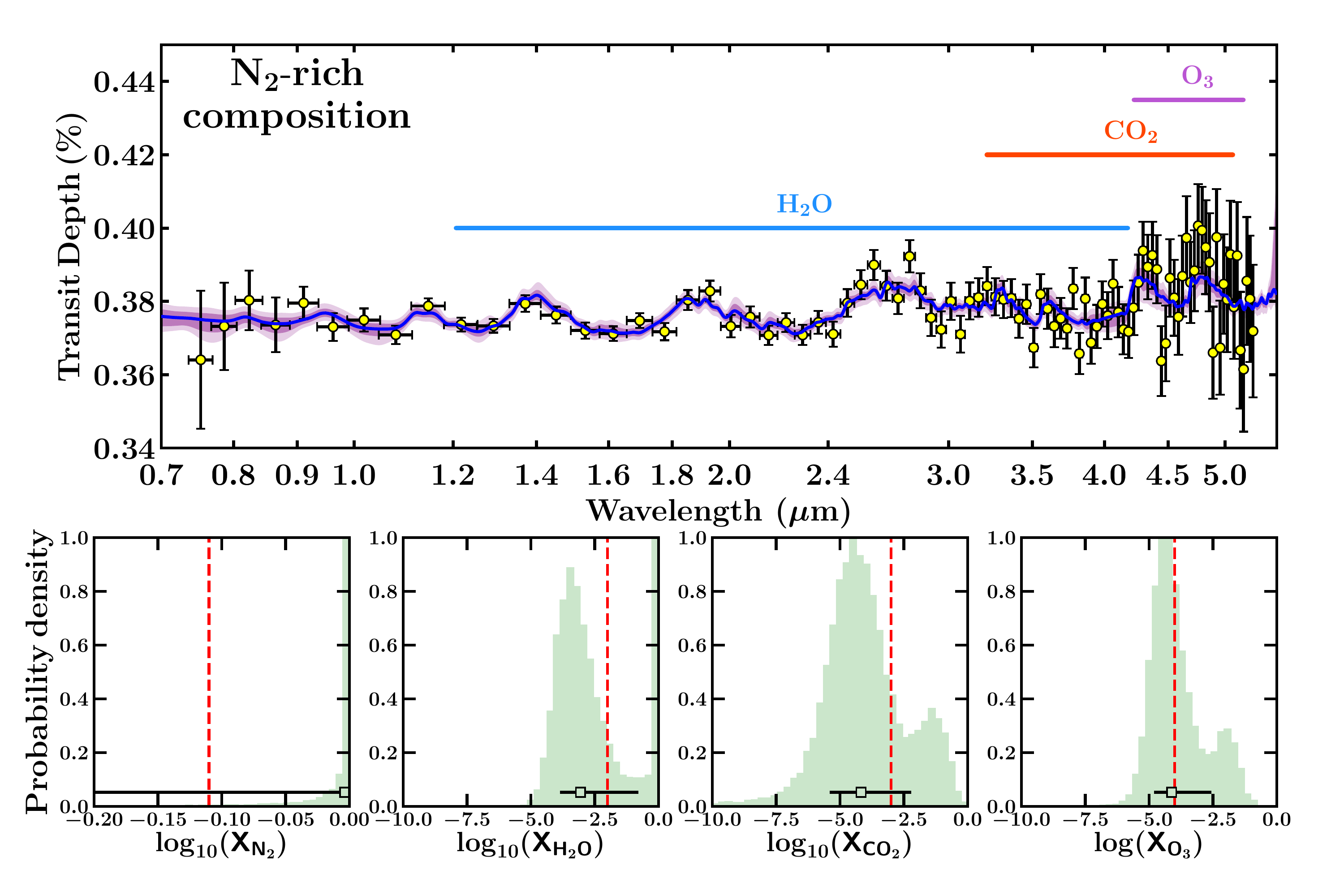}
\centering
\caption[TRAPPIST_N2_rich]{ Retrieval of synthetic TRAPPIST-1~d observations for a N$_2$-rich atmosphere with enhanced O$_3$. Top: Synthetic observations (gold markers, black error bars) and median retrieved model (blue line) for NIRSpec prism. Shaded purple areas indicate 1 and 2$\sigma$ confidence intervals. Blue, red, and purple horizontal lines show, respectively, the approximate wavelength regions where H$_2$O, CO$_2$, and O$_3$ spectral features are expected. Bottom: Posterior distributions for N$_2$, H$_2$O, CO$_2$, and O$_3$. Vertical dashed lines show the true input values.} \label{fig:trapn2}
\end{figure*} 

Besides demonstrating \textsl{Aurora}'s capabilities, we explore the number of JWST transits required for the spectroscopic observations to provide chemical detections and abundance constraints of TRAPPIST-1~d's atmosphere. We consider observations from 1, 2, 3, 5 and 10 JWST transits. We consider 2 transits motivated by the upcoming GTO 1201 program, and an upper limit at 10 transits based on estimates for characterizing the TRAPPIST-1 system from \citet{Batalha2018} (see section \ref{subsec:discussion_se}). We empirically find that 10 transits with JWST-NIRSpec prism can provide chemical constraints of $\sim1$~dex or better and robust detections ($\gtrsim3\sigma$) for multiple chemical species in both the CO$_2$-rich and N$_2$-rich model atmospheres. The synthetic observations are generated using PANDEXO \citep{Batalha2017b} for the NIRSpec prism using subarray SUB512. The synthetic observations include Gaussian white noise. Additional details are described in appendix \ref{app:pandexo_trappist}.

The synthetic observations are generated using models that consider CO$_2$–-CO$_2$ and N$_2$–-N$_2$ collision induced absorption, as well as Rayleigh-scattering due to N$_2$, CO$_2$, and H$_2$O. The models in the retrievals do not assume a H-rich atmosphere and consider absorption due to H$_2$O, CO$_2$, N$_2$, O$_2$, and O$_3$. Including the effects of H$_2$–-H$_2$ and H$_2$–-He CIA, and following the description for non H-rich atmospheres in section \ref{subsec:reparameterization}, the retrieval considers a total of 6 chemical components: H$_2$+He, H$_2$O, CO$_2$, N$_2$, O$_2$, and O$_3$. The model also considers one parameter for an isothermal temperature profile, and one parameter for $P_{\rm{ref}}$. We do not consider the full cloud and haze prescriptions presented in this work due to the lack of observations in the optical required to robustly constrain the presence of hazes. However, we consider the presence of inhomogeneous clouds using two parameters: one for the cloud cover ($\phi_{\mathrm{clouds}}$) and one for the cloud top pressure ($\log_{10}$($P_{\text{cloud}}$)). In summary, the retrieval model has a total of 10 parameters: 6 parameters for the chemical components, 1 parameter for the isothermal temperature, 1 parameter for the reference pressure, and 2 parameters for the presence of inhomogeneous clouds.

We begin by analysing the results from our exploration of the number of transits required to characterise TRAPPIST-1~d's atmosphere. Figure \ref{fig:detection_matrix} shows a summary of the chemical detections for the various numbers of transits considered. We perform this exploration following a conservative approach in which any model preference $<2\sigma$ does not constitute a detection (red squares), model preferences $>2\sigma$ and $<3\sigma$ are suggestive of a chemical detection (yellow squares), and model preferences $>3\sigma$ may be considered detections (green squares). Our search suggests that for a CO$_2$-rich clear atmosphere, 10 transits with JWST-NIRSpec will be able to provide detections of CO$_2$ and H$_2$O. Likewise, for an N$_2$-rich clear atmosphere 10 transits would provide detections of H$_2$O, and possibly O$_3$ if present at enhanced levels (10-100$\times$ Earth levels) as assumed in the input model described above. For the N$_2$-rich atmosphere, although N$_2$ is found to be the main atmospheric component, its lack of spectral features makes its robust detection difficult.

However, considering clear atmospheres only results in optimistic estimates that may be revised when considering the presence of clouds and hazes. We consider observing the cloudy cases described above using the same number of transits (right columns of Figure \ref{fig:detection_matrix}). As expected, the presence of a cloud deck mutes several of the spectral features resulting in weaker chemical detections or non-detections.

Next, we present the retrieved constraints using \textsl{Aurora} for the cases with the strongest chemical detections, i.e., 10 transits for a cloud-free model, starting with a clear CO$_2$-rich atmosphere. The retrieved chemical abundances of interest and retrieved spectrum are shown in Figure \ref{fig:trapco2} along with the synthetic observations. The retrieved abundances for the species included in the true input model are $\log_{10}(X_{\text{H}_2+\text{He}})=-5.53 ^{+ 3.01 }_{- 3.02 }$, $\log_{10}(X_{\text{H}_2\text{O}})=-1.64 ^{+ 0.63 }_{- 0.92 }$, $\log_{10}(X_{\text{CO}_2})=-0.07 ^{+ 0.06 }_{- 0.78 }$, and $\log_{10}(X_{\text{N}_2})=-4.27 ^{+ 3.35 }_{- 3.76 }$. The retrieved values are consistent within $\sim1\sigma$ of the true model input values. \textsl{Aurora} is capable of accurately distinguishing the main constituent of the modelled CO$_2$-rich atmosphere, with the posterior distribution of CO$_2$ corresponding to high abundances. Furthermore, the precisions on the retrieved H$_2$O abundance is $\lesssim$1~dex, comparable to current chemical constraints for gas giants \citep[e.g.,][]{Welbanks2019b}.

Although the input model corresponds to a cloud free atmosphere, we consider in our retrieval model the possibility of inhomogenous cloud cover. The cloud parameterization retrieves a cloud cover of $\phi_{\mathrm{clouds}}=0.37 ^{+ 0.39 }_{- 0.25 }$ and $\log_{10}$($P_{\text{cloud}}$)$=0.09 ^{+ 1.22 }_{- 3.08 }$, consistent with a clear atmosphere (e.g., relatively small cloud fraction cover) with the gray cloud deck placed below the expected photosphere (e.g., not muting spectral features) in agreement with the input model. Overall, these results indicate that the chemical characterisation of a CO$_2$-rich cloud-free atmosphere is possible with 10 JWST-NIRSpec transits. Under these conditions, \textsl{Aurora} is capable of detecting the main component of the atmosphere (CO$_2$) at 4.7$\sigma$ and the trace gas (H$_2$O) at 4.3$\sigma$.

We briefly mention the results from considering the more challenging scenario of a cloudy CO$_2$-rich atmosphere. As described above, we consider scenarios with 100\% cloudy atmospheres with cloud top pressures of 0.1 bar and 1 mbar. While both cases indicate a CO$_2$ rich atmosphere, the muted spectral features result in weaker detections of the main atmospheric constituent (e.g., 4.3$\sigma$ for 0.1 bar and $\sim2\sigma$ for 1 mbar, see Figure \ref{fig:detection_matrix}). Although not resulting in strong detections, \textsl{Aurora} can correctly identify that the observations correspond to a cloudy atmosphere, with the 1 mbar case retrieving $\phi_{\mathrm{clouds}}=0.82 ^{+ 0.13 }_{- 0.25 }$ and $\log_{10}$($P_{\text{cloud}}$)$=-2.80 ^{+ 1.25 }_{- 1.24 }$ (i.e., relatively high cloud cover fraction and a top cloud deck pressure above the expected photosphere). In agreement with other studies (see section \ref{subsec:discussion_se}), our results suggest that robustly characterising the atmosphere of a cloudy rocky exoplanet will need more than 10 JWST-NIRSpec transits.

Next, we present the results from retrieving the clear N$_2$-rich atmosphere with 10 JWST-NIRSpec transits. \textsl{Aurora} retrieves abundances of $\log_{10}(X_{\text{N}_2})=-0.0037^{+ 0.0035 }_{- 2.2195 }$, $\log_{10}(X_{\text{H}_2\text{O}})=-3.06 ^{+ 2.27 }_{- 0.81 }$, $\log_{10}(X_{\text{CO}_2})=-4.18 ^{+ 1.97 }_{- 1.23 }$, and $\log_{10}(X_{\text{O}_3})=-4.12 ^{+ 1.56 }_{- 0.68 }$, for the species with detection significances $\gtrsim$2$\sigma$. The retrieved values are consistent within $1\sigma$ of the input parameter despite the white noise in the observations. The retrieved cloud parameters are $\phi_{\mathrm{clouds}}=0.51 ^{+ 0.31 }_{- 0.34 }$ and $\log_{10}$($P_{\text{cloud}}$)$=-0.34 ^{+ 1.12 }_{- 1.86 }$, consistent with a cloud-free atmosphere due to the retrieved cloud deck top pressure being below the expected photosphere. Figure \ref{fig:trapn2} presents the synthetic observations as well as the retrieved spectrum and the posterior distributions for the chemical species of interest.

Although the retrieval indicates that N$_2$ is the main component of the atmosphere, the lack of strong spectral features results in a moderate detection of N$_2$ with 10 JWST-NIRSpec transits. Besides identifying the main component of the atmosphere, \textsl{Aurora} is able to detect H$_2$O (6.1$\sigma$) and O$_3$ ($\gtrsim$2$\sigma$) with 10 transits and provide constraints in their abundances to a precision of $\sim$1.5~dex. Although optimistic, these estimates present a tantalising prospect for the detection of possible biosignatures in habitable zone rocky exoplanets. If present in $\sim$10-100$\times$ higher abundance than present-day Earth stratospheric levels \citep[e.g.,][see section \ref{subsec:discussion_se}]{Anderson1987}, 10 NIRSpec prism transits could provide an initial indication of O$_3$ in TRAPPIST-1~d.

As performed with the CO$_2$-rich case, we investigate the effect of a cloud deck at 0.1 bar and 1 mbar on the estimates above. As shown in Figure \ref{fig:detection_matrix}, the presence of a cloud deck results in weaker or no chemical detections. Nonetheless, the retrieved cloud parameters are mostly consistent with the input cloudy models, with the 1 mbar case retrieving $\phi_{\mathrm{clouds}}=0.89 ^{+ 0.07 }_{- 0.12 }$ and $\log_{10}$($P_{\text{cloud}}$)$=-4.62 ^{+ 1.07 }_{- 0.87 }$. In agreement with the CO$_2$-rich case, and as expected, this N$_2$-rich case suggests that a cloudy atmosphere is more difficult to characterise than a clear-atmosphere.

The number of transits with JWST required for the chemical characterisation of rocky exoplanets can vary depending on the system parameters, instrument of choice, and the desired precision on the retrieved atmospheric properties. As such, our result of 10 JWST-NIRSpec transits is specific to the cases considered here. We discuss additional considerations that could revise these results as well as future considerations in section \ref{subsec:discussion_se}.

\section{Summary and Discussion}
\label{sec:conclusions}
In this work we introduce \textsl{Aurora}, a next-generation atmospheric retrieval framework for transmission spectra of a wide range of exoplanets: from gas giants with H-rich atmospheres to rocky exoplanets with secondary non H-rich atmospheres. \textsl{Aurora} retains the capabilities from previous retrieval codes and presents advancements for the analysis of future observations. These key advancements are:

\begin{itemize}
    \item \textsl{Aurora} can retrieve the bulk composition of any exoplanet atmosphere without the assumptions of a H-rich atmosphere. The retrieved parameter estimates are robust for H-rich and H-poor atmospheres. We demonstrate this on current and synthetic observations of hot Jupiters, mini Neptunes and rocky exoplanets.
    \item We introduce a new generalised treatment for inhomogeneous clouds and hazes. The new cloud and haze prescription explores a larger parameter space compared to previous treatments. Our new approach mitigates some biases and limitations in previous prescriptions and is robust when assuming a H-rich atmosphere or not.
    \item \textsl{Aurora} incorporates in its retrieval framework the next generation of nested sampling algorithms. These are highly adaptable and designed for handling highly degenerate problems and problems of higher dimensionality. This advancement is key in the development of multidimensional retrieval techniques and alleviates some of their computational needs.  
    \item \textsl{Aurora} has a modular structure designed to evolve with the needs of future spectroscopic observations. The new modular capabilities include:
    \begin{itemize}
        \item Noise modelling capabilities beyond the traditional assumed independently distributed Gaussian errors. These include the ability to retrieve an inflation for the standard deviation of observations and consider correlated noise using Gaussian processes.
        \item Forward models considering the effects of light refraction, forward scattering, and Mie-scattering due to condensates.
    \end{itemize}
\end{itemize}

In this work we have validated \textsl{Aurora}'s Bayesian retrieval framework using up-to-date existing observations of the hot Jupiter HD~209458~b and the mini-Neptune K2-18b. We further validate \textsl{Aurora}'s retrieval framework using HST-STIS and JWST-NIRSpec synthetic observations for K2-18b, and JWST-NIRSpec synthetic observations for the rocky exoplanet TRAPPIST-1~d. Our results highlight 4 findings:

\begin{itemize}
    \item For hot Jupiters, the retrieved parameter estimates are robust against assumptions of a H-rich atmosphere or not. The cloud and haze prescription introduced in this work results in a higher model evidence than previous inhomogeneous cloud and haze prescriptions when applied to current observations of the well-studied hot Jupiter HD~209458~b.
    \item For current observations of mini Neptunes, we have demonstrated that the atmosphere of K2-18b is H-rich. Furthermore, the nested sampling algorithms included in \textsl{Aurora} retrieve almost identical parameter estimates. The retrieved properties of K2-18b are consistent with previous results.
    \item For future observations of mini Neptunes with JWST, abundance estimates could result in precisions of $\sim 0.5$~dex or better. Abundance estimates obtained with JWST observations can be robust even in the absence of observations in the optical wavelengths, for relatively low cloud/haze covers as for the case of K2-18b.
    \item For future observations of rocky exoplanets, \textsl{Aurora} can robustly identify their dominant atmospheric composition as well as reliably detect and constrain the abundance of trace gases. For example, 10 JWST transits of TRAPPIST-1~d could enable clear detections and abundance constraints of H$_2$O in a cloud-free N$_2$-rich or CO$_2$-rich atmosphere. Furthermore, 10 JWST transits could enable initial indications of O$_3$ if present at enhanced levels ($\sim10-100\times$ present day Earth's stratospheric abundances) in a cloud-free N$_2$-rich atmosphere.
\end{itemize}

We discuss the implications of our results for the analysis of current and future observations of hot Jupiters, mini Neptunes and rocky exoplanets.

\subsection{Constraining the Composition of Mini-Neptunes}

Recent spectroscopic observations \citep[e.g.,][]{Benneke2019a, Benneke2019b, Tsiaras2019} have demonstrated that mini-Neptunes are advantageous targets in the search of H$_2$O vapour and other possible molecular features in low-mass exoplanets. The lack of an analogue for this type of planet in our solar system, represents a unique opportunity to learn about the diversity of planet configurations and compositions. Straddling the gap between terrestrial planets and ice giants, it is not always clear whether the atmospheres of some of these planets are H-rich or not. Similar to our approach in section \ref{subsubsec:k218b_xhrich}, we suggest that the interpretation of future observations of mini-Neptunes and possible super-Earths should begin by not assuming a bulk atmospheric composition. One should perform an agnostic retrieval first, and retrieve the main atmospheric constituent. Then, and if the data suggests the planet's atmosphere is H-rich, a second retrieval assuming an H-rich atmospheric composition could be informative and should be performed. In this context the two retrieval approaches should be seen as complementary and informative. 

Considerations about the presence of clouds and hazes in these mini-Neptunes remain to be explored. The possible absence of clouds in the observable atmosphere of temperate planets like K2-18b, as presented here and in agreement with previous studies \citep{Welbanks2019b, Madhusudhan2020}, represents a surprise when compared to hotter cloudy planets like GJ~1214b \citep{Kreidberg2014a}. On the other hand, the possibility of Mie scattering clouds as recently argued in the atmosphere of GJ~3470b \citep{Benneke2019a} could indicate a complex diversity in the presence of condensates in this type of planets. 

Future studies could investigate the need for more complex cloud and haze models in retrievals when interpreting observations of mini-Neptunes. For instance, while \citet{Benneke2019a} develop and implement a new Mie-scattering cloud parameterization for atmospheric retrievals in order to explain the observations of GJ~3470b, \citet{Welbanks2019b} explain the same observations without invoking Mie-scattering and implementing previous prescriptions for inhomogenous clouds and hazes. Investigating the performance of those cloud prescriptions and the one introduced here could elucidate whether the apparent drop in transit depth in the spectro-photometric observations of GJ~3470b indeed requires invoking Mie-scattering particles. Furthermore, incorporating the Mie-scattering module available in \textsl{Aurora} into its retrieval framework, could provide further insights into the atmospheric nature of this and other planets. 

Likewise, future studies may investigate how the limitations of previous inhomogenous cloud and haze prescriptions unveiled in this work affect recent studies that investigate the influence of cloud model choices on retrieval solutions \citep[e.g.,][]{Barstow2020a}. A full discussion of the different degeneracies and biases in different prescriptions for the clouds and hazes is beyond the scope of this work and reserved for future investigations.

Nonetheless, in order to robustly constrain the presence and properties of clouds and hazes, spectroscopic observations in the optical wavelengths are essential \citep[e.g.,][]{Line2016, Welbanks2019a}. While the results in section \ref{subsubsec:future_k218b} suggest that state-of-the-art spectroscopic observations in the optical with HST-STIS may not be precise enough to robustly constraint the properties of cloud and hazes for the marginal cloud/haze cover in our K2-18b test case, achievable constrains with HST-STIS for instances with enhanced cloud and haze cover and for other mini-Neptunes remain to be explored. Future studies could also explore constraints on the properties of clouds and hazes using observations from multiple instruments on JWST, HST, and ground based facilities.

\subsection{Constraining the Composition of Rocky Exoplanets}
\label{subsec:discussion_se}

Constraining the chemical composition of rocky exoplanet with heavy mean molecular weight atmospheres, needs dedicated observational efforts. Exploratory studies using synthetic observations, as performed here and other studies discussed below, can inform the requirements for future observational campaigns. On the number of transits required for the characterisation of a TRAPPIST-1~d like planet, our results are broadly consistent with previous studies of the TRAPPIST-1 planets \citep[e.g.,][]{Morley2017, Krissansen-Totton2018, Batalha2018, Lustig-Yaeger2019, Wunderlich2019}. 

For instance \citet{Batalha2018}, using models for TRAPPIST-1~f that consider the presence of a gray cloud deck and an information content based approach \citep{Batalha2017a}, find that the NIRSpec prism could detect and constrain the dominant atmospheric absorber in H$_2$O-rich and CO$_2$-rich atmospheres of rocky exoplanets by the 10th transit to uncertainties smaller than 0.5~dex. \citet{Batalha2018} argue that if the dominant absorber has not been observed by the 10th transit it is unlikely that more transits could provide more information. Naturally, our suggestion of characterising a rocky exoplanet using 10 JWST transits is valid only if a featureless spectrum has been ruled out using the first few transits. 

If not pursuing a robust chemical characterisation, a fewer number of transits could help identify spectroscopic features and reject a featureless spectrum. Our results broadly agree with the study of \citet{Lustig-Yaeger2019} who find that two NIRSpec prism transits are enough to rule out a featureless spectrum for TRAPPIST-1~d, although they use a signal to noise metric and we use a Bayesian detection significance metric. Additionally, \citet{Lustig-Yaeger2019} employ atmospheric models from \citet{Lincowski2018} considering self-consistent atmospheric compositions, and find that CO$_2$ could be weakly detected in TRAPPIST-1~d using one transit of JWST-NIRSpec prism in a variety of O$_2$ and CO$_2$ rich atmospheres. While based on simpler atmospheric models and limited to the CO$_2$-rich composition, our results suggest that for a cloud free atmosphere two transits of JWST-NIRSpec prism may suffice to provide initial detections ($\sim2\sigma$) of the main atmospheric component in TRAPPIST-1~d.

Nevertheless, our results are limited to the specific model considerations we have investigated. Modifying our model assumptions of a cloud-free atmosphere, limited number of absorbers, lack of stellar contamination, isothermal temperature profile, amongst others, could result in a larger number of transits required for the desired atmospheric constraints. For instance, using more complex general circulation models (GCM), \citet{Komacek2020} demonstrate that $\sim$10 NIRSpec prism transits would be required to detect H$_2$O vapour in the atmosphere of a terrestrial-size exoplanet orbiting a late-type M dwarf when ignoring the effect of clouds and using a similar signal to noise metric to \citet{Lustig-Yaeger2019}. This result is broadly consistent with our estimate and that of other studies \citep[e.g.,][]{Batalha2018}. However, when the effect of clouds is considered, \citet{Komacek2020} find that 63 or more transits are required to detect water. Their results, and our exploration of cloudy models in CO$_2$-rich and N$_2$-rich atmospheres, suggest that the presence of clouds may significantly increase the number of transits required to detect water features.

Similarly, considering the effect of stellar contamination could significantly affect our interpretations. Recently, \citet{Rackham2018} argue that the stellar contamination impact in the transmission spectra of the TRAPPIST-1 planets can be comparable to or larger than the signal produced by an atmospheric feature. In that case, not accounting for stellar contamination could result in a false positive and be a limiting factor in obtaining reliable abundance constraints. Future studies could investigate the effect of stellar contamination in retrievals \citep[e.g., as in][]{Pinhas2018} for super-Earths/mini-Neptunes using the stellar heterogeneity module included in \textsl{Aurora} and revisit our reported estimates.

Furthermore, considering more complex atmospheric compositions with multiple absorbers as investigated by \citet{Morley2017} could better inform our estimates. In their study, \citet{Morley2017} use radiative-convective models of the TRAPPIST-1 planets assuming Earth-like, Venus-like, and Titan-like atmospheres to determine the number of NIRSpec/NIRISS transit observations required to rule out a flat spectrum at $\sim5\sigma$ confidence. Their results suggest that as few as 13 transits could rule out a flat spectrum for a Venus-like atmospheric composition on TRAPPIST-1~d. While our results seem to be broadly consistent with those of \citet{Morley2017}, albeit more optimistic, the impact of non-isothermal profiles and other chemical compositions on our results remains to be investigated. 

Lastly, the number of transits required for characterising H$_2$O and CO$_2$ may not be representative of the requirements for detecting and characterising other chemical species, including possible biosignatures. For instance, \citet{Barstow2016} investigate the number of transits required to detect O$_3$ in the atmosphere of TRAPPIST-1~d. Their study assumes an atmospheric chemistry identical to Earth's present day atmosphere and employs an optimal estimation retrieval algorithm with isothermal models with clouds deep in the atmosphere where they do not have a significant effect on the spectrum. Their results suggest that present-day Earth levels of O$_3$ would be detectable with 30 transits of NIRSpec prism and MIRI Low-Resolution Spectrometer. Our results suggest that 10 transits of TRAPPIST-1~d with NIRSpec prism could provide initial indications of O$_3$ in a N$_2$-rich cloud-free atmosphere if present at enhanced abundances \citep[$\sim10\mathrm{-}100\times$ present-day stratospheric Earth levels, e.g.,][]{Anderson1987, Barstow2016, Barstow2016b}. Future studies using \textsl{Aurora} could further investigate the requirements for the detection and robust characterisation of O$_3$ and other possible biosignatures \citep[e.g.,][]{Krissansen-Totton2018, Wunderlich2019}.

The characterisation of rocky exoplanets with JWST remains an attractive avenue in the search for atmospheric features in habitable zone planets and the search for possible bio-signatures. Although our results indicate that precise abundance constraints will be possible with the upcoming generation of telescopes, several outstanding considerations mentioned above need to be explored. Particularly the presence of clouds, hazes, and stellar contamination may present a significant hindrance in the characterisation of rocky exoplanets. If true, temperate cloud-free sub-Neptunes like K2-18b may be the best targets for atmospheric characterisation of low-mass exoplanets.

\newpage
\subsection{On Multidimensional Effects}
\label{subsec:limitations}

Modelling the presence of inhomogeneities in the atmospheric properties of a planet requires models beyond 1-dimensional considerations. Depending on their degree of inhomogeneity, these irregularities can potentially affect the retrieved atmospheric properties. For instance, in this work we have explored how inhomogeneities in cloud and haze cover affect the retrieved chemical abundances when assuming or not a H-rich atmosphere. \textsl{Aurora} currently employs a combination of 1-dimensional models to capture the multidimensional effect of inhomogeneous cloud and haze cover. Nonetheless, other heterogeneities and their multidimensional nature can also affect the retrieved atmospheric properties. Recent studies have explored possible limitations of 1-dimensional retrievals in the context of transmission spectroscopy \citep[e.g.,][]{Changeat2019, MacDonald2020, Pluriel2020}.

For instance, compositional differences in the atmospheric chemistry of exoplanets is an effect largely expected for ultra hot Jupiters (UHJs) with day-side temperatures $\gtrsim2200$~K \citep[e.g.,][]{Arcangeli2018,Parmentier2018}. These highly irradiated, tidally locked planets, can exhibit large contrasts between the atmospheric temperature of their day-side and their night-side, which in turn can result in strong variations in their atmospheric composition. Retrievals of UHJs should not assume a homogeneous chemical composition or temperature structure. \textsl{Aurora}'s retrieval framework is currently designed for planets without strong temperature inhomogeneities across the terminator (e.g., low or moderately irradiated planets). However, we have designed \textsl{Aurora} with the implementation of multidimensional effects in mind.

Our current retrieval framework can be readily generalised to incorporate multi-dimensional P-T profiles and non-uniform mixing ratios, just as we have done for clouds and hazes. The inclusion of new nested sampling algorithms, optimised for the treatment of high-dimensional parameter spaces and highly degenerate solutions, aids our retrieval framework in these future developments. Future works can expand the current retrieval framework to consider the effects necessary for the appropriate study of UHJs. Nevertheless, these considerations are not imperative for the planets considered in this study where large compositional/temperature gradients between the day-side and night-side of a planet are not expected.

\newpage
\subsection{Concluding Remarks} 

Currently over 50 transiting exoplanets have been observed with transmission spectroscopy and nearly 20 chemical species have been detected in exoplanetary atmospheres \citep[e.g.,][]{Madhusudhan2019}. While observations of hot gas giants with H-rich atmospheres have been the most abundant, advancements in observing facilities \citep[e.g.,][]{Gillon2011,Gillon2017}, large observing campaigns \citep[e.g.,][]{Kreidberg2014a,Benneke2019b}, as well as the so called M-dwarf opportunity \citep[e.g.,][]{Scalo2007, Charbonneau2007} have allowed for tantalising transmission spectra of mini-Neptunes and super-Earths. The imminent launch of JWST and the continuous observing efforts with HST and ground facilities promise to reveal many more spectra of transiting exoplanets, including the prospect of spectral features in non H-rich atmospheres. From hot Jupiters to temperate low-mass exoplanets, their spectra could provide further insights into their formation paths, possible trends in compositions and maybe even their prospects for habitability. It is in this context that retrieval capabilities like \textsl{Aurora} could play an important role in accurate interpretation of spectroscopic observations. 

\acknowledgments
{L.W. thanks the Gates Cambridge Trust for support towards his doctoral studies. This research is made open access thanks to the Bill \& Melinda Gates Foundation. This work was performed using resources provided by the Cambridge Service for Data Driven Discovery (CSD3) operated by the University of Cambridge Research Computing Service (\url{www.csd3.cam.ac.uk}), provided by Dell EMC and Intel using Tier-2 funding from the Engineering and Physical Sciences Research Council (capital grant EP/P020259/1), and DiRAC funding from the Science and Technology Facilities Council (\url{www.dirac.ac.uk}). This research has made use of the NASA Exoplanet Archive, which is operated by the California Institute of Technology, under contract with the National Aeronautics and Space Administration under the Exoplanet Exploration Program. L.W. thanks Siddharth Gandhi and Arazi Pinhas for their help with the AURA retrieval code which is the progenitor of \textsl{Aurora} and for various helpful discussions. L.W. thanks Peter McGill for helpful discussions, in particular on Gaussian processes. L.W. thanks Anjali Piette for helpful discussions, in particular on sources of opacity. L.W. thanks Matthew Nixon for helpful discussions and his input generating Figure \ref{fig:trappist_hab}. We thank the anonymous referee for their helpful comments.}

\software{Astropy \citep{Astropy2018}, Celerite \citep{celerite}, Dynesty \citep{Speagle2020}, George \citep{george}, Matplotlib \citep{Matplotlib}, MultiNest \citep{Feroz2009, Feroz2013}, Numpy \citep{numpy}, Pandexo \citep{Batalha2017b}, PyMultiNest \citep{Buchner2014}, PyPolyChord \citep{Handley2015a, Handley2015b}, SciPy \citep{scipy}.}

\clearpage
\newpage
\appendix
\restartappendixnumbering
\onecolumngrid 
\section{Priors Used in This Work}

\begin{deluxetable}{c|c|c}[h]
\tablecaption{Parameters and priors used in this work.
\label{table:priors}}
\tablecolumns{3}
\tablewidth{\columnwidth}
\tablehead{
\colhead{Parameter} & \colhead{Prior distribution} & \colhead{Prior range}
}
\startdata
$X_i$ & Log-uniform & \makecell{$10^{-12}\text{-}10^{0}$ for H-rich retrievals of HD~209485b (section \ref{subsec:cloud_validation})  \\$10^{-12}\text{-}10^{-0.3}$ for H-rich retrievals of K2-18b (section \ref{subsec:samplers_results} and \ref{subsubsec:k218b_xhrich})} \\ \hline
$z_i$ & Uniform & \makecell{$-4.61\text{-}23.03$ for non H-rich retrievals with 6 chemical components (section \ref{subsubsec:trappist1d}) \\$-3.45\text{-}24.18$ for non H-rich retrievals with 8 chemical components (section \ref{subsubsec:k218b_xhrich}) \\$-3.07\text{-}24.56$ for non H-rich retrievals with 9 chemical components (section \ref{subsec:cloud_validation})  \\ $-2.76\text{-}24.87$ for non H-rich retrievals with 10 chemical components  (section \ref{subsubsec:k218b_xhrich}) }\\ \hline
$T_0$ & Uniform &  \makecell{$0\text{-}300$~K for retrievals of K2-18b \\$0\text{-}400$~K for retrievals of TRAPPIST-1~d \\$800\text{-}1550$~K for retrievals of HD~209458~b} \\ \hline
$\alpha_{1,2}$  &  Uniform & $0.02\text{-}2.00$ K$^{-1/2}$\\ \hline
$P_{1,2,\text{cloud},\text{ref}}$ & Log-uniform & $10^{-6}\text{-}10^{2}$ bar\\ \hline
$P_{3 }$ & Log-uniform & $10^{-2}\text{-}10^{2}$ bar\\ \hline
$a$ & Log-uniform  & $10^{-4}\text{-}10^{10}$\\ \hline
$\gamma$ & Uniform & -$20\text{-}2$\\ \hline
 \makecell{$\phi_{\mathrm{clouds}}$\\$\phi_{\mathrm{hazes}}$\\$\phi_{\mathrm{clouds+hazes}}$} & Uniform & $0\text{-}1$
\enddata 
\tablecomments{The priors for the compositional parameters ($z_i$) are reported here to two decimal places and calculated for the purposes of this table. The compositional parameter priors are automatically calculated by \textsl{Aurora} given the condition that they must span the range $10^{-12}\text{-}10^{0}$ in the $X_i$ space and that they must satisfy the conditions explained in section \ref{subsec:reparameterization}. }
\end{deluxetable}

\onecolumngrid 
\section{ Additional Forward Scattering and Refraction Models for a Rocky Exoplanet.}
\label{app:trappist_fwd_models}

\begin{figure}
  \centering
  \begin{tabular}[b]{@{}p{0.43\textwidth}@{}}
    \centering\includegraphics[width=\linewidth]{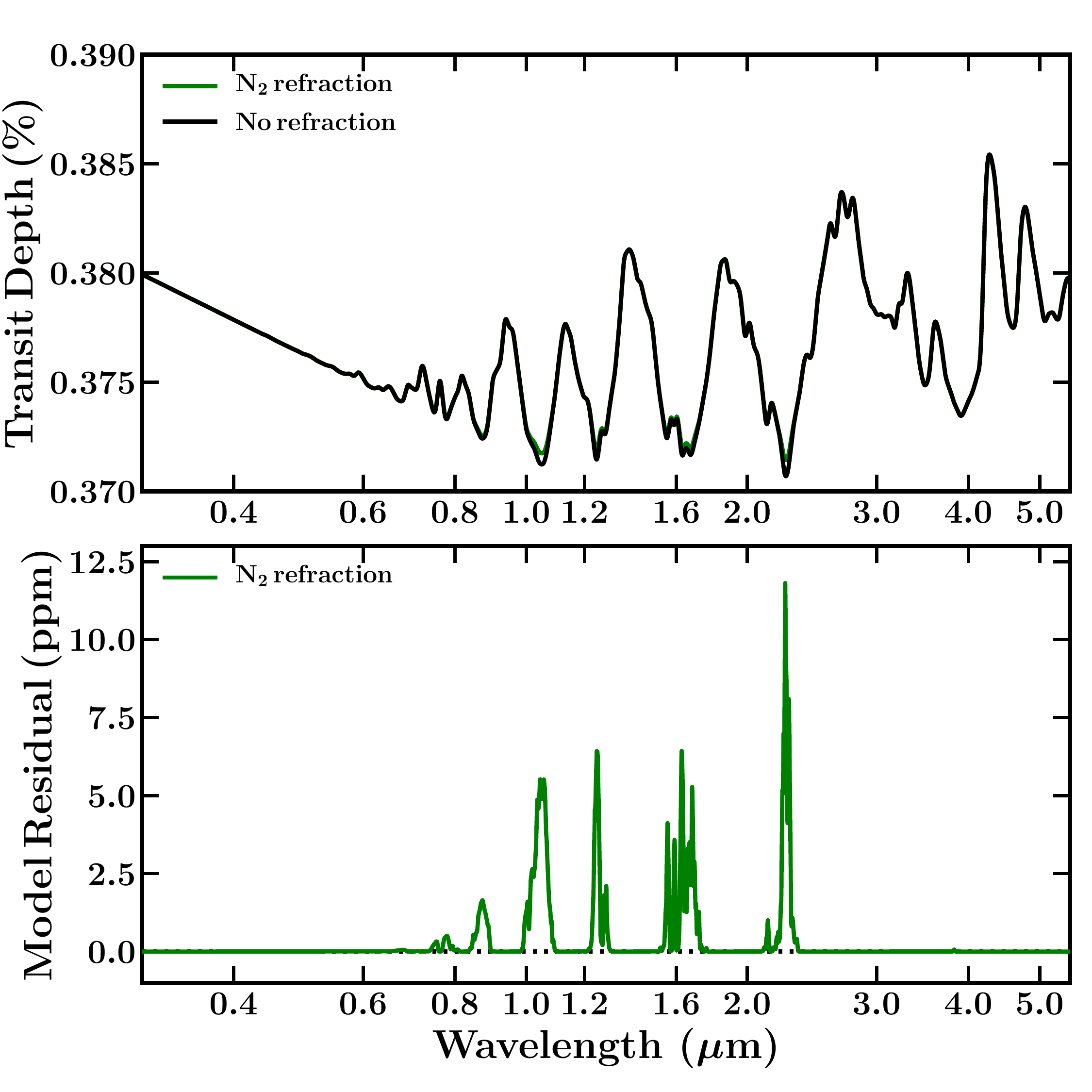} \\
    \centering\small (a) Wavelength-dependent refraction
  \end{tabular}%
  \quad
  \begin{tabular}[b]{@{}p{0.43\textwidth}@{}}
    \centering\includegraphics[width=\linewidth]{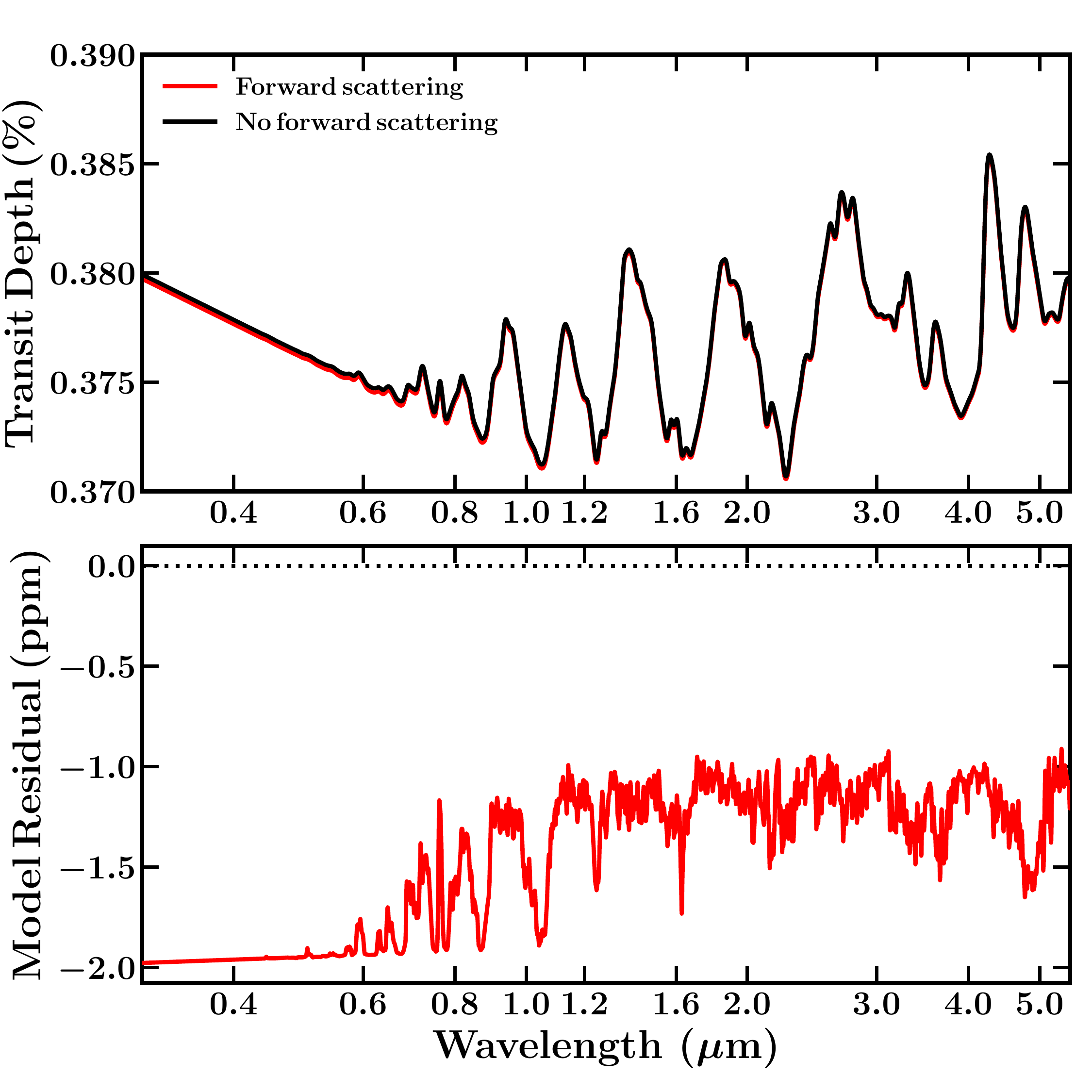} \\
    \centering\small  (b) Forward scattering
  \end{tabular}
  \caption{ (a) Effects of wavelength-dependent refraction and (b) forward scattering for a TRAPPIST-1~d-like rocky exoplanet. The top panels show the forward model without these effects considered (black), the model with N$_2$ refraction (green), and the model considering forward scattering (red). The bottom panels show the model residuals.}
  \label{fig:trappist_fwd_models}
\end{figure}

For completion, we consider the effects of wavelength-dependent refraction and forward scattering presented in section \ref{subsubsec:refraction_scattering} on the rocky exoplanet TRAPPIST-1~d. We consider the same N$_2$-rich atmospheric model used in section \ref{subsubsec:trappist1d}. Namely, we consider an atmosphere composed of 77.51\% N$_2$, 21.38\% O$_2$, 1\% H$_2$O, 0.1\% CO$_2$, and 0.01\% O$_3$. The model spectra are generated at a constant resolution of R=5000 between $0.3$-$5.5\mu$m, and smoothed for the purposes of Figure \ref{fig:trappist_fwd_models}.

When considering the effects of wavelength-dependent refraction, we specifically include N$_2$ refraction as our atmospheric model is N$_2$-rich. The forward model considering the effects of forward scattering uses the same assumptions as in section \ref{subsubsec:refraction_scattering} (i.e., $g=0.95$ and $\tilde{\omega}_o=1$) and the planet-star orbital separation reported by \citet{Agol2020}. Figure \ref{fig:trappist_fwd_models} shows these forward models and their respective model residuals.

These model considerations have a stronger effect in this rocky exoplanet test case than in the mini-Neptune test case presented in section \ref{subsubsec:refraction_scattering}. The impact of N$_2$ refraction results in a model difference of $\sim12$~ppm. On the other hand, forward scattering results in a model difference $\lesssim2$~ppm. The impact of these considerations is $\sim3\times$ stronger in this TRAPPIST-1~d case than in the K2-18b case presented in the main text. Nonetheless, these results are dependent on the assumed synthetic model considerations and should not be generalised to other test cases beyond the ones considered here.

\newpage

\onecolumngrid 
\section{Validation of Aurora on HD~209458~b}

\begin{rotatetable}
\begin{deluxetable}{lcccccccc}
\tablecaption{Retrieved parameters for the spectrum of HD~209458~b using different cloud and haze models, as well as assuming a H-rich atmosphere or not as explained in section \ref{subsec:cloud_validation}
\label{table:HD209485b_validation}}
\tablewidth{700pt}
\tabletypesize{\scriptsize}
\movetableright=1pt
\tablehead{
\colhead{Parameter}  & \colhead{Model 0} & \colhead{Model 0} & \colhead{Model 1} & \colhead{Model 1} & \colhead{Model 2}& \colhead{Model 2} & \colhead{Model 3}& \colhead{Model 3} \\
\colhead{}  & \colhead{H-rich} & \colhead{no H-rich} & \colhead{H-rich} & \colhead{no H-rich}  & \colhead{H-rich} & \colhead{no H-rich}  & \colhead{H-rich} & \colhead{no H-rich}
}
\startdata
$\log_{10} \left(X_{\textnormal{H}_2+\textnormal{He}}\right)$& N/A &$-6.03 ^{+ 1.38 }_{- 2.18 }\times10^{-06}$ & N/A &$-1.99 ^{+ 0.97 }_{- 2.84 }\times10^{-05}$& N/A&$-9.52 ^{+ 7.49}_{- 120.49}\times10^{-05}$ & N/A& $-1.94 ^{+ 1.68 }_{- 34.22}\times10^{-04}$ \\
 $\log_{10} \left(X_{\textnormal{H}_2\textnormal{O}}\right)$ & $-4.98 ^{+ 0.13 }_{- 0.14 }$ &$-4.99 ^{+ 0.13 }_{- 0.14 }$&$-4.53 ^{+ 0.34 }_{- 0.28 }$&$-4.48 ^{+ 0.29 }_{- 0.27 }$ &$-4.06 ^{+ 0.73 }_{- 0.51 }$&$-4.00 ^{+ 0.68 }_{- 0.51 }$ &$-3.89 ^{+ 0.78 }_{- 0.62 }$&$-3.81 ^{+ 0.73 }_{- 0.63 }$ \\
$\log_{10} \left(X_{\textnormal{Na}}\right)$&  $-6.75 ^{+ 0.20 }_{- 0.21 }$&$-6.74 ^{+ 0.20 }_{- 0.22 }$  &$-5.43 ^{+ 0.62 }_{- 0.49 }$ &$-5.35 ^{+ 0.54 }_{- 0.43 }$&$-4.34 ^{+ 1.55 }_{- 0.99 }$ &$-4.23 ^{+ 1.45 }_{- 0.99 }$&$-3.95 ^{+ 1.71 }_{- 1.27 }$&$-3.80 ^{+ 1.57 }_{- 1.27 }$  \\
$\log_{10} \left(X_{\textnormal{K}}\right)$&  $-7.94 ^{+ 0.24 }_{- 0.25 }$ &$-7.92 ^{+ 0.23 }_{- 0.26 }$&$-6.96 ^{+ 0.63 }_{- 0.51 }$&$-6.90 ^{+ 0.54 }_{- 0.45 }$&$-6.04 ^{+ 1.44 }_{- 0.96 }$&$-5.96 ^{+ 1.40 }_{- 0.93 }$ &$-5.73 ^{+ 1.65 }_{- 1.15 }$&$-5.58 ^{+ 1.51 }_{- 1.18 }$  \\
$\log_{10} \left(X_{\textnormal{CH}_4}\right)$& $-6.25 ^{+ 0.58 }_{- 3.44 }$&$-6.13 ^{+ 0.47 }_{- 2.45 }$&$-8.62 ^{+ 2.11 }_{- 2.23 }$&$-8.09 ^{+ 1.62 }_{- 1.84 }$ &$-8.28 ^{+ 2.28 }_{- 2.44 }$ &$-7.71 ^{+ 1.83 }_{- 2.14 }$&$-8.19 ^{+ 2.33 }_{- 2.52 }$ &$-7.56 ^{+ 1.88 }_{- 2.23 }$\\
$\log_{10} \left(X_{\textnormal{NH}_3}\right)$ &   $-5.93 ^{+ 0.11 }_{- 0.13 }$&$-5.92 ^{+ 0.11 }_{- 0.13 }$ &$-6.57 ^{+ 0.80 }_{- 3.48 }$&$-6.42 ^{+ 0.67 }_{- 2.80 }$&$-6.40 ^{+ 1.10 }_{- 3.59 }$ &$-6.18 ^{+ 0.96 }_{- 2.93 }$&$-6.77 ^{+ 1.50 }_{- 3.48 }$ &$-6.33 ^{+ 1.17 }_{- 2.95 }$ \\
$\log_{10} \left(X_{\textnormal{HCN}}\right)$ &  $-9.83 ^{+ 1.97 }_{- 1.47 }$&$-9.02 ^{+ 1.64 }_{- 1.46 }$  &$-9.44 ^{+ 2.26 }_{- 1.74 }$&$-8.76 ^{+ 1.71 }_{- 1.50 }$&$-9.37 ^{+ 2.37 }_{- 1.76 }$ &$-8.50 ^{+ 2.00 }_{- 1.75 }$&$-9.25 ^{+ 2.44 }_{- 1.87 }$ &$-8.39 ^{+ 2.12 }_{- 1.79 }$ \\
$\log_{10} \left(X_{\textnormal{CO}}\right)$& $-7.08 ^{+ 1.43 }_{- 1.65 }$&$-6.76 ^{+ 1.26 }_{- 1.43 }$ &$-6.33 ^{+ 1.62 }_{- 1.85 }$&$-6.16 ^{+ 1.42 }_{- 1.49 }$&$-5.93 ^{+ 1.79 }_{- 1.98 }$&$-5.62 ^{+ 1.66 }_{- 1.76 }$&$-5.76 ^{+ 1.90 }_{- 2.02 }$&$-5.38 ^{+ 1.70 }_{- 1.87 }$  \\
$\log_{10} \left(X_{\textnormal{CO}_2}\right)$&  $-10.19 ^{+ 1.47 }_{- 1.22 }$&$-9.42 ^{+ 1.17 }_{- 1.21 }$&$-9.94 ^{+ 1.64 }_{- 1.40 }$ &$-9.22 ^{+ 1.23 }_{- 1.22 }$&$-9.69 ^{+ 1.80 }_{- 1.55 }$&$-9.02 ^{+ 1.51 }_{- 1.40 }$ &$-9.65 ^{+ 1.85 }_{- 1.61 }$&$-8.89 ^{+ 1.60 }_{- 1.49 }$ \\
 \hline
$T_0$ (K) &  $810.55 ^{+ 14.87 }_{- 7.71 }$&$810.36 ^{+ 14.43 }_{- 7.51 }$ &$972.42 ^{+ 265.03 }_{- 126.67 }$ &$984.32 ^{+ 225.59 }_{- 125.41 }$ &$1064.53 ^{+ 283.25 }_{- 195.12 }$ &$1062.14 ^{+ 282.43 }_{- 192.28 }$&$1081.01 ^{+ 277.22 }_{- 202.50 }$  &$1076.33 ^{+ 267.44 }_{- 197.43 }$  \\
$\alpha_1$ & $1.74 ^{+ 0.19 }_{- 0.32 }$&$1.74 ^{+ 0.18 }_{- 0.32 }$ &$1.17 ^{+ 0.54 }_{- 0.52 }$&$1.17 ^{+ 0.49 }_{- 0.46 }$&$1.10 ^{+ 0.59 }_{- 0.53 }$ &$1.11 ^{+ 0.58 }_{- 0.53 }$&$1.04 ^{+ 0.62 }_{- 0.54 }$ &$1.06 ^{+ 0.61 }_{- 0.54 }$   \\
$\alpha_2$ & $1.48 ^{+ 0.37 }_{- 0.65 }$&$1.47 ^{+ 0.37 }_{- 0.63 }$&$0.97 ^{+ 0.68 }_{- 0.58 }$ &$1.03 ^{+ 0.57 }_{- 0.54 }$ &$1.00 ^{+ 0.65 }_{- 0.59 }$ &$1.01 ^{+ 0.64 }_{- 0.59 }$&$0.99 ^{+ 0.67 }_{- 0.59 }$&$0.99 ^{+ 0.65 }_{- 0.59 }$    \\
$\log_{10}$($P_{1}$) (bar) & $-1.78 ^{+ 1.93 }_{- 1.94 }$&$-1.78 ^{+ 1.96 }_{- 1.88 }$&$-1.61 ^{+ 1.68 }_{- 1.72 }$&$-1.51 ^{+ 1.46 }_{- 1.51 }$&$-1.56 ^{+ 1.66 }_{- 1.76 }$&$-1.57 ^{+ 1.66 }_{- 1.74 }$&$-1.61 ^{+ 1.67 }_{- 1.77 }$&$-1.62 ^{+ 1.68 }_{- 1.73 }$ \\
$\log_{10}$($P_{2}$) (bar)&  $-3.89 ^{+ 1.83 }_{- 1.40 }$&$-3.90 ^{+ 1.85 }_{- 1.35 }$&$-4.13 ^{+ 1.92 }_{- 1.29 }$ &$-4.01 ^{+ 1.60 }_{- 1.26 }$&$-4.07 ^{+ 1.92 }_{- 1.34 }$&$-4.11 ^{+ 1.91 }_{- 1.28 }$&$-4.14 ^{+ 1.92 }_{- 1.30 }$&$-4.09 ^{+ 1.84 }_{- 1.31 }$ \\
$\log_{10}$($P_{3}$) (bar) & $0.52 ^{+ 1.06 }_{- 1.67 }$&$0.46 ^{+ 1.09 }_{- 1.66 }$ &$0.62 ^{+ 0.96 }_{- 1.32 }$ &$0.67 ^{+ 0.84 }_{- 1.11 }$ &$0.64 ^{+ 0.95 }_{- 1.37 }$ &$0.62 ^{+ 0.94 }_{- 1.33 }$&$0.59 ^{+ 0.97 }_{- 1.34 }$&$0.60 ^{+ 0.95 }_{- 1.34 }$ \\
 \hline
$\log_{10}$($P_{\rm{ref}}$) (bar)&  $-0.71 ^{+ 0.06 }_{- 0.06 }$&$-0.72 ^{+ 0.06 }_{- 0.05 }$&$-2.64 ^{+ 0.66 }_{- 0.63 }$ &$-2.70 ^{+ 0.55 }_{- 0.56 }$&$-3.15 ^{+ 0.79 }_{- 0.73 }$&$-3.18 ^{+ 0.76 }_{- 0.71 }$&$-3.33 ^{+ 0.79 }_{- 0.64 }$&$-3.37 ^{+ 0.80 }_{- 0.63 }$ \\
 \hline
$\log_{10}$(a)&  N/A  & N/A &$4.35 ^{+ 0.71 }_{- 1.01 }$&$4.49 ^{+ 0.56 }_{- 0.65 }$ &$2.88 ^{+ 0.91 }_{- 0.85 }$&$2.96 ^{+ 0.85 }_{- 0.85 }$&$3.28 ^{+ 1.01 }_{- 1.13 }$ &$3.34 ^{+ 0.92 }_{- 1.07 }$  \\
$\gamma$ & N/A& N/A&$-14.04 ^{+ 4.53 }_{- 3.94 }$ &$-14.54 ^{+ 3.85 }_{- 3.31 }$ &$-16.57 ^{+ 3.15 }_{- 2.37 }$ &$-16.60 ^{+ 2.93 }_{- 2.29 }$&$-16.15 ^{+ 3.36 }_{- 2.60 }$ &$-16.27 ^{+ 3.25 }_{- 2.50 }$  \\
$\log_{10}$($P_{\text{cloud}}$) (bar) & N/A& N/A&$-4.41 ^{+ 0.80 }_{- 0.57 }$&$-4.47 ^{+ 0.56 }_{- 0.50 }$ &$-4.60 ^{+ 0.99 }_{- 0.91 }$&$-4.65 ^{+ 0.97 }_{- 0.86 }$ &$-4.72 ^{+ 0.99 }_{- 0.84 }$ &$-4.74 ^{+ 0.99 }_{- 0.82 }$\\
$\phi_{\mathrm{clouds}}$ & N/A& N/A & N/A& N/A &$0.60 ^{+ 0.07 }_{- 0.11 }$ &$0.59 ^{+ 0.07 }_{- 0.11 }$&$0.34 ^{+ 0.18 }_{- 0.20 }$  &$0.35 ^{+ 0.17 }_{- 0.20 }$\\
$\phi_{\mathrm{hazes}}$ & N/A& N/A& N/A& N/A&$0.30 ^{+ 0.09 }_{- 0.07 }$ &$0.30 ^{+ 0.08 }_{- 0.07 }$ &$0.27 ^{+ 0.09 }_{- 0.10 }$&$0.27 ^{+ 0.09 }_{- 0.09 }$ \\
$\phi_{\mathrm{cloud+hazes}}$  & N/A& N/A&$0.54 ^{+ 0.09 }_{- 0.12 }$ &$0.53 ^{+ 0.08 }_{- 0.10 }$& N/A& N/A &$0.24 ^{+ 0.19 }_{- 0.16 }$ &$0.24 ^{+ 0.19 }_{- 0.15 }$\\
\hline
$\log$($\mathcal{Z}$)  &  949.74 & 946.68 & 957.70 & 951.35 & 958.14 & 955.38 &958.40&955.74
\enddata
\tablecomments{N/A means that the parameter not considered in the model by construction.}
\end{deluxetable}
\end{rotatetable}

\section{Procedure for Simulating JWST Observations}
\subsection{PANDEXO Input for K2-18b}
\label{app:pandexo_k218b}
The assumed stellar spectrum for K2-18 is interpolated from the Phoenix Stellar Atlas \citep{Husser2013} by PANDEXO assuming $T_{\mathrm{eff}}$= 3457 \citep{Benneke2019b}, [Fe/H]=0.12 \citep{Sarkis2018}, $\log_{10}$(g)=4.858 cgs \citep{Crossfield2016}, R$_{\text{star}}= 0.4445 R_{\odot}$ \citep{Benneke2019b}, and a K-band magnitude of 8.899. We assume a transit duration of 2.663 hours \citep{Benneke2017}. The in transit and out of transit fluxes are computed using this stellar spectrum and a constant transit depth model assuming a planet radius of 2.61 R$_\Earth$ \citep{Benneke2019b}.

Our simulated observations assume a saturation limit of 80\% full well and an extremely optimistic noise floor of 5~ppm. Our assumed noise floor is much smaller than expectations from \citet{Greene2016} and \citet{Beichman2018}. The upcoming JWST Cycle 1 will better inform the existence of a possible systematic noise floor. Nonetheless, given that the noise in our synthetic observations at the native resolution of the gratings is much higher than noise floor expectations, our choice of 5~ppm has little impact on our results. We generate these synthetic observations for 1 transit and the number of groups per integration suggested by PANDEXO following the \textit{optimize} option. The number of groups per integration is 13, 14, and 25 for G140H, G235H, and G395H respectively. Once the observations are simulated using PANDEXO, we proceed bin every two pixels to obtain individual spectral resolution elements. Subsequently, we remove any resolution elements that fall within the gaps in the spectral configurations chosen; namely 1.31-1.35 $\mu$m for G140H/F100LP, 2.20-2.27 $\mu$m for G235H/F170LP and 3.72-3.82 $\mu$m for G395H/F290LP. After this, we remove any outlier elements with noise levels above 500~ppm for G140H and G235H, and 400~ppm for G395H. We proceed to bin the observations every 40 resolution elements, remove any binned elements that fall within the gaps in the spectral configurations chosen, and obtain an average resolution of R$\sim$70. The synthetic observations have a mean precision of $\sim 26$~ppm, $\sim 30$~ppm, and $\sim 39$~ppm, for G140H, G235H, and G395H, respectively.

\subsection{PANDEXO Input for TRAPPIST-1~d}
\label{app:pandexo_trappist}
The assumed stellar spectrum is interpolated from the Phoenix Stellar Atlas assuming $T_{\mathrm{eff}}$= 2566~K \citep{Agol2020}, [Fe/H]=0.04, R$_{\text{star}}= 0.1170 R_{\odot}$ \citep{Gillon2017}, $\log_{10}$(g)=5.2396 cgs \citep{Agol2020} and a K-band magnitude of 10.296. We assume a transit duration of 48.87 minutes \citep{Agol2020}. The in transit and out of transit fluxes are computed using this stellar spectrum and a a constant transit depth model assuming a planet radius of 0.788 R$_\Earth$ \citep{Agol2020}.

We maintain the assumption of a saturation limit of 80\% full well and an optimistic noise floor of 5~ppm. We generate the observations for 10 NIRSpec prism transits and 2 groups per integration. The instrument configuration for PANDEXO uses aperture S1600A1 and subarray SUB512. We bin every two pixels to obtain individual resolution elements. We remove any elements with precisions larger than 300~ppm, leaving elements $\gtrsim0.7\mu$m and $\lesssim5.3\mu$m.

\clearpage
\onecolumngrid 
\section{Posterior Distributions for the Retrieval of K2-18b Using Synthetic Observations}
\label{app:k218b_posterior}
\begin{figure*}[b!]
\includegraphics[width=\textwidth]{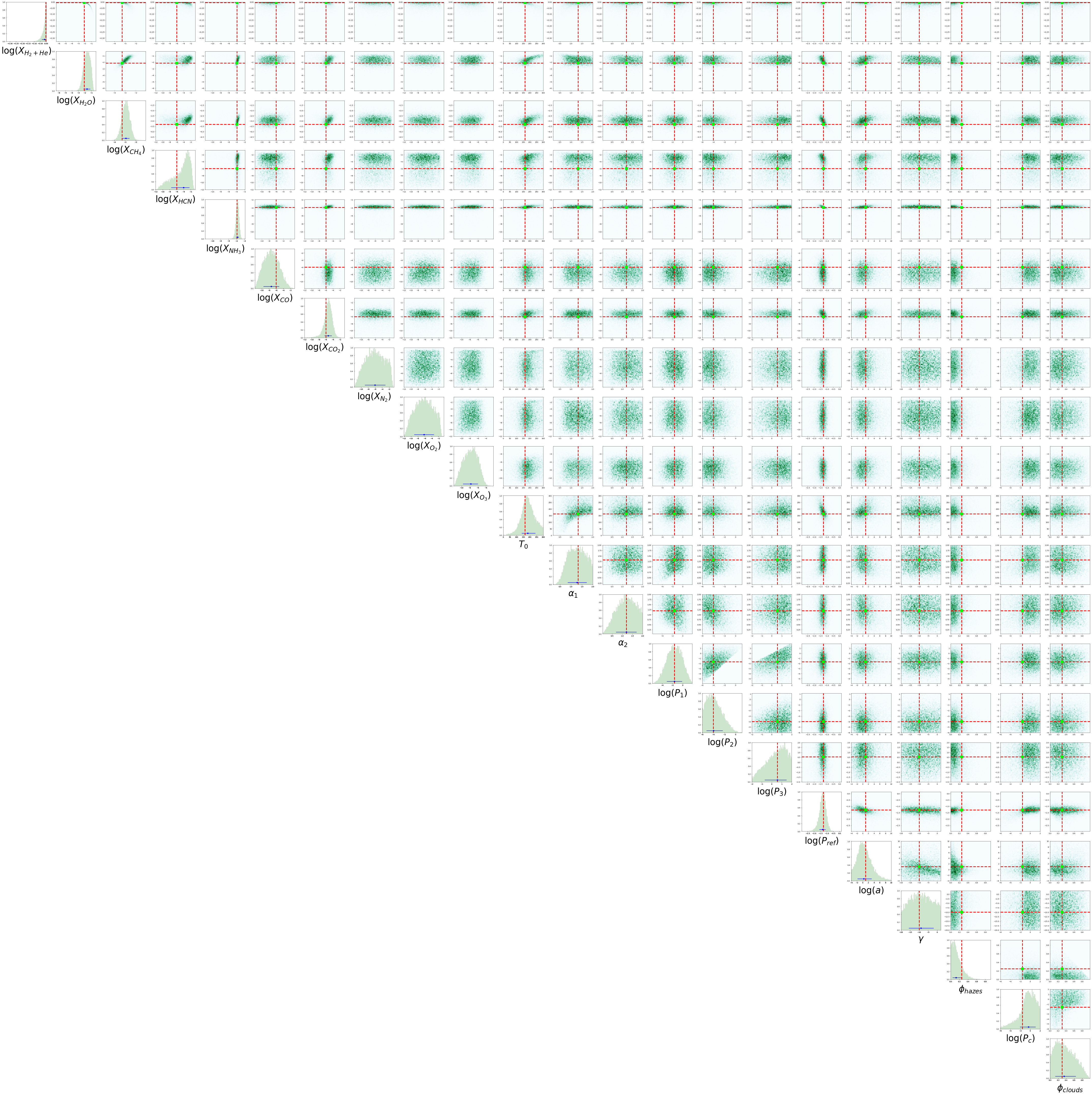}

    \def\arraystretch{1.2}
    \setlength{\arrayrulewidth}{1.0pt}
    \centering
    \footnotesize{
    \vspace{-10.0cm}\hspace{-11.5cm}\begin{tabular}{ccc}
    \hline
    Parameter & Input value &Median $^{+1\sigma}_{-1\sigma}$ \\
    \hline
    $\log_{10} \left(X_{\textnormal{H}_2+\textnormal{He}}\right)$ & $-0.004$ & $-0.013 ^{+ 0.010 }_{- 0.024 } $\\
    $\log_{10} \left( X_{\textnormal{H}_2\textnormal{O}}\right)$  &$-2.10$ &$-1.66 ^{+ 0.50 }_{- 0.55 }$\\
    $\log_{10} \left(X_{\textnormal{CH}_4}\right)$ &$-3.30$ &$-2.94 ^{+ 0.35 }_{- 0.37 }$\\
    $\log_{10} \left(X_{\textnormal{HCN}}\right)$ &$-6.00$ &$-4.04 ^{+ 1.64 }_{- 3.54 }$\\
    $\log_{10} \left(X_{\textnormal{NH}_3}\right)$ &$-3.90$ &$-3.79 ^{+ 0.36 }_{- 0.40 }$\\
    $\log_{10} \left(X_{\textnormal{CO}}\right)$ &$-6.00$ &$-7.41 ^{+ 2.22 }_{- 2.18 }$\\
    $\log_{10} \left(X_{\textnormal{CO}_2}\right)$ &$-6.00$ &$-5.32 ^{+ 0.97 }_{- 1.11 }$\\
    $\log_{10} \left(X_{\textnormal{N}_2}\right)$ & N/A &$-6.04 ^{+ 3.02 }_{- 2.92 }$\\
    $\log_{10} \left(X_{\textnormal{O}_2}\right)$ &N/A &$-6.23 ^{+ 2.90 }_{- 2.86 }$\\
    $\log_{10} \left(X_{\textnormal{O}_3}\right)$ &N/A&$-7.75 ^{+ 1.78 }_{- 1.93 } $\\
    T$_0$ (K) &$162.12$ &$ 184.35 ^{+ 55.83 }_{- 46.66 }  $\\
    $\alpha_1$ &$1.32$ &$ 1.28 ^{+ 0.44 }_{- 0.43 }  $\\
    $\alpha_2$ &$1.20$ &$1.20 ^{+ 0.50 }_{- 0.52 }  $\\
    $\log_{10}$($\text{P}_{1}$) (bar) &$-1.64$ &$-1.62 ^{+ 1.51 }_{- 1.51 } $\\
    $\log_{10}$($\text{P}_{2}$) (bar) &$-4.00$ &$-3.98 ^{+ 1.63 }_{- 1.26 }$\\
    $\log_{10}$($\text{P}_{3}$) (bar) &$0.56$ &$0.56 ^{+ 0.92 }_{- 1.27 } $\\
    $\log_{10}$($\text{P}_{ref}$) (bar) &$-1.27$ &$-1.32 ^{+ 0.24 }_{- 0.26 }$\\
    $\log_{10}$(a) &$1.0$ &$0.39 ^{+ 2.78 }_{- 2.22 }$\\
    $\gamma$ &$-10.0$ &$ -9.03 ^{+ 7.03 }_{- 6.67 } $\\
    $\phi_{\mathrm{hazes}}$ &$0.25$ &$0.12 ^{+ 0.14 }_{- 0.08 }  $\\
    $\log_{10}$($\text{P}_{\text{cloud}}$) (bar) &$-1.60$ &$-0.35 ^{+ 1.43 }_{- 1.77 }$\\
    $\phi_{\mathrm{clouds}}$ &$0.30$ &$0.35 ^{+ 0.29 }_{- 0.22 }$\\

    \hline
    \end{tabular}}
    \vspace{0.5cm}
    
\caption[k218b_stis_jwst_1]{Posterior distribution for the retrieval of synthetic HST-STIS and JWST-NIRSpec observations of K2-18b (see section \ref{subsubsec:future_k218b}). The retrieval does not assume a H-rich atmosphere. The input parameters for the true model are shown using vertical dashed red lines. Additional examples of posterior distributions are available online {\url{https://osf.io/3m94k}}. } \label{fig:k218b_stis_jwst_1}
\end{figure*} 

\clearpage

\bibliographystyle{aasjournal}
\bibliography{biblio}

\begin{thebibliography}{}
\expandafter\ifx\csname natexlab\endcsname\relax\def\natexlab#1{#1}\fi
\providecommand{\url}[1]{\href{#1}{#1}}
\providecommand{\dodoi}[1]{doi:~\href{http://doi.org/#1}{\nolinkurl{#1}}}
\providecommand{\doeprint}[1]{\href{http://ascl.net/#1}{\nolinkurl{http://ascl.net/#1}}}
\providecommand{\doarXiv}[1]{\href{https://arxiv.org/abs/#1}{\nolinkurl{https://arxiv.org/abs/#1}}}

\bibitem[{{Adams} {et~al.}(2019){Adams}, {Gao}, {de Pater}, \&
  {Morley}}]{Adams2019}
{Adams}, D., {Gao}, P., {de Pater}, I., \& {Morley}, C.~V. 2019, \apj, 874, 61,
  \dodoi{10.3847/1538-4357/ab074c}

\bibitem[{{Agol} {et~al.}(2020){Agol}, {Dorn}, {Grimm}, {Turbet}, {Ducrot},
  {Delrez}, {Gillon}, {Demory}, {Burdanov}, {Barkaoui}, {Benkhaldoun},
  {Bolmont}, {Burgasser}, {Carey}, {de Wit}, {Fabrycky}, {Foreman-Mackey},
  {Haldemann}, {Hernandez}, {Ingalls}, {Jehin}, {Langford}, {Leconte},
  {Lederer}, {Luger}, {Malhotra}, {Meadows}, {Morris}, {Pozuelos}, {Queloz},
  {Raymond}, {Selsis}, {Sestovic}, {Triaud}, \& {Van Grootel}}]{Agol2020}
{Agol}, E., {Dorn}, C., {Grimm}, S.~L., {et~al.} 2020, arXiv e-prints,
  arXiv:2010.01074.
\newblock \doarXiv{2010.01074}

\bibitem[{Aitchison(1986)}]{Aitchison1986}
Aitchison, J. 1986, The Statistical Analysis of Compositional Data, Monographs
  on Statistics and Applied Probability (Springer Netherlands)

\bibitem[{Aitchison \& J.~Egozcue(2005)}]{Aitchison2005}
Aitchison, J., \& J.~Egozcue, J. 2005, Mathematical Geology, 37, 829,
  \dodoi{10.1007/s11004-005-7383-7}

\bibitem[{{Al-Refaie} {et~al.}(2019){Al-Refaie}, {Changeat}, {Waldmann}, \&
  {Tinetti}}]{Al-Refaie2019}
{Al-Refaie}, A.~F., {Changeat}, Q., {Waldmann}, I.~P., \& {Tinetti}, G. 2019,
  arXiv e-prints, arXiv:1912.07759.
\newblock \doarXiv{1912.07759}

\bibitem[{{Allard} {et~al.}(2016){Allard}, {Spiegelman}, \&
  {Kielkopf}}]{Allard2016}
{Allard}, N.~F., {Spiegelman}, F., \& {Kielkopf}, J.~F. 2016, \aap, 589, A21,
  \dodoi{10.1051/0004-6361/201628270}

\bibitem[{{Allard} {et~al.}(2019){Allard}, {Spiegelman}, {Leininger}, \&
  {Molliere}}]{Allard2019}
{Allard}, N.~F., {Spiegelman}, F., {Leininger}, T., \& {Molliere}, P. 2019,
  \aap, 628, A120, \dodoi{10.1051/0004-6361/201935593}

\bibitem[{{Ambikasaran} {et~al.}(2015){Ambikasaran}, {Foreman-Mackey},
  {Greengard}, {Hogg}, \& {O'Neil}}]{george}
{Ambikasaran}, S., {Foreman-Mackey}, D., {Greengard}, L., {Hogg}, D.~W., \&
  {O'Neil}, M. 2015, IEEE Transactions on Pattern Analysis and Machine
  Intelligence, 38, 252, \dodoi{10.1109/TPAMI.2015.2448083}

\bibitem[{Anderson(1987)}]{Anderson1987}
Anderson, J.~G. 1987, Annual Review of Physical Chemistry, 38, 489,
  \dodoi{10.1146/annurev.pc.38.100187.002421}

\bibitem[{{Arcangeli} {et~al.}(2018){Arcangeli}, {D{\'e}sert}, {Line}, {Bean},
  {Parmentier}, {Stevenson}, {Kreidberg}, {Fortney}, {Mansfield}, \&
  {Showman}}]{Arcangeli2018}
{Arcangeli}, J., {D{\'e}sert}, J.-M., {Line}, M.~R., {et~al.} 2018, \apjl, 855,
  L30, \dodoi{10.3847/2041-8213/aab272}

\bibitem[{{Asplund} {et~al.}(2009){Asplund}, {Grevesse}, {Sauval}, \&
  {Scott}}]{Asplund2009}
{Asplund}, M., {Grevesse}, N., {Sauval}, A.~J., \& {Scott}, P. 2009, \araa, 47,
  481, \dodoi{10.1146/annurev.astro.46.060407.145222}

\bibitem[{{Astropy Collaboration} {et~al.}(2018){Astropy Collaboration},
  {Price-Whelan}, {Sip{\H{o}}cz}, {G{\"u}nther}, {Lim}, {Crawford}, {Conseil},
  {Shupe}, {Craig}, {Dencheva}, {Ginsburg}, {VanderPlas}, {Bradley},
  {P{\'e}rez-Su{\'a}rez}, {de Val-Borro}, {Aldcroft}, {Cruz}, {Robitaille},
  {Tollerud}, {Ardelean}, {Babej}, {Bach}, {Bachetti}, {Bakanov}, {Bamford},
  {Barentsen}, {Barmby}, {Baumbach}, {Berry}, {Biscani}, {Boquien}, {Bostroem},
  {Bouma}, {Brammer}, {Bray}, {Breytenbach}, {Buddelmeijer}, {Burke},
  {Calderone}, {Cano Rodr{\'\i}guez}, {Cara}, {Cardoso}, {Cheedella}, {Copin},
  {Corrales}, {Crichton}, {D'Avella}, {Deil}, {Depagne}, {Dietrich}, {Donath},
  {Droettboom}, {Earl}, {Erben}, {Fabbro}, {Ferreira}, {Finethy}, {Fox},
  {Garrison}, {Gibbons}, {Goldstein}, {Gommers}, {Greco}, {Greenfield},
  {Groener}, {Grollier}, {Hagen}, {Hirst}, {Homeier}, {Horton}, {Hosseinzadeh},
  {Hu}, {Hunkeler}, {Ivezi{\'c}}, {Jain}, {Jenness}, {Kanarek}, {Kendrew},
  {Kern}, {Kerzendorf}, {Khvalko}, {King}, {Kirkby}, {Kulkarni}, {Kumar},
  {Lee}, {Lenz}, {Littlefair}, {Ma}, {Macleod}, {Mastropietro}, {McCully},
  {Montagnac}, {Morris}, {Mueller}, {Mumford}, {Muna}, {Murphy}, {Nelson},
  {Nguyen}, {Ninan}, {N{\"o}the}, {Ogaz}, {Oh}, {Parejko}, {Parley}, {Pascual},
  {Patil}, {Patil}, {Plunkett}, {Prochaska}, {Rastogi}, {Reddy Janga},
  {Sabater}, {Sakurikar}, {Seifert}, {Sherbert}, {Sherwood-Taylor}, {Shih},
  {Sick}, {Silbiger}, {Singanamalla}, {Singer}, {Sladen}, {Sooley},
  {Sornarajah}, {Streicher}, {Teuben}, {Thomas}, {Tremblay}, {Turner},
  {Terr{\'o}n}, {van Kerkwijk}, {de la Vega}, {Watkins}, {Weaver}, {Whitmore},
  {Woillez}, {Zabalza}, \& {Astropy Contributors}}]{Astropy2018}
{Astropy Collaboration}, {Price-Whelan}, A.~M., {Sip{\H{o}}cz}, B.~M., {et~al.}
  2018, \aj, 156, 123, \dodoi{10.3847/1538-3881/aabc4f}

\bibitem[{Barbary(2105)}]{nestle}
Barbary, K. 2105, Nestle sampling library,
  \url{https://github.com/kbarbary/nestle},  GitHub

\bibitem[{{Barber} {et~al.}(2014){Barber}, {Strange}, {Hill}, {Polyansky},
  {Mellau}, {Yurchenko}, \& {Tennyson}}]{Barber2014}
{Barber}, R.~J., {Strange}, J.~K., {Hill}, C., {et~al.} 2014, \mnras, 437,
  1828, \dodoi{10.1093/mnras/stt2011}

\bibitem[{{Barstow}(2020)}]{Barstow2020a}
{Barstow}, J.~K. 2020, arXiv e-prints, arXiv:2002.02945.
\newblock \doarXiv{2002.02945}

\bibitem[{{Barstow} {et~al.}(2014){Barstow}, {Aigrain}, {Irwin}, {Hackler},
  {Fletcher}, {Lee}, \& {Gibson}}]{Barstow2014}
{Barstow}, J.~K., {Aigrain}, S., {Irwin}, P.~G.~J., {et~al.} 2014, \apj, 786,
  154, \dodoi{10.1088/0004-637X/786/2/154}

\bibitem[{{Barstow} {et~al.}(2016){Barstow}, {Aigrain}, {Irwin}, {Kendrew}, \&
  {Fletcher}}]{Barstow2016b}
{Barstow}, J.~K., {Aigrain}, S., {Irwin}, P.~G.~J., {Kendrew}, S., \&
  {Fletcher}, L.~N. 2016, \mnras, 458, 2657, \dodoi{10.1093/mnras/stw489}

\bibitem[{{Barstow} {et~al.}(2017){Barstow}, {Aigrain}, {Irwin}, \&
  {Sing}}]{Barstow2017}
{Barstow}, J.~K., {Aigrain}, S., {Irwin}, P.~G.~J., \& {Sing}, D.~K. 2017,
  \apj, 834, 50, \dodoi{10.3847/1538-4357/834/1/50}

\bibitem[{{Barstow} \& {Irwin}(2016)}]{Barstow2016}
{Barstow}, J.~K., \& {Irwin}, P.~G.~J. 2016, \mnras, 461, L92,
  \dodoi{10.1093/mnrasl/slw109}

\bibitem[{{Batalha} {et~al.}(2018){Batalha}, {Lewis}, {Line}, {Valenti}, \&
  {Stevenson}}]{Batalha2018}
{Batalha}, N.~E., {Lewis}, N.~K., {Line}, M.~R., {Valenti}, J., \& {Stevenson},
  K. 2018, \apjl, 856, L34, \dodoi{10.3847/2041-8213/aab896}

\bibitem[{{Batalha} {et~al.}(2019){Batalha}, {Lewis}, {Fortney}, {Batalha},
  {Kempton}, {Lewis}, \& {Line}}]{Batalha2019}
{Batalha}, N.~E., {Lewis}, T., {Fortney}, J.~J., {et~al.} 2019, \apjl, 885,
  L25, \dodoi{10.3847/2041-8213/ab4909}

\bibitem[{{Batalha} \& {Line}(2017)}]{Batalha2017a}
{Batalha}, N.~E., \& {Line}, M.~R. 2017, \aj, 153, 151,
  \dodoi{10.3847/1538-3881/aa5faa}

\bibitem[{{Batalha} {et~al.}(2017){Batalha}, {Mandell}, {Pontoppidan},
  {Stevenson}, {Lewis}, {Kalirai}, {Earl}, {Greene}, {Albert}, \&
  {Nielsen}}]{Batalha2017b}
{Batalha}, N.~E., {Mandell}, A., {Pontoppidan}, K., {et~al.} 2017, \pasp, 129,
  064501, \dodoi{10.1088/1538-3873/aa65b0}

\bibitem[{{Beichman} \& {Greene}(2018)}]{Beichman2018}
{Beichman}, C.~A., \& {Greene}, T.~P. 2018, {Observing Exoplanets with the
  James Webb Space Telescope} (Springer International Publishing), 85,
  \dodoi{10.1007/978-3-319-55333-7_85}

\bibitem[{{Benneke} \& {Seager}(2012)}]{Benneke2012}
{Benneke}, B., \& {Seager}, S. 2012, \apj, 753, 100,
  \dodoi{10.1088/0004-637X/753/2/100}

\bibitem[{{Benneke} \& {Seager}(2013)}]{Benneke2013}
---. 2013, \apj, 778, 153, \dodoi{10.1088/0004-637X/778/2/153}

\bibitem[{{Benneke} {et~al.}(2017){Benneke}, {Werner}, {Petigura}, {Knutson},
  {Dressing}, {Crossfield}, {Schlieder}, {Livingston}, {Beichman},
  {Christiansen}, {Krick}, {Gorjian}, {Howard}, {Sinukoff}, {Ciardi}, \&
  {Akeson}}]{Benneke2017}
{Benneke}, B., {Werner}, M., {Petigura}, E., {et~al.} 2017, \apj, 834, 187,
  \dodoi{10.3847/1538-4357/834/2/187}

\bibitem[{{Benneke} {et~al.}(2019{\natexlab{a}}){Benneke}, {Knutson},
  {Lothringer}, {Crossfield}, {Moses}, {Morley}, {Kreidberg}, {Fulton},
  {Dragomir}, {Howard}, {Wong}, {D{\'e}sert}, {McCullough}, {Kempton},
  {Fortney}, {Gilliland }, {Deming}, \& {Kammer}}]{Benneke2019a}
{Benneke}, B., {Knutson}, H.~A., {Lothringer}, J., {et~al.} 2019{\natexlab{a}},
  Nature Astronomy, 3, 813, \dodoi{10.1038/s41550-019-0800-5}

\bibitem[{{Benneke} {et~al.}(2019{\natexlab{b}}){Benneke}, {Wong}, {Piaulet},
  {Knutson}, {Crossfield}, {Lothringer}, {Morley}, {Gao}, {Greene}, {Dressing},
  {Dragomir}, {Howard}, {McCullough}, {Fortney}, \& {Fraine}}]{Benneke2019b}
{Benneke}, B., {Wong}, I., {Piaulet}, C., {et~al.} 2019{\natexlab{b}}, arXiv
  e-prints, arXiv:1909.04642.
\newblock \doarXiv{1909.04642}

\bibitem[{{Blecic}(2016)}]{Blecic2016}
{Blecic}, J. 2016, arXiv e-prints, arXiv:1604.02692.
\newblock \doarXiv{1604.02692}

\bibitem[{{Blecic} {et~al.}(2017){Blecic}, {Dobbs-Dixon}, \&
  {Greene}}]{Blecic2017}
{Blecic}, J., {Dobbs-Dixon}, I., \& {Greene}, T. 2017, \apj, 848, 127,
  \dodoi{10.3847/1538-4357/aa8171}

\bibitem[{{Breiman} {et~al.}(1984){Breiman}, {Friedman}, {Stone}, \&
  {Olshen}}]{Breiman1984}
{Breiman}, L., {Friedman}, J., {Stone}, C.~J., \& {Olshen}, R.~A. 1984,
  Classification and Regression Trees (Taylor \& Francis)

\bibitem[{{Brogi} \& {Line}(2019)}]{Brogi2019}
{Brogi}, M., \& {Line}, M.~R. 2019, \aj, 157, 114,
  \dodoi{10.3847/1538-3881/aaffd3}

\bibitem[{{Bruno} {et~al.}(2020){Bruno}, {Lewis}, {Alam}, {L{\'o}pez-Morales},
  {Barstow}, {Wakeford}, {Sing}, {Henry}, {Ballester}, {Bourrier}, {Buchhave},
  {Cohen}, {Mikal-Evans}, {Garc{\'\i}a Mu{\~n}oz}, {Lavvas}, \&
  {Sanz-Forcada}}]{Bruno2020}
{Bruno}, G., {Lewis}, N.~K., {Alam}, M.~K., {et~al.} 2020, \mnras, 491, 5361,
  \dodoi{10.1093/mnras/stz3194}

\bibitem[{{Buchner} {et~al.}(2014){Buchner}, {Georgakakis}, {Nandra}, {Hsu},
  {Rangel}, {Brightman}, {Merloni}, {Salvato}, {Donley}, \&
  {Kocevski}}]{Buchner2014}
{Buchner}, J., {Georgakakis}, A., {Nandra}, K., {et~al.} 2014, \aap, 564, A125,
  \dodoi{10.1051/0004-6361/201322971}

\bibitem[{{Burningham} {et~al.}(2017){Burningham}, {Marley}, {Line}, {Lupu},
  {Visscher}, {Morley}, {Saumon}, \& {Freedman}}]{Burningham2017}
{Burningham}, B., {Marley}, M.~S., {Line}, M.~R., {et~al.} 2017, \mnras, 470,
  1177, \dodoi{10.1093/mnras/stx1246}

\bibitem[{{Caldas} {et~al.}(2019){Caldas}, {Leconte}, {Selsis}, {Waldmann},
  {Bord{\'e}}, {Rocchetto}, \& {Charnay}}]{Caldas2019}
{Caldas}, A., {Leconte}, J., {Selsis}, F., {et~al.} 2019, \aap, 623, A161,
  \dodoi{10.1051/0004-6361/201834384}

\bibitem[{{Changeat} \& {Al-Refaie}(2020)}]{Changeat2020b}
{Changeat}, Q., \& {Al-Refaie}, A.~F. 2020, arXiv e-prints, arXiv:2006.14237.
\newblock \doarXiv{2006.14237}

\bibitem[{{Changeat} {et~al.}(2019){Changeat}, {Edwards}, {Waldmann}, \&
  {Tinetti}}]{Changeat2019}
{Changeat}, Q., {Edwards}, B., {Waldmann}, I.~P., \& {Tinetti}, G. 2019, \apj,
  886, 39, \dodoi{10.3847/1538-4357/ab4a14}

\bibitem[{{Changeat} {et~al.}(2020){Changeat}, {Keyte}, {Waldmann}, \&
  {Tinetti}}]{Changeat2020a}
{Changeat}, Q., {Keyte}, L., {Waldmann}, I.~P., \& {Tinetti}, G. 2020, \apj,
  896, 107, \dodoi{10.3847/1538-4357/ab8f8b}

\bibitem[{{Charbonneau} {et~al.}(2000){Charbonneau}, {Brown}, {Latham}, \&
  {Mayor}}]{Charbonneau2000}
{Charbonneau}, D., {Brown}, T.~M., {Latham}, D.~W., \& {Mayor}, M. 2000, \apjl,
  529, L45, \dodoi{10.1086/312457}

\bibitem[{{Charbonneau} {et~al.}(2002){Charbonneau}, {Brown}, {Noyes}, \&
  {Gilliland}}]{Charbonneau2002}
{Charbonneau}, D., {Brown}, T.~M., {Noyes}, R.~W., \& {Gilliland}, R.~L. 2002,
  \apj, 568, 377, \dodoi{10.1086/338770}

\bibitem[{{Charbonneau} \& {Deming}(2007)}]{Charbonneau2007}
{Charbonneau}, D., \& {Deming}, D. 2007, arXiv e-prints, arXiv:0706.1047.
\newblock \doarXiv{0706.1047}

\bibitem[{Chayes(1960)}]{Chayes1960}
Chayes, F. 1960, Journal of Geophysical Research (1896-1977), 65, 4185,
  \dodoi{10.1029/JZ065i012p04185}

\bibitem[{{Chen} {et~al.}(2018){Chen}, {Pall{\'e}}, {Welbanks},
  {Prieto-Arranz}, {Madhusudhan}, {Gandhi}, {Casasayas-Barris}, {Murgas},
  {Nortmann}, {Crouzet}, {Parviainen}, \& {Gandolfi}}]{Chen2018}
{Chen}, G., {Pall{\'e}}, E., {Welbanks}, L., {et~al.} 2018, \aap, 616, A145,
  \dodoi{10.1051/0004-6361/201833033}

\bibitem[{{Cobb} {et~al.}(2019){Cobb}, {Himes}, {Soboczenski}, {Zorzan},
  {O'Beirne}, {G{\"u}ne{\textcommabelow s} Baydin}, {Gal}, {Domagal-Goldman},
  {Arney}, {Angerhausen}, \& {2018 NASA FDL Astrobiology Team}}]{Cobb2019}
{Cobb}, A.~D., {Himes}, M.~D., {Soboczenski}, F., {et~al.} 2019, \aj, 158, 33,
  \dodoi{10.3847/1538-3881/ab2390}

\bibitem[{{Col{\'o}n} {et~al.}(2020){Col{\'o}n}, {Kreidberg}, {Welbanks},
  {Line}, {Madhusudhan}, {Beatty}, {Tamburo}, {Stevenson}, {Mandell},
  {Rodriguez}, {Barclay}, {Lopez}, {Stassun}, {Angerhausen}, {Fortney},
  {James}, {Pepper}, {Ahlers}, {Plavchan}, {Awiphan}, {Kotnik}, {McLeod},
  {Murawski}, {Chotani}, {LeBrun}, {Matzko}, {Rea}, {Vidaurri}, {Webster},
  {Williams}, {Cox}, {Tan}, \& {Gilbert}}]{Colon2020}
{Col{\'o}n}, K.~D., {Kreidberg}, L., {Welbanks}, L., {et~al.} 2020, \aj, 160,
  280, \dodoi{10.3847/1538-3881/abc1e9}

\bibitem[{{Crossfield} {et~al.}(2016){Crossfield}, {Ciardi}, {Petigura},
  {Sinukoff}, {Schlieder}, {Howard}, {Beichman}, {Isaacson}, {Dressing},
  {Christiansen}, {Fulton}, {L{\'e}pine}, {Weiss}, {Hirsch}, {Livingston},
  {Baranec}, {Law}, {Riddle}, {Ziegler}, {Howell}, {Horch}, {Everett}, {Teske},
  {Martinez}, {Obermeier}, {Benneke}, {Scott}, {Deacon}, {Aller}, {Hansen},
  {Mancini}, {Ciceri}, {Brahm}, {Jord{\'a}n}, {Knutson}, {Henning}, {Bonnefoy},
  {Liu}, {Crepp}, {Lothringer}, {Hinz}, {Bailey}, {Skemer}, \&
  {Defrere}}]{Crossfield2016}
{Crossfield}, I. J.~M., {Ciardi}, D.~R., {Petigura}, E.~A., {et~al.} 2016,
  \apjs, 226, 7, \dodoi{10.3847/0067-0049/226/1/7}

\bibitem[{{Cubillos}(2016)}]{Cubillos2016}
{Cubillos}, P.~E. 2016, arXiv e-prints, arXiv:1604.01320.
\newblock \doarXiv{1604.01320}

\bibitem[{{Dalgarno} \& {Williams}(1962)}]{Dalgarno1962}
{Dalgarno}, A., \& {Williams}, D.~A. 1962, \apj, 136, 690,
  \dodoi{10.1086/147428}

\bibitem[{{Damiano} \& {Hu}(2020)}]{Damiano2020}
{Damiano}, M., \& {Hu}, R. 2020, \aj, 159, 175,
  \dodoi{10.3847/1538-3881/ab79a5}

\bibitem[{{de Wit} \& {Seager}(2013)}]{DeWit2013}
{de Wit}, J., \& {Seager}, S. 2013, Science, 342, 1473,
  \dodoi{10.1126/science.1245450}

\bibitem[{{Deming} {et~al.}(2013){Deming}, {Wilkins}, {McCullough}, {Burrows},
  {Fortney}, {Agol}, {Dobbs-Dixon}, {Madhusudhan}, {Crouzet}, {Desert},
  {Gilliland}, {Haynes}, {Knutson}, {Line}, {Magic}, {Mand ell}, {Ranjan},
  {Charbonneau}, {Clampin}, {Seager}, \& {Showman}}]{Deming2013}
{Deming}, D., {Wilkins}, A., {McCullough}, P., {et~al.} 2013, \apj, 774, 95,
  \dodoi{10.1088/0004-637X/774/2/95}

\bibitem[{{Evans} {et~al.}(2017){Evans}, {Sing}, {Kataria}, {Goyal}, {Nikolov},
  {Wakeford}, {Deming}, {Marley}, {Amundsen}, {Ballester}, {Barstow},
  {Ben-Jaffel}, {Bourrier}, {Buchhave}, {Cohen}, {Ehrenreich}, {Garc{\'\i}a
  Mu{\~n}oz}, {Henry}, {Knutson}, {Lavvas}, {Lecavelier Des Etangs}, {Lewis},
  {L{\'o}pez-Morales}, {Mandell}, {Sanz-Forcada}, {Tremblin}, \&
  {Lupu}}]{Evans2017}
{Evans}, T.~M., {Sing}, D.~K., {Kataria}, T., {et~al.} 2017, \nat, 548, 58,
  \dodoi{10.1038/nature23266}

\bibitem[{{Feng} {et~al.}(2020){Feng}, {Line}, \& {Fortney}}]{Feng2020}
{Feng}, Y.~K., {Line}, M.~R., \& {Fortney}, J.~J. 2020, arXiv e-prints,
  arXiv:2006.11442.
\newblock \doarXiv{2006.11442}

\bibitem[{{Feroz} {et~al.}(2009){Feroz}, {Hobson}, \& {Bridges}}]{Feroz2009}
{Feroz}, F., {Hobson}, M.~P., \& {Bridges}, M. 2009, \mnras, 398, 1601,
  \dodoi{10.1111/j.1365-2966.2009.14548.x}

\bibitem[{{Feroz} {et~al.}(2013){Feroz}, {Hobson}, {Cameron}, \&
  {Pettitt}}]{Feroz2013}
{Feroz}, F., {Hobson}, M.~P., {Cameron}, E., \& {Pettitt}, A.~N. 2013, arXiv
  e-prints, arXiv:1306.2144.
\newblock \doarXiv{1306.2144}

\bibitem[{{Feroz} {et~al.}(2019){Feroz}, {Hobson}, {Cameron}, \&
  {Pettitt}}]{Feroz2019}
---. 2019, The Open Journal of Astrophysics, 2, 10,
  \dodoi{10.21105/astro.1306.2144}

\bibitem[{{Fisher} \& {Heng}(2018)}]{Fisher2018}
{Fisher}, C., \& {Heng}, K. 2018, \mnras, 481, 4698,
  \dodoi{10.1093/mnras/sty2550}

\bibitem[{{Fisher} \& {Heng}(2019)}]{Fisher2019}
---. 2019, \apj, 881, 25, \dodoi{10.3847/1538-4357/ab29e8}

\bibitem[{{Fisher} {et~al.}(2020){Fisher}, {Hoeijmakers}, {Kitzmann},
  {M{\'a}rquez-Neila}, {Grimm}, {Sznitman}, \& {Heng}}]{Fisher2020}
{Fisher}, C., {Hoeijmakers}, H.~J., {Kitzmann}, D., {et~al.} 2020, \aj, 159,
  192, \dodoi{10.3847/1538-3881/ab7a92}

\bibitem[{{Foreman-Mackey} {et~al.}(2017){Foreman-Mackey}, {Agol}, {Angus}, \&
  {Ambikasaran}}]{celerite}
{Foreman-Mackey}, D., {Agol}, E., {Angus}, R., \& {Ambikasaran}, S. 2017, AJ,
  154, 220, \dodoi{10.3847/1538-3881/aa9332}

\bibitem[{{Foreman-Mackey} {et~al.}(2013){Foreman-Mackey}, {Hogg}, {Lang}, \&
  {Goodman}}]{Foremanmackey2013}
{Foreman-Mackey}, D., {Hogg}, D.~W., {Lang}, D., \& {Goodman}, J. 2013, \pasp,
  125, 306, \dodoi{10.1086/670067}

\bibitem[{{Foreman-Mackey} {et~al.}(2015){Foreman-Mackey}, {Montet}, {Hogg},
  {Morton}, {Wang}, \& {Sch{\"o}lkopf}}]{Foremanmackey2015}
{Foreman-Mackey}, D., {Montet}, B.~T., {Hogg}, D.~W., {et~al.} 2015, \apj, 806,
  215, \dodoi{10.1088/0004-637X/806/2/215}

\bibitem[{{Gandhi} \& {Madhusudhan}(2017)}]{Gandhi2017}
{Gandhi}, S., \& {Madhusudhan}, N. 2017, \mnras, 472, 2334,
  \dodoi{10.1093/mnras/stx1601}

\bibitem[{{Gandhi} \& {Madhusudhan}(2018)}]{Gandhi2018}
---. 2018, \mnras, 474, 271, \dodoi{10.1093/mnras/stx2748}

\bibitem[{{Gandhi} {et~al.}(2019){Gandhi}, {Madhusudhan}, {Hawker}, \&
  {Piette}}]{Gandhi2019}
{Gandhi}, S., {Madhusudhan}, N., {Hawker}, G., \& {Piette}, A. 2019, arXiv
  e-prints, arXiv:1910.14042.
\newblock \doarXiv{1910.14042}

\bibitem[{{Gandhi} {et~al.}(2020){Gandhi}, {Brogi}, {Yurchenko}, {Tennyson},
  {Coles}, {Webb}, {Birkby}, {Guilluy}, {Hawker}, {Madhusudhan}, {Bonomo}, \&
  {Sozzetti}}]{Gandhi2020}
{Gandhi}, S., {Brogi}, M., {Yurchenko}, S.~N., {et~al.} 2020, \mnras, 495, 224,
  \dodoi{10.1093/mnras/staa981}

\bibitem[{{Gillon} {et~al.}(2011){Gillon}, {Jehin}, {Magain}, {Chantry},
  {Hutsem{\'e}kers}, {Manfroid}, {Queloz}, \& {Udry}}]{Gillon2011}
{Gillon}, M., {Jehin}, E., {Magain}, P., {et~al.} 2011, in European Physical
  Journal Web of Conferences, Vol.~11, European Physical Journal Web of
  Conferences, 06002, \dodoi{10.1051/epjconf/20101106002}

\bibitem[{{Gillon} {et~al.}(2017){Gillon}, {Triaud}, {Demory}, {Jehin}, {Agol},
  {Deck}, {Lederer}, {de Wit}, {Burdanov}, {Ingalls}, {Bolmont}, {Leconte},
  {Raymond}, {Selsis}, {Turbet}, {Barkaoui}, {Burgasser}, {Burleigh}, {Carey},
  {Chaushev}, {Copperwheat}, {Delrez}, {Fernand es}, {Holdsworth}, {Kotze},
  {Van Grootel}, {Almleaky}, {Benkhaldoun}, {Magain}, \& {Queloz}}]{Gillon2017}
{Gillon}, M., {Triaud}, A. H.~M.~J., {Demory}, B.-O., {et~al.} 2017, \nat, 542,
  456, \dodoi{10.1038/nature21360}

\bibitem[{{Greene} {et~al.}(2016){Greene}, {Line}, {Montero}, {Fortney},
  {Lustig-Yaeger}, \& {Luther}}]{Greene2016}
{Greene}, T.~P., {Line}, M.~R., {Montero}, C., {et~al.} 2016, \apj, 817, 17,
  \dodoi{10.3847/0004-637X/817/1/17}

\bibitem[{{Guzm{\'a}n-Mesa} {et~al.}(2020){Guzm{\'a}n-Mesa}, {Kitzmann},
  {Fisher}, {Burgasser}, {Hoeijmakers}, {M{\'a}rquez-Neila}, {Grimm},
  {Mandell}, {Sznitman}, \& {Heng}}]{Guzman-Mesa2020}
{Guzm{\'a}n-Mesa}, A., {Kitzmann}, D., {Fisher}, C., {et~al.} 2020, \aj, 160,
  15, \dodoi{10.3847/1538-3881/ab9176}

\bibitem[{{Handley} {et~al.}(2015{\natexlab{a}}){Handley}, {Hobson}, \&
  {Lasenby}}]{Handley2015a}
{Handley}, W.~J., {Hobson}, M.~P., \& {Lasenby}, A.~N. 2015{\natexlab{a}},
  \mnras, 450, L61, \dodoi{10.1093/mnrasl/slv047}

\bibitem[{{Handley} {et~al.}(2015{\natexlab{b}}){Handley}, {Hobson}, \&
  {Lasenby}}]{Handley2015b}
---. 2015{\natexlab{b}}, \mnras, 453, 4384, \dodoi{10.1093/mnras/stv1911}

\bibitem[{{Harris} {et~al.}(2020){Harris}, {Jarrod Millman}, {van der Walt},
  {Gommers}, {Virtanen}, {Cournapeau}, {Wieser}, {Taylor}, {Berg}, {Smith},
  {Kern}, {Picus}, {Hoyer}, {van Kerkwijk}, {Brett}, {Haldane}, {Fern{\'a}ndez
  del R{\'\i}o}, {Wiebe}, {Peterson}, {G{\'e}rard-Marchant}, {Sheppard},
  {Reddy}, {Weckesser}, {Abbasi}, {Gohlke}, \& {Oliphant}}]{numpy}
{Harris}, C.~R., {Jarrod Millman}, K., {van der Walt}, S.~J., {et~al.} 2020,
  arXiv e-prints, arXiv:2006.10256.
\newblock \doarXiv{2006.10256}

\bibitem[{{Hayes} {et~al.}(2020){Hayes}, {Kerins}, {Awiphan}, {McDonald},
  {Morgan}, {Chuanraksasat}, {Komonjinda}, {Sanguansak}, {Kittara}, \&
  {SPEARNet Collaboration}}]{Hayes2020}
{Hayes}, J.~J.~C., {Kerins}, E., {Awiphan}, S., {et~al.} 2020, \mnras, 494,
  4492, \dodoi{10.1093/mnras/staa978}

\bibitem[{{Henry} {et~al.}(2000){Henry}, {Marcy}, {Butler}, \&
  {Vogt}}]{Henry2000}
{Henry}, G.~W., {Marcy}, G.~W., {Butler}, R.~P., \& {Vogt}, S.~S. 2000, \apjl,
  529, L41, \dodoi{10.1086/312458}

\bibitem[{{Henyey} \& {Greenstein}(1941)}]{Henyey1941}
{Henyey}, L.~G., \& {Greenstein}, J.~L. 1941, \apj, 93, 70,
  \dodoi{10.1086/144246}

\bibitem[{{Higson} {et~al.}(2019){Higson}, {Handley}, {Hobson}, \&
  {Lasenby}}]{Higson2019}
{Higson}, E., {Handley}, W., {Hobson}, M., \& {Lasenby}, A. 2019, Statistics
  and Computing, 29, 891, \dodoi{10.1007/s11222-018-9844-0}

\bibitem[{Hunter(2007)}]{Matplotlib}
Hunter, J.~D. 2007, Computing in Science \& Engineering, 9, 90,
  \dodoi{10.1109/MCSE.2007.55}

\bibitem[{{Husser} {et~al.}(2013){Husser}, {Wende-von Berg}, {Dreizler},
  {Homeier}, {Reiners}, {Barman}, \& {Hauschildt}}]{Husser2013}
{Husser}, T.~O., {Wende-von Berg}, S., {Dreizler}, S., {et~al.} 2013, \aap,
  553, A6, \dodoi{10.1051/0004-6361/201219058}

\bibitem[{{Irwin} {et~al.}(2014){Irwin}, {Barstow}, {Bowles}, {Fletcher},
  {Aigrain}, \& {Lee}}]{Irwin2014}
{Irwin}, P.~G.~J., {Barstow}, J.~K., {Bowles}, N.~E., {et~al.} 2014, \icarus,
  242, 172, \dodoi{10.1016/j.icarus.2014.08.005}

\bibitem[{{Irwin} \& {Dyudina}(2002)}]{Irwin2002}
{Irwin}, P.~G.~J., \& {Dyudina}, U. 2002, \icarus, 156, 52,
  \dodoi{10.1006/icar.2001.6773}

\bibitem[{{Irwin} {et~al.}(2020){Irwin}, {Parmentier}, {Taylor}, {Barstow},
  {Aigrain}, {Lee}, \& {Garland }}]{Irwin2020}
{Irwin}, P. G.~J., {Parmentier}, V., {Taylor}, J., {et~al.} 2020, \mnras, 493,
  106, \dodoi{10.1093/mnras/staa238}

\bibitem[{{Irwin} {et~al.}(2001){Irwin}, {Weir}, {Taylor}, {Calcutt}, \&
  {Carlson}}]{Irwin2001}
{Irwin}, P.~G.~J., {Weir}, A.~L., {Taylor}, F.~W., {Calcutt}, S.~B., \&
  {Carlson}, R.~W. 2001, \icarus, 149, 397, \dodoi{10.1006/icar.2000.6542}

\bibitem[{{Irwin} {et~al.}(2008){Irwin}, {Teanby}, {de Kok}, {Fletcher},
  {Howett}, {Tsang}, {Wilson}, {Calcutt}, {Nixon}, \& {Parrish}}]{Irwin2008}
{Irwin}, P.~G.~J., {Teanby}, N.~A., {de Kok}, R., {et~al.} 2008, \jqsrt, 109,
  1136, \dodoi{10.1016/j.jqsrt.2007.11.006}

\bibitem[{{Iyer} \& {Line}(2020)}]{Iyer2020}
{Iyer}, A.~R., \& {Line}, M.~R. 2020, \apj, 889, 78,
  \dodoi{10.3847/1538-4357/ab612e}

\bibitem[{{Karman} {et~al.}(2019){Karman}, {Gordon}, {van der Avoird},
  {Baranov}, {Boulet}, {Drouin}, {Groenenboom}, {Gustafsson}, {Hartmann},
  {Kurucz}, {Rothman}, {Sun}, {Sung}, {Thalman}, {Tran}, {Wishnow},
  {Wordsworth}, {Vigasin}, {Volkamer}, \& {van der Zande}}]{Karman2019}
{Karman}, T., {Gordon}, I.~E., {van der Avoird}, A., {et~al.} 2019, \icarus,
  328, 160, \dodoi{10.1016/j.icarus.2019.02.034}

\bibitem[{Kass \& Raftery(1995)}]{Kass1995}
Kass, R.~E., \& Raftery, A.~E. 1995, Journal of the American Statistical
  Association, 90, 773

\bibitem[{{Kitzmann} {et~al.}(2020){Kitzmann}, {Heng}, {Oreshenko}, {Grimm},
  {Apai}, {Bowler}, {Burgasser}, \& {Marley}}]{Kitzmann2020}
{Kitzmann}, D., {Heng}, K., {Oreshenko}, M., {et~al.} 2020, \apj, 890, 174,
  \dodoi{10.3847/1538-4357/ab6d71}

\bibitem[{{Komacek} {et~al.}(2020){Komacek}, {Fauchez}, {Wolf}, \&
  {Abbot}}]{Komacek2020}
{Komacek}, T.~D., {Fauchez}, T.~J., {Wolf}, E.~T., \& {Abbot}, D.~S. 2020,
  \apjl, 888, L20, \dodoi{10.3847/2041-8213/ab6200}

\bibitem[{{Kopparapu}(2013)}]{Kopparapu2013}
{Kopparapu}, R.~K. 2013, \apjl, 767, L8, \dodoi{10.1088/2041-8205/767/1/L8}

\bibitem[{{Kreidberg} {et~al.}(2014{\natexlab{a}}){Kreidberg}, {Bean},
  {D{\'e}sert}, {Line}, {Fortney}, {Madhusudhan}, {Stevenson}, {Showman},
  {Charbonneau}, {McCullough}, {Seager}, {Burrows}, {Henry}, {Williamson},
  {Kataria}, \& {Homeier}}]{Kreidberg2014b}
{Kreidberg}, L., {Bean}, J.~L., {D{\'e}sert}, J.-M., {et~al.}
  2014{\natexlab{a}}, \apjl, 793, L27, \dodoi{10.1088/2041-8205/793/2/L27}

\bibitem[{{Kreidberg} {et~al.}(2014{\natexlab{b}}){Kreidberg}, {Bean},
  {D{\'e}sert}, {Benneke}, {Deming}, {Stevenson}, {Seager}, {Berta-Thompson},
  {Seifahrt}, \& {Homeier}}]{Kreidberg2014a}
---. 2014{\natexlab{b}}, \nat, 505, 69, \dodoi{10.1038/nature12888}

\bibitem[{{Krissansen-Totton} {et~al.}(2018){Krissansen-Totton}, {Garland},
  {Irwin}, \& {Catling}}]{Krissansen-Totton2018}
{Krissansen-Totton}, J., {Garland}, R., {Irwin}, P., \& {Catling}, D.~C. 2018,
  \aj, 156, 114, \dodoi{10.3847/1538-3881/aad564}

\bibitem[{{Lacy} \& {Burrows}(2020)}]{Lacy2020}
{Lacy}, B.~I., \& {Burrows}, A.~S. 2020, arXiv e-prints, arXiv:2006.06899.
\newblock \doarXiv{2006.06899}

\bibitem[{{Lavie} {et~al.}(2017){Lavie}, {Mendon{\c{c}}a}, {Mordasini},
  {Malik}, {Bonnefoy}, {Demory}, {Oreshenko}, {Grimm}, {Ehrenreich}, \&
  {Heng}}]{Lavie2017}
{Lavie}, B., {Mendon{\c{c}}a}, J.~M., {Mordasini}, C., {et~al.} 2017, \aj, 154,
  91, \dodoi{10.3847/1538-3881/aa7ed8}

\bibitem[{{Lavvas} {et~al.}(2019){Lavvas}, {Koskinen}, {Steinrueck},
  {Garc{\'\i}a Mu{\~n}oz}, \& {Showman}}]{Lavvas2019}
{Lavvas}, P., {Koskinen}, T., {Steinrueck}, M.~E., {Garc{\'\i}a Mu{\~n}oz}, A.,
  \& {Showman}, A.~P. 2019, \apj, 878, 118, \dodoi{10.3847/1538-4357/ab204e}

\bibitem[{{Lecavelier Des Etangs} {et~al.}(2008){Lecavelier Des Etangs},
  {Pont}, {Vidal-Madjar}, \& {Sing}}]{Lecavelier2008a}
{Lecavelier Des Etangs}, A., {Pont}, F., {Vidal-Madjar}, A., \& {Sing}, D.
  2008, \aap, 481, L83, \dodoi{10.1051/0004-6361:200809388}

\bibitem[{{Lee} {et~al.}(2012){Lee}, {Fletcher}, \& {Irwin}}]{Lee2012}
{Lee}, J.~M., {Fletcher}, L.~N., \& {Irwin}, P.~G.~J. 2012, \mnras, 420, 170,
  \dodoi{10.1111/j.1365-2966.2011.20013.x}

\bibitem[{{Lee} {et~al.}(2013){Lee}, {Heng}, \& {Irwin}}]{Lee2013}
{Lee}, J.-M., {Heng}, K., \& {Irwin}, P. G.~J. 2013, \apj, 778, 97,
  \dodoi{10.1088/0004-637X/778/2/97}

\bibitem[{{Lincowski} {et~al.}(2018){Lincowski}, {Meadows}, {Crisp},
  {Robinson}, {Luger}, {Lustig-Yaeger}, \& {Arney}}]{Lincowski2018}
{Lincowski}, A.~P., {Meadows}, V.~S., {Crisp}, D., {et~al.} 2018, \apj, 867,
  76, \dodoi{10.3847/1538-4357/aae36a}

\bibitem[{{Line} {et~al.}(2014{\natexlab{a}}){Line}, {Fortney}, {Marley}, \&
  {Sorahana}}]{Line2014b}
{Line}, M.~R., {Fortney}, J.~J., {Marley}, M.~S., \& {Sorahana}, S.
  2014{\natexlab{a}}, \apj, 793, 33, \dodoi{10.1088/0004-637X/793/1/33}

\bibitem[{{Line} {et~al.}(2014{\natexlab{b}}){Line}, {Knutson}, {Wolf}, \&
  {Yung}}]{Line2014a}
{Line}, M.~R., {Knutson}, H., {Wolf}, A.~S., \& {Yung}, Y.~L.
  2014{\natexlab{b}}, \apj, 783, 70, \dodoi{10.1088/0004-637X/783/2/70}

\bibitem[{{Line} \& {Parmentier}(2016)}]{Line2016}
{Line}, M.~R., \& {Parmentier}, V. 2016, \apj, 820, 78,
  \dodoi{10.3847/0004-637X/820/1/78}

\bibitem[{{Line} {et~al.}(2015){Line}, {Teske}, {Burningham}, {Fortney}, \&
  {Marley}}]{Line2015}
{Line}, M.~R., {Teske}, J., {Burningham}, B., {Fortney}, J.~J., \& {Marley},
  M.~S. 2015, \apj, 807, 183, \dodoi{10.1088/0004-637X/807/2/183}

\bibitem[{{Line} {et~al.}(2013){Line}, {Wolf}, {Zhang}, {Knutson}, {Kammer},
  {Ellison}, {Deroo}, {Crisp}, \& {Yung}}]{Line2013a}
{Line}, M.~R., {Wolf}, A.~S., {Zhang}, X., {et~al.} 2013, \apj, 775, 137,
  \dodoi{10.1088/0004-637X/775/2/137}

\bibitem[{{Lupu} {et~al.}(2016){Lupu}, {Marley}, {Lewis}, {Line}, {Traub}, \&
  {Zahnle}}]{Lupu2016}
{Lupu}, R.~E., {Marley}, M.~S., {Lewis}, N., {et~al.} 2016, \aj, 152, 217,
  \dodoi{10.3847/0004-6256/152/6/217}

\bibitem[{{Lustig-Yaeger} {et~al.}(2019){Lustig-Yaeger}, {Meadows}, \&
  {Lincowski}}]{Lustig-Yaeger2019}
{Lustig-Yaeger}, J., {Meadows}, V.~S., \& {Lincowski}, A.~P. 2019, \aj, 158,
  27, \dodoi{10.3847/1538-3881/ab21e0}

\bibitem[{{MacDonald} {et~al.}(2020){MacDonald}, {Goyal}, \&
  {Lewis}}]{MacDonald2020}
{MacDonald}, R.~J., {Goyal}, J.~M., \& {Lewis}, N.~K. 2020, \apjl, 893, L43,
  \dodoi{10.3847/2041-8213/ab8238}

\bibitem[{{MacDonald} \& {Madhusudhan}(2017)}]{MacDonald2017}
{MacDonald}, R.~J., \& {Madhusudhan}, N. 2017, \mnras, 469, 1979,
  \dodoi{10.1093/mnras/stx804}

\bibitem[{{Madhusudhan}(2012)}]{Madhusudhan2012}
{Madhusudhan}, N. 2012, \apj, 758, 36, \dodoi{10.1088/0004-637X/758/1/36}

\bibitem[{{Madhusudhan}(2018)}]{Madhusudhan2018}
---. 2018, {Atmospheric Retrieval of Exoplanets} (Springer), 104,
  \dodoi{10.1007/978-3-319-55333-7_104}

\bibitem[{{Madhusudhan}(2019)}]{Madhusudhan2019}
---. 2019, arXiv e-prints, arXiv:1904.03190.
\newblock \doarXiv{1904.03190}

\bibitem[{{Madhusudhan} {et~al.}(2014){Madhusudhan}, {Crouzet}, {McCullough},
  {Deming}, \& {Hedges}}]{Madhusudhan2014a}
{Madhusudhan}, N., {Crouzet}, N., {McCullough}, P.~R., {Deming}, D., \&
  {Hedges}, C. 2014, \apjl, 791, L9, \dodoi{10.1088/2041-8205/791/1/L9}

\bibitem[{{Madhusudhan} {et~al.}(2020){Madhusudhan}, {Nixon}, {Welbanks},
  {Piette}, \& {Booth}}]{Madhusudhan2020}
{Madhusudhan}, N., {Nixon}, M.~C., {Welbanks}, L., {Piette}, A. A.~A., \&
  {Booth}, R.~A. 2020, \apjl, 891, L7, \dodoi{10.3847/2041-8213/ab7229}

\bibitem[{{Madhusudhan} \& {Seager}(2009)}]{Madhusudhan2009}
{Madhusudhan}, N., \& {Seager}, S. 2009, \apj, 707, 24,
  \dodoi{10.1088/0004-637X/707/1/24}

\bibitem[{{Madhusudhan} \& {Seager}(2011)}]{Madhusudhan2011}
---. 2011, \apj, 729, 41, \dodoi{10.1088/0004-637X/729/1/41}

\bibitem[{{Mai} \& {Line}(2019)}]{Mai2019}
{Mai}, C., \& {Line}, M.~R. 2019, \apj, 883, 144,
  \dodoi{10.3847/1538-4357/ab3e6d}

\bibitem[{{M{\'a}rquez-Neila} {et~al.}(2018){M{\'a}rquez-Neila}, {Fisher},
  {Sznitman}, \& {Heng}}]{Marquez-Neila2018}
{M{\'a}rquez-Neila}, P., {Fisher}, C., {Sznitman}, R., \& {Heng}, K. 2018,
  Nature Astronomy, 2, 719, \dodoi{10.1038/s41550-018-0504-2}

\bibitem[{{Min} {et~al.}(2020){Min}, {Ormel}, {Chubb}, {Helling}, \&
  {Kawashima}}]{Min2020}
{Min}, M., {Ormel}, C.~W., {Chubb}, K., {Helling}, C., \& {Kawashima}, Y. 2020,
  arXiv e-prints, arXiv:2006.12821.
\newblock \doarXiv{2006.12821}

\bibitem[{{Molli{\`e}re} {et~al.}(2019){Molli{\`e}re}, {Wardenier}, {van
  Boekel}, {Henning}, {Molaverdikhani}, \& {Snellen}}]{Molliere2019}
{Molli{\`e}re}, P., {Wardenier}, J.~P., {van Boekel}, R., {et~al.} 2019, \aap,
  627, A67, \dodoi{10.1051/0004-6361/201935470}

\bibitem[{{Morley} {et~al.}(2017){Morley}, {Kreidberg}, {Rustamkulov},
  {Robinson}, \& {Fortney}}]{Morley2017}
{Morley}, C.~V., {Kreidberg}, L., {Rustamkulov}, Z., {Robinson}, T., \&
  {Fortney}, J.~J. 2017, \apj, 850, 121, \dodoi{10.3847/1538-4357/aa927b}

\bibitem[{{Moses} {et~al.}(2013){Moses}, {Line}, {Visscher}, {Richardson},
  {Nettelmann}, {Fortney}, {Barman}, {Stevenson}, \& {Madhusudhan}}]{Moses2013}
{Moses}, J.~I., {Line}, M.~R., {Visscher}, C., {et~al.} 2013, \apj, 777, 34,
  \dodoi{10.1088/0004-637X/777/1/34}

\bibitem[{{Nayak} {et~al.}(2017){Nayak}, {Lupu}, {Marley}, {Fortney},
  {Robinson}, \& {Lewis}}]{Nayak2017}
{Nayak}, M., {Lupu}, R., {Marley}, M.~S., {et~al.} 2017, \pasp, 129, 034401,
  \dodoi{10.1088/1538-3873/129/973/034401}

\bibitem[{{Nikolov} {et~al.}(2018){Nikolov}, {Sing}, {Fortney}, {Goyal},
  {Drummond}, {Evans}, {Gibson}, {De Mooij}, {Rustamkulov}, {Wakeford},
  {Smalley}, {Burgasser}, {Hellier}, {Helling}, {Mayne}, {Madhusudhan},
  {Kataria}, {Baines}, {Carter}, {Ballester}, {Barstow}, {McCleery}, \&
  {Spake}}]{Nikolov2018}
{Nikolov}, N., {Sing}, D.~K., {Fortney}, J.~J., {et~al.} 2018, \nat, 557, 526,
  \dodoi{10.1038/s41586-018-0101-7}

\bibitem[{{Nixon} \& {Madhusudhan}(2020)}]{Nixon2020}
{Nixon}, M.~C., \& {Madhusudhan}, N. 2020, \mnras, 496, 269,
  \dodoi{10.1093/mnras/staa1150}

\bibitem[{{Ohno} \& {Kawashima}(2020)}]{Ohno2020}
{Ohno}, K., \& {Kawashima}, Y. 2020, arXiv e-prints, arXiv:2005.08880.
\newblock \doarXiv{2005.08880}

\bibitem[{{Parmentier} {et~al.}(2018){Parmentier}, {Line}, {Bean}, {Mansfield},
  {Kreidberg}, {Lupu}, {Visscher}, {D{\'e}sert}, {Fortney}, {Deleuil},
  {Arcangeli}, {Showman}, \& {Marley}}]{Parmentier2018}
{Parmentier}, V., {Line}, M.~R., {Bean}, J.~L., {et~al.} 2018, \aap, 617, A110,
  \dodoi{10.1051/0004-6361/201833059}

\bibitem[{Pawlowsky-Glahn \& Buccianti(2011)}]{Pawlowsky-Glahn2011}
Pawlowsky-Glahn, V., \& Buccianti, A. 2011, Compositional data analysis (Wiley
  Online Library)

\bibitem[{Pearson(1897)}]{Pearson1897}
Pearson, K. 1897, Proceedings of the Royal Society of London, 60, 489,
  \dodoi{10.1098/rspl.1896.0076}

\bibitem[{{Piette} \& {Madhusudhan}(2020)}]{Piette2020}
{Piette}, A. A.~A., \& {Madhusudhan}, N. 2020, arXiv e-prints,
  arXiv:2007.15004.
\newblock \doarXiv{2007.15004}

\bibitem[{{Pinhas} \& {Madhusudhan}(2017)}]{Pinhas2017}
{Pinhas}, A., \& {Madhusudhan}, N. 2017, \mnras, 471, 4355,
  \dodoi{10.1093/mnras/stx1849}

\bibitem[{{Pinhas} {et~al.}(2019){Pinhas}, {Madhusudhan}, {Gandhi}, \&
  {MacDonald}}]{Pinhas2019}
{Pinhas}, A., {Madhusudhan}, N., {Gandhi}, S., \& {MacDonald}, R. 2019, \mnras,
  482, 1485, \dodoi{10.1093/mnras/sty2544}

\bibitem[{{Pinhas} {et~al.}(2018){Pinhas}, {Rackham}, {Madhusudhan}, \&
  {Apai}}]{Pinhas2018}
{Pinhas}, A., {Rackham}, B.~V., {Madhusudhan}, N., \& {Apai}, D. 2018, \mnras,
  480, 5314, \dodoi{10.1093/mnras/sty2209}

\bibitem[{{Pluriel} {et~al.}(2020){Pluriel}, {Zingales}, {Leconte}, \&
  {Parmentier}}]{Pluriel2020}
{Pluriel}, W., {Zingales}, T., {Leconte}, J., \& {Parmentier}, V. 2020, \aap,
  636, A66, \dodoi{10.1051/0004-6361/202037678}

\bibitem[{{Pont} {et~al.}(2008){Pont}, {Knutson}, {Gilliland}, {Moutou}, \&
  {Charbonneau}}]{Pont2008}
{Pont}, F., {Knutson}, H., {Gilliland}, R.~L., {Moutou}, C., \& {Charbonneau},
  D. 2008, \mnras, 385, 109, \dodoi{10.1111/j.1365-2966.2008.12852.x}

\bibitem[{{Rackham} {et~al.}(2017){Rackham}, {Espinoza}, {Apai},
  {L{\'o}pez-Morales}, {Jord{\'a}n}, {Osip}, {Lewis}, {Rodler}, {Fraine},
  {Morley}, \& {Fortney}}]{Rackham2017}
{Rackham}, B., {Espinoza}, N., {Apai}, D., {et~al.} 2017, \apj, 834, 151,
  \dodoi{10.3847/1538-4357/aa4f6c}

\bibitem[{{Rackham} {et~al.}(2018){Rackham}, {Apai}, \&
  {Giampapa}}]{Rackham2018}
{Rackham}, B.~V., {Apai}, D., \& {Giampapa}, M.~S. 2018, \apj, 853, 122,
  \dodoi{10.3847/1538-4357/aaa08c}

\bibitem[{{Rasmussen} \& {Williams}(2006)}]{Rasmussen2006}
{Rasmussen}, C.~E., \& {Williams}, C. K.~I. 2006, {Gaussian Processes for
  Machine Learning} (MIT press Cambridge, MA)

\bibitem[{{Richard} {et~al.}(2012){Richard}, {Gordon}, {Rothman}, {Abel},
  {Frommhold}, {Gustafsson}, {Hartmann}, {Hermans}, {Lafferty}, {Orton},
  {Smith}, \& {Tran}}]{Richard2012}
{Richard}, C., {Gordon}, I.~E., {Rothman}, L.~S., {et~al.} 2012, \jqsrt, 113,
  1276, \dodoi{10.1016/j.jqsrt.2011.11.004}

\bibitem[{{Robinson} {et~al.}(2017){Robinson}, {Fortney}, \&
  {Hubbard}}]{Robinson2017}
{Robinson}, T.~D., {Fortney}, J.~J., \& {Hubbard}, W.~B. 2017, \apj, 850, 128,
  \dodoi{10.3847/1538-4357/aa951e}

\bibitem[{{Rodgers}(2000)}]{Rodgers2000}
{Rodgers}, C.~D. 2000, {Inverse Methods for Atmospheric Sounding: Theory and
  Practice} (WORLD SCIENTIFIC), \dodoi{10.1142/3171}

\bibitem[{{Rothman} {et~al.}(2010){Rothman}, {Gordon}, {Barber}, {Dothe},
  {Gamache}, {Goldman}, {Perevalov}, {Tashkun}, \& {Tennyson}}]{Rothman2010}
{Rothman}, L.~S., {Gordon}, I.~E., {Barber}, R.~J., {et~al.} 2010, \jqsrt, 111,
  2139, \dodoi{10.1016/j.jqsrt.2010.05.001}

\bibitem[{{Sarkis} {et~al.}(2018){Sarkis}, {Henning}, {K{\"u}rster},
  {Trifonov}, {Zechmeister}, {Tal-Or}, {Anglada-Escud{\'e}}, {Hatzes},
  {Lafarga}, {Dreizler}, {Ribas}, {Caballero}, {Reiners}, {Mallonn}, {Morales},
  {Kaminski}, {Aceituno}, {Amado}, {B{\'e}jar}, {Hagen}, {Jeffers},
  {Quirrenbach}, {Launhardt}, {Marvin}, \& {Montes}}]{Sarkis2018}
{Sarkis}, P., {Henning}, T., {K{\"u}rster}, M., {et~al.} 2018, \aj, 155, 257,
  \dodoi{10.3847/1538-3881/aac108}

\bibitem[{{Scalo} {et~al.}(2007){Scalo}, {Kaltenegger}, {Segura}, {Fridlund},
  {Ribas}, {Kulikov}, {Grenfell}, {Rauer}, {Odert}, {Leitzinger}, {Selsis},
  {Khodachenko}, {Eiroa}, {Kasting}, \& {Lammer}}]{Scalo2007}
{Scalo}, J., {Kaltenegger}, L., {Segura}, A.~G., {et~al.} 2007, Astrobiology,
  7, 85, \dodoi{10.1089/ast.2006.0125}

\bibitem[{{Seager} \& {Deming}(2010)}]{Seager2010}
{Seager}, S., \& {Deming}, D. 2010, \araa, 48, 631,
  \dodoi{10.1146/annurev-astro-081309-130837}

\bibitem[{{Seager} \& {Sasselov}(2000)}]{Seager2000}
{Seager}, S., \& {Sasselov}, D.~D. 2000, \apj, 537, 916, \dodoi{10.1086/309088}

\bibitem[{{Sedaghati} {et~al.}(2017){Sedaghati}, {Boffin}, {MacDonald},
  {Gandhi}, {Madhusudhan}, {Gibson}, {Oshagh}, {Claret}, \&
  {Rauer}}]{Sedaghati2017}
{Sedaghati}, E., {Boffin}, H. M.~J., {MacDonald}, R.~J., {et~al.} 2017, \nat,
  549, 238, \dodoi{10.1038/nature23651}

\bibitem[{{Seidel} {et~al.}(2020){Seidel}, {Ehrenreich}, {Pino}, {Bourrier},
  {Lavie}, {Allart}, {Wyttenbach}, \& {Lovis}}]{Seidel2020}
{Seidel}, J.~V., {Ehrenreich}, D., {Pino}, L., {et~al.} 2020, \aap, 633, A86,
  \dodoi{10.1051/0004-6361/201936892}

\bibitem[{{Shardanand} \& {Rao}(1977)}]{Rao1977}
{Shardanand}, \& {Rao}, A.~D.~P. 1977, {Absolute Rayleigh scattering cross
  sections of gases and freons of stratospheric interest in the visible and
  ultraviolet regions}, NASA technical note (National Aeronautics and Space
  Administration)

\bibitem[{{Sing} {et~al.}(2016){Sing}, {Fortney}, {Nikolov}, {Wakeford},
  {Kataria}, {Evans}, {Aigrain}, {Ballester}, {Burrows}, {Deming},
  {D{\'e}sert}, {Gibson}, {Henry}, {Huitson}, {Knutson}, {Lecavelier Des
  Etangs}, {Pont}, {Showman}, {Vidal-Madjar}, {Williamson}, \&
  {Wilson}}]{Sing2016}
{Sing}, D.~K., {Fortney}, J.~J., {Nikolov}, N., {et~al.} 2016, \nat, 529, 59,
  \dodoi{10.1038/nature16068}

\bibitem[{Sivia \& Skilling(2006)}]{Sivia2006}
Sivia, D., \& Skilling, J. 2006, Data analysis: a Bayesian tutorial (OUP
  Oxford)

\bibitem[{{Skilling}(2004)}]{Skilling2004}
{Skilling}, J. 2004, in American Institute of Physics Conference Series, Vol.
  735, American Institute of Physics Conference Series, ed. R.~{Fischer},
  R.~{Preuss}, \& U.~V. {Toussaint}, 395--405, \dodoi{10.1063/1.1835238}

\bibitem[{Skilling(2006)}]{Skilling2006}
Skilling, J. 2006, Bayesian Anal., 1, 833, \dodoi{10.1214/06-BA127}

\bibitem[{Sneep \& Ubachs(2005)}]{Sneep2005}
Sneep, M., \& Ubachs, W. 2005, Journal of Quantitative Spectroscopy and
  Radiative Transfer, 92, 293 ,
  \dodoi{https://doi.org/10.1016/j.jqsrt.2004.07.025}

\bibitem[{{Soboczenski} {et~al.}(2018){Soboczenski}, {Himes}, {O'Beirne},
  {Zorzan}, {Gunes Baydin}, {Cobb}, {Gal}, {Angerhausen}, {Mascaro}, {Arney},
  \& {Domagal-Goldman}}]{Soboczenski2018}
{Soboczenski}, F., {Himes}, M.~D., {O'Beirne}, M.~D., {et~al.} 2018, arXiv
  e-prints, arXiv:1811.03390.
\newblock \doarXiv{1811.03390}

\bibitem[{{Speagle}(2020)}]{Speagle2020}
{Speagle}, J.~S. 2020, \mnras, 493, 3132, \dodoi{10.1093/mnras/staa278}

\bibitem[{Tanner(1949)}]{Tanner1949}
Tanner, J.~M. 1949, Journal of Applied Physiology, 2, 1,
  \dodoi{10.1152/jappl.1949.2.1.1}

\bibitem[{{Tegmark} {et~al.}(2004){Tegmark}, {Strauss}, {Blanton}, {Abazajian},
  {Dodelson}, {Sandvik}, {Wang}, {Weinberg}, {Zehavi}, {Bahcall}, {Hoyle},
  {Schlegel}, {Scoccimarro}, {Vogeley}, {Berlind}, {Budavari}, {Connolly},
  {Eisenstein}, {Finkbeiner}, {Frieman}, {Gunn}, {Hui}, {Jain}, {Johnston},
  {Kent}, {Lin}, {Nakajima}, {Nichol}, {Ostriker}, {Pope}, {Scranton},
  {Seljak}, {Sheth}, {Stebbins}, {Szalay}, {Szapudi}, {Xu}, {Annis},
  {Brinkmann}, {Burles}, {Castand er}, {Csabai}, {Loveday}, {Doi}, {Fukugita},
  {Gillespie}, {Hennessy}, {Hogg}, {Ivezi{\'c}}, {Knapp}, {Lamb}, {Lee},
  {Lupton}, {McKay}, {Kunszt}, {Munn}, {O'Connell}, {Peoples}, {Pier},
  {Richmond}, {Rockosi}, {Schneider}, {Stoughton}, {Tucker}, {vanden Berk},
  {Yanny}, \& {York}}]{Tegmark2004}
{Tegmark}, M., {Strauss}, M.~A., {Blanton}, M.~R., {et~al.} 2004, \prd, 69,
  103501, \dodoi{10.1103/PhysRevD.69.103501}

\bibitem[{Thalman {et~al.}(2014)Thalman, Zarzana, Tolbert, \&
  Volkamer}]{Thalman2014}
Thalman, R., Zarzana, K.~J., Tolbert, M.~A., \& Volkamer, R. 2014, Journal of
  Quantitative Spectroscopy and Radiative Transfer, 147, 171 ,
  \dodoi{https://doi.org/10.1016/j.jqsrt.2014.05.030}

\bibitem[{{Trotta}(2008)}]{Trotta2008}
{Trotta}, R. 2008, Contemporary Physics, 49, 71,
  \dodoi{10.1080/00107510802066753}

\bibitem[{{Trotta}(2017)}]{Trotta2017}
---. 2017, arXiv e-prints, arXiv:1701.01467.
\newblock \doarXiv{1701.01467}

\bibitem[{{Tsiaras} {et~al.}(2019){Tsiaras}, {Waldmann}, {Tinetti}, {Tennyson},
  \& {Yurchenko}}]{Tsiaras2019}
{Tsiaras}, A., {Waldmann}, I.~P., {Tinetti}, G., {Tennyson}, J., \&
  {Yurchenko}, S.~N. 2019, Nature Astronomy, 3, 1086,
  \dodoi{10.1038/s41550-019-0878-9}

\bibitem[{{Virtanen} {et~al.}(2020){Virtanen}, {Gommers}, {Oliphant},
  {Haberland}, {Reddy}, {Cournapeau}, {Burovski}, {Peterson}, {Weckesser},
  {Bright}, {van der Walt}, {Brett}, {Wilson}, {Jarrod Millman}, {Mayorov},
  {Nelson}, {Jones}, {Kern}, {Larson}, {Carey}, {Polat}, {Feng}, {Moore}, {Vand
  erPlas}, {Laxalde}, {Perktold}, {Cimrman}, {Henriksen}, {Quintero}, {Harris},
  {Archibald}, {Ribeiro}, {Pedregosa}, {van Mulbregt}, \&
  {Contributors}}]{scipy}
{Virtanen}, P., {Gommers}, R., {Oliphant}, T.~E., {et~al.} 2020, Nature
  Methods, 17, 261, \dodoi{https://doi.org/10.1038/s41592-019-0686-2}

\bibitem[{{von Essen} {et~al.}(2019){von Essen}, {Mallonn}, {Welbanks},
  {Madhusudhan}, {Pinhas}, {Bouy}, \& {Weis Hansen}}]{vonEssen2019}
{von Essen}, C., {Mallonn}, M., {Welbanks}, L., {et~al.} 2019, \aap, 622, A71,
  \dodoi{10.1051/0004-6361/201833837}

\bibitem[{{Wakeford} {et~al.}(2017){Wakeford}, {Sing}, {Kataria}, {Deming},
  {Nikolov}, {Lopez}, {Tremblin}, {Amundsen}, {Lewis}, {Mandell}, {Fortney},
  {Knutson}, {Benneke}, \& {Evans}}]{Wakeford2017}
{Wakeford}, H.~R., {Sing}, D.~K., {Kataria}, T., {et~al.} 2017, Science, 356,
  628, \dodoi{10.1126/science.aah4668}

\bibitem[{{Wakeford} {et~al.}(2018){Wakeford}, {Sing}, {Deming}, {Lewis},
  {Goyal}, {Wilson}, {Barstow}, {Kataria}, {Drummond}, {Evans}, {Carter},
  {Nikolov}, {Knutson}, {Ballester}, \& {Mand ell}}]{Wakeford2018}
{Wakeford}, H.~R., {Sing}, D.~K., {Deming}, D., {et~al.} 2018, \aj, 155, 29,
  \dodoi{10.3847/1538-3881/aa9e4e}

\bibitem[{{Waldmann}(2016)}]{Waldmann2016}
{Waldmann}, I.~P. 2016, \apj, 820, 107, \dodoi{10.3847/0004-637X/820/2/107}

\bibitem[{{Waldmann} {et~al.}(2015{\natexlab{a}}){Waldmann}, {Rocchetto},
  {Tinetti}, {Barton}, {Yurchenko}, \& {Tennyson}}]{Waldmann2015b}
{Waldmann}, I.~P., {Rocchetto}, M., {Tinetti}, G., {et~al.} 2015{\natexlab{a}},
  \apj, 813, 13, \dodoi{10.1088/0004-637X/813/1/13}

\bibitem[{{Waldmann} {et~al.}(2015{\natexlab{b}}){Waldmann}, {Tinetti},
  {Rocchetto}, {Barton}, {Yurchenko}, \& {Tennyson}}]{Waldmann2015a}
{Waldmann}, I.~P., {Tinetti}, G., {Rocchetto}, M., {et~al.} 2015{\natexlab{b}},
  \apj, 802, 107, \dodoi{10.1088/0004-637X/802/2/107}

\bibitem[{{Welbanks} \& {Madhusudhan}(2019)}]{Welbanks2019a}
{Welbanks}, L., \& {Madhusudhan}, N. 2019, \aj, 157, 206,
  \dodoi{10.3847/1538-3881/ab14de}

\bibitem[{{Welbanks} {et~al.}(2019){Welbanks}, {Madhusudhan}, {Allard},
  {Hubeny}, {Spiegelman}, \& {Leininger}}]{Welbanks2019b}
{Welbanks}, L., {Madhusudhan}, N., {Allard}, N.~F., {et~al.} 2019, \apjl, 887,
  L20, \dodoi{10.3847/2041-8213/ab5a89}

\bibitem[{{Woitke} {et~al.}(2018){Woitke}, {Helling}, {Hunter}, {Millard},
  {Turner}, {Worters}, {Blecic}, \& {Stock}}]{Woitke2018}
{Woitke}, P., {Helling}, C., {Hunter}, G.~H., {et~al.} 2018, \aap, 614, A1,
  \dodoi{10.1051/0004-6361/201732193}

\bibitem[{{Wunderlich} {et~al.}(2019){Wunderlich}, {Godolt}, {Grenfell},
  {St{\"a}dt}, {Smith}, {Gebauer}, {Schreier}, {Hedelt}, \&
  {Rauer}}]{Wunderlich2019}
{Wunderlich}, F., {Godolt}, M., {Grenfell}, J.~L., {et~al.} 2019, \aap, 624,
  A49, \dodoi{10.1051/0004-6361/201834504}

\bibitem[{{Wyttenbach} {et~al.}(2015){Wyttenbach}, {Ehrenreich}, {Lovis},
  {Udry}, \& {Pepe}}]{Wyttenbach2015}
{Wyttenbach}, A., {Ehrenreich}, D., {Lovis}, C., {Udry}, S., \& {Pepe}, F.
  2015, \aap, 577, A62, \dodoi{10.1051/0004-6361/201525729}

\bibitem[{{Yurchenko} {et~al.}(2011){Yurchenko}, {Barber}, \&
  {Tennyson}}]{Yurchenko2011}
{Yurchenko}, S.~N., {Barber}, R.~J., \& {Tennyson}, J. 2011, \mnras, 413, 1828,
  \dodoi{10.1111/j.1365-2966.2011.18261.x}

\bibitem[{{Yurchenko} \& {Tennyson}(2014)}]{Yurchenko2014}
{Yurchenko}, S.~N., \& {Tennyson}, J. 2014, \mnras, 440, 1649,
  \dodoi{10.1093/mnras/stu326}

\bibitem[{{Zalesky} {et~al.}(2019){Zalesky}, {Line}, {Schneider}, \&
  {Patience}}]{Zalesky2019}
{Zalesky}, J.~A., {Line}, M.~R., {Schneider}, A.~C., \& {Patience}, J. 2019,
  \apj, 877, 24, \dodoi{10.3847/1538-4357/ab16db}

\bibitem[{{Zhang} {et~al.}(2020){Zhang}, {Chachan}, {Kempton}, {Knutson},
  {Wenjun}, \& {Chang}}]{Zhang2020}
{Zhang}, M., {Chachan}, Y., {Kempton}, E. M.~R., {et~al.} 2020, arXiv e-prints,
  arXiv:2004.09513.
\newblock \doarXiv{2004.09513}

\bibitem[{{Zhang} {et~al.}(2019){Zhang}, {Chachan}, {Kempton}, \&
  {Knutson}}]{Zhang2019}
{Zhang}, M., {Chachan}, Y., {Kempton}, E. M.~R., \& {Knutson}, H.~A. 2019,
  \pasp, 131, 034501, \dodoi{10.1088/1538-3873/aaf5ad}

\bibitem[{{Zhang}(2020)}]{ZhangXi2020}
{Zhang}, X. 2020, arXiv e-prints, arXiv:2006.13384.
\newblock \doarXiv{2006.13384}

\bibitem[{{Zingales} \& {Waldmann}(2018)}]{Zingales2018}
{Zingales}, T., \& {Waldmann}, I.~P. 2018, \aj, 156, 268,
  \dodoi{10.3847/1538-3881/aae77c}

\end{thebibliography}

\end{document}